\documentclass[a4paper,11pt]{article}
\pdfoutput=1 

\usepackage{jheppub} 

\usepackage[T1]{fontenc} 
\usepackage[utf8]{inputenc}
\usepackage{hyperref}
\usepackage{bm}
\usepackage{xcolor}
\usepackage{multirow}
\newcommand{\pr}[2]{\frac{\partial #1}{\partial #2}}
\newcommand{\prtot}[2]{\frac{d #1}{d #2}}
\newcommand{\abs}[1]{\lvert #1 \rvert}
\def\L{\Lambda}

\DeclareUnicodeCharacter{2212}{-}

\title{\boldmath Transverse $\Lambda$ polarization in $e^+e^-$ processes within a TMD factorization approach and the polarizing fragmentation function
}

\author[a,b,1]{Umberto D'Alesio,\note{Corresponding author.}}
\author[c]{Leonard Gamberg,}
\author[b]{Francesco Murgia}
\author[a,b]{and Marco Zaccheddu}

\affiliation[a]{Dipartimento di Fisica, Università di Cagliari, Cittadella Univ., I-09042 Monserrato (CA), Italy}
\affiliation[b]{INFN, Sezione di Cagliari, Cittadella Univ., I-09042 Monserrato (CA), Italy }
\affiliation[c]{Division of Science, Penn State Berks, Reading, PA 19610, USA }

\emailAdd{umberto.dalesio@ca.infn.it}
\emailAdd{lpg10@psu.edu}
\emailAdd{francesco.murgia@ca.infn.it}
\emailAdd{marco.zaccheddu@ca.infn.it}

\abstract{We perform a re-analysis of Belle data for the transverse $\Lambda$ and $\bar\Lambda$ polarization in $e^+ e^-$ annihilation processes within a transverse momentum dependent (TMD) factorization approach.
We consider two data sets, one referring to the associated production of $\Lambda$'s with a light unpolarized hadron in an almost back-to-back configuration, and one for the inclusive $\Lambda$ production, with the reconstruction of the thrust axis.
We adopt the Collins-Soper-Sterman framework and employ the recent formulation on the factorization of single-inclusive hadron production.
This extends a previous phenomenological study carried out in a more simplified TMD approach, leading to a new extraction of the polarizing fragmentation function (FF). While confirming several features of the previous analysis, here we include the proper QCD scale dependence of this TMD FF and carefully exploit the role of its nonperturbative component. The compatibility, within a unique TMD factorization scheme, of double and single-inclusive hadron production is discussed in detail. Moreover,  we consider another data set for inclusive $\Lambda$ production, at much larger energies, from the OPAL Collaboration, in order to test the consistency of the entire approach. Finally, we elaborate on some fundamental issues like the role of $SU(2)$ isospin symmetry and the heavy quark contributions.}

\begin{document}
\maketitle
\flushbottom

\section{Introduction}
\label{sec:introduction}

One of the outstanding challenges in QCD theory and phenomenology has been to describe and predict the observed large transverse single spin asymmetries (TSSAs)  in deep inelastic scattering (DIS) processes using factorization theorems derived within perturbative QCD.

During the mid 70s, significant TSSAs in inclusive pion production in proton-proton collisions, at center of mass (cm) energy of few GeV, were observed at the Argonne Laboratory synchrotron~\cite{Dick:1975ty,Klem:1976ui,Dragoset:1978gg},
while during the same period at Fermilab, $\Lambda$-hyperons produced in unpolarized proton-nuclear collisions at $\sqrt{s}\approx$ 24 GeV and moderate transverse momentum, $P_T$ (below $\sim 1.5$~GeV), displayed large transverse polarization with respect to the production plane~\cite{Bunce:1976yb,Schachinger:1978qs,Heller:1978ty}. Large TSSAs (approx 30-40\% ) continued to be observed in the 90s at Fermilab~\cite{Adams:1991cs,Adams:1991ru,Adams:1991rx,Bravar:1996ki} in pseudo-scalar meson production with center of mass energies of $\sqrt{s}\approx 20$~GeV.
These results were confirmed by the STAR, PHENIX and BRAHMS Collaborations at the
Relativistic Heavy Ion Collider (RHIC), at cm energies up to  $\sqrt{s}\approx 500$~GeV covering a wide range in Feynman $x_F=2 P_L/\sqrt{s}$ (where $P_L$ is the longitudinal momentum of the final hadron) and $P_T$~\cite{Adams:2003fx,Adams:2006uz,PHENIX:2003qdw,Adler:2005in,:2008mi,BRAHMS:2007tyt}.  
More striking, is the transverse polarization data for $\Lambda$-hyperon production from {\em unpolarized} hadron collisions.  Along with followup experiments from
Fermi-lab~\cite{Lundberg:1989hw,Ramberg:1994tk},
fixed target measurements of this reaction were reported by the NA48 Collaboration~\cite{Fanti:1998px} and the HERA-B Collaboration~\cite{Abt:2006da}. At CERN a $\L$ polarization of approximately 35\% was also measured in $pp$ collisions at the ISR at  cm energies of $\sqrt{s}=52$ and 63~GeV~\cite{Erhan:1979xm}.

From the theory side we know that large transverse polarization effects cannot be explained within the collinear QCD factorization at leading twist which, at large enough $P_T$, predicts negligible values~\cite{Kane:1978nd}. Presumably then hyperon polarization, observed in unpolarized collisions, necessarily has to originate from nonperturbative effects during the hadronization process, as they are produced from parity conserving strong interaction and in turn undergo self-analyzing weak decay. For this reason a study of the $\L$ polarization enables us to obtain important information on this nonperturbative mechanism.


Within the context of QCD factorization theorems the hadronization of partons is described in terms of nonperturbative matrix elements of QCD operators which can be fitted to experimental data. This is a challenging endeavour on the basis of data taken from nucleon-nuclear scattering experiments, since these processes are mediated solely by the strong force where an analytical description is complicated due to competing effects that enter QCD factorization formulas for spin dependent $pp$ and/or $pA$ reactions.

A simplification emerges for processes involving electromagnetic interactions such as in semi-inclusive deep inelastic scattering (SIDIS) where polarized $\L$'s can be produced in $ep\rightarrow e\L + X$ or in quasi-real photoproduction. Such experimental studies were carried out by the HERMES collaboration~\cite{HERMES:2006lro, HERMES:2007fpi,HERMES:2014fmx} as well as in neutrino-nucleon scattering by the NOMAD Collaboration~\cite{NOMAD:2000wdf,NOMAD:2001iup}.

Probably the most direct process to gain access to (un)polarized $\L$ fragmentation functions (FFs) is single and/or semi-inclusive hadron production electron-positron annihilation (SIA). Recently, the Belle Collaboration~\cite{Guan:2018ckx} has collected data for the transverse polarization of $\Lambda$-hyperons produced together with light mesons in an almost back-to-back configuration as well as for Lambda’s inclusively produced, where the $\Lambda$ transverse momentum is measured with respect to the thrust axis.  Also, data on polarized $\Lambda$ fragmentation in this reaction has been provided by the OPAL Collaboration at LEP~\cite{OPAL:1997oem}. This measurement was performed on the $Z$-pole, that is at the cm energy equal to the mass of the $Z$ boson.

In this paper, we investigate associated production of transversely polarized $\Lambda$-hyperons in the process $e^+ e^-\rightarrow \Lambda^\uparrow (\pi /K) + X$
as well as its inclusive production in $e^+ e^-\rightarrow \Lambda^\uparrow (\rm{jet}) + X$.
We present a renewed analysis of Belle data by exploiting the CSS evolution equations and the recent theory developments on the factorization of single-inclusive hadron production in  $e^+ e^-$ annihilation processes~\cite{Kang:2020yqw,Gamberg:2021iat}.

A critical issue in
the first  phenomenological studies~\cite{DAlesio:2020wjq,Callos:2020qtu} of the $\Lambda$ polarizing FFs from the analysis of Belle data is that the
extraction of these  TMD FFs  was carried out at fixed scale ($Q=10.58\,\text{GeV}$).  Thus these studies do not  employ  TMD evolution.   Moreover, and more relevant, in Ref.~\cite{DAlesio:2020wjq} we used a simplified and phenomenological model to study the transversely polarized $\Lambda$  produced in a single-inclusive process\footnote{We notice that a first attempt, within the same simplified model, to describe the transverse $\Lambda$ polarization in unpolarized $pp$ collisions was carried out in Ref.~\cite{Anselmino:2000vs}.}, due to the lack of a generalized TMD factorization formalism.
Indeed,  this approach does not enable one to make predictions at different energy scales. For that, one should employ  TMD factorization theorems, and consequently  proper
evolution equations for both the double and the single-inclusive hadron production cross sections.

In fact, unlike the cross section for double-hadron production in $e^+e^-$ annihilation processes, only recently new advancements in the TMD factorization of the cross section for single-inclusive production processes have appeared. Among them, are the works of Refs.~\cite{Makris:2020ltr,Kang:2020yqw}, where the factorization has been formulated within an effective theory context, and where a first phenomenological analysis of the Belle data based on the latter paper was carried out in~\cite{Gamberg:2021iat}. Moreover, a Collins-Soper-Sterman (CSS) formalism has been adopted in Refs.~\cite{Boglione:2020auc,Boglione:2020cwn,Boglione:2021wov}.

The purpose of this study is therefore to present a renewed analysis of Belle data by exploiting the TMD framework in its full glory, paying special attention to scale evolution effects and to the nonperturbative component of the polarizing FF. We will also touch a couple of fundamental issues, namely the $SU(2)$ isospin symmetry and the role of heavy quark contributions.

The paper is organised as follows: in Section~\ref{sec:2_h_prod} we present the main  formulas and the cross section for the production of a transversely polarized spin-$1/2$ hadron, in association with a light hadron, in $e^+e^-$ annihilation processes, and their expressions in the impact parameter space. Then in Section~\ref{sec:CSS_RG_eqs} we show how these convolutions can be treated within TMD factorization, by employing the CSS evolution equations. In Section~\ref{sec:1-h_prod} we summarize some useful results already presented in Ref.~\cite{Kang:2020yqw}, giving expressions for the cross sections for single-inclusive hadron production and for the transverse polarization.
All these results will be exploited to re-analyze the Belle data in Section~\ref{sec:analysis}, where we show the outcomes of the fits for the double-hadron production data alone and the combined fit of both data set, discussing our main findings. Here we also consider the role of OPAL data, checking our predictions (based only on Belle data) against them or including them in a global fit. Lastly, in Section~\ref{sec:conclusions} we collect our concluding remarks. 
\section{Double-hadron production}
\label{sec:2_h_prod}
We start considering the process $e^+e^-\to h_1(S_1)\, h_2(S_2) +X$ where $h_1$ and $h_2$ are two spin-1/2 hadrons, with spin polarization vectors $S_{1,2}$ and masses $M_{1,2}$, produced almost back-to-back in the center-of-mass frame of the incoming leptons. For more details we refer to Ref.~\cite{DAlesio:2021dcx}.

\subsection{Kinematics and cross section}
\label{subsec:kin}

\begin{figure}[!th]
\centering
\includegraphics[width=13cm]{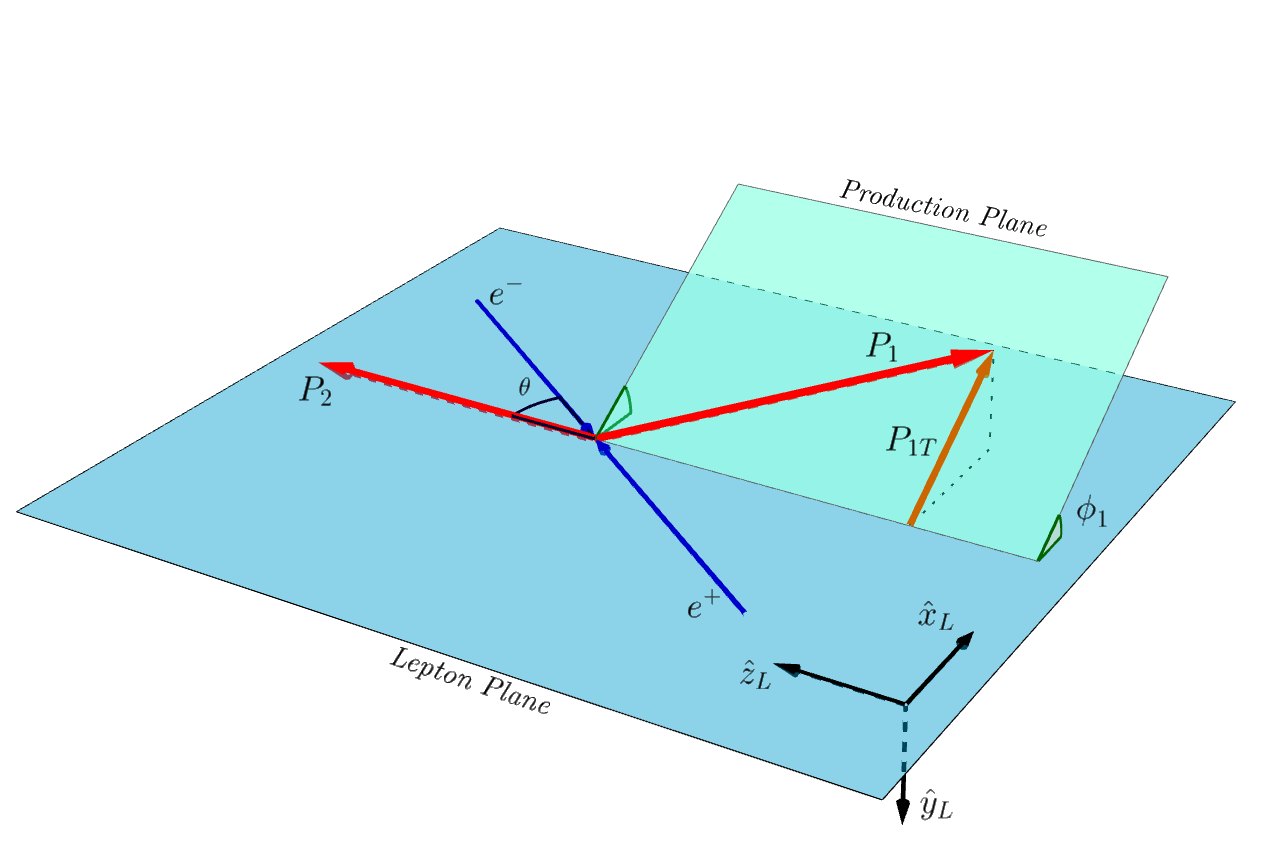}
    \caption{Kinematics for the process $e^+e^-\to h_1\, h_2 +X$ in the hadron-frame configuration.}
    \label{fig:kin2_2}
\end{figure}

We adopt the following kinematics set-up: we fix the $\hat{z}_L$ axis (in a \emph{laboratory} frame) along the momentum of the second hadron, $h_2$, with the first one, $h_1$, moving in the opposite hemisphere, with a small transverse momentum $\bm{P}_{1T}$ with respect to the second hadron direction. This is illustrated in Fig.~\ref{fig:kin2_2}, where $(\hat{\bm{x}}_L,\hat{\bm{y}}_L,\hat{\bm{z}}_L)$ are the unit vectors in the laboratory frame and $\bm{P}_{1}$ and $\bm{P}_{2}$ are the momenta of, respectively, the first and second hadron. %

We notice that the frame adopted here has the unit vectors, $
\hat{y}_L, \hat{z}_L$, inverted with respect to those of the hadron frame used in~\cite{DAlesio:2021dcx}. This choice allows us to employ directly the convolutions adopted in Ref.~\cite{Boer:1997mf} (see also Ref.~\cite{Pitonyak:2013dsu}), with a direct connection with the convolutions in $\bm{b}_T$-space (see below).

We can define two planes: the \emph{Lepton Plane}, determined by the leptons and the hadron $h_2$ momenta, and the \emph{Production Plane}, determined by the momenta of the two observed hadrons, $h_{1,2}$, at an angle $\phi_1$ with respect to the \emph{Lepton Plane}.
%
%
In this configuration, referred to as the hadron frame, one measures only the momenta of the two hadrons and the azimuthal distribution of the hadron $h_1$ around the hadron $h_2$ direction. No information on the original quark-antiquark direction is required.

From the theoretical point of view, it is however more convenient to adopt yet a different frame, where the two hadrons are back to back, along a new $\hat{\bm{z}}$ axis, and the hadron transverse unbalance is now carried out by the virtual photon. All details of this transformation are given in Appendix~\ref{apx:rotations}. In this frame the differential cross section can be expressed as~\cite{Boer:1997mf}
\begin{equation}
     \frac{d\sigma^{e^+e^-\to h_1(S_1) h_2(S_2)\, X}}{2dy dz_1dz_2d^2\bm{q}_{T}}\,,
\label{eq:xsec2h}
\end{equation}
where $\bm{q}_{T}$ is the transverse momentum of the virtual photon. This is related to the transverse momentum of the first hadron as follows:
\begin{equation}
    \bm{P}_{1T} = - z_1 \bm{q}_T\,.
\label{eq:qt_P1T}
\end{equation}
The two scaling variables in Eq.~(\ref{eq:xsec2h}) are the usual light-cone momentum fractions $z_{1,2}$ of the final-state hadrons, defined as
\begin{equation}
    z_1 = \frac{P^-_{1}}{k^-}, \qquad z_2 = \frac{P^+_{2}}{p^+}\,,
\label{eq:light_con_def}
\end{equation}
where $k$ and $p$ are the four-momenta of the quark and the antiquark, fragmenting with a certain transverse momentum $\bm{k}_\perp$ and $\bm{p}_\perp$, respectively, into the hadron $h_1$ and $h_2$. 
These scaling variables are in turn directly related to another set of variables, usually adopted in experimental analyses: the energy fraction
\begin{equation}
 z_{h}  = \frac{E_h}{E_q} =\frac{2E_h}{Q} \simeq z \Bigg( 1 + \frac{M^2_{h}}{ z^2 {Q}^2}\Bigg)\,,
\label{eq:en_fract}
\end{equation}
and the momentum fraction
\begin{equation}
    \begin{split}
        z_{p} =\frac{|\bm{P}_h|}{E_q} =\frac{2|\bm{P}_h|}{Q} \simeq  z \Bigg( 1 - \frac{M^2_{h}}{ z^2 {Q}^2}\Bigg)\,,
    \end{split}
\label{eq:long_fract}
\end{equation}
where $Q$ is the cm energy of the process, $Q^2 = q^2$, with $q$ being the four momentum of the virtual photon. The last equalities are obtained neglecting powers of $\bm{k}_\perp^2/Q^2$. Notice that the variable $z_h$, usually defined also as an invariant,  $z_h= 2P_h\cdot q/ Q^2$, coincides with the energy fraction above in the hadron frame.

Finally, the fraction $y$ is defined as $y=P_2\cdot k_{e^+} /P_2\cdot q$ (with $k_{e^+}$ being the $e^+$ momentum), that in the hadron frame reduces to
\begin{equation}
y = \Big(1 - \sqrt{1-4M_2^2/z_{h_2}^2Q^2}\cos\theta\Big)/2 \simeq (1 - \cos\theta)/2\,,
\end{equation}
where $\theta$ is the angle between the hadron 2 momentum and the incoming lepton directions (see Fig.~\ref{fig:kin2_2}).
Notice that in all relevant variables we will keep kinematic corrections in $M^2/Q^2$, useful for the study of massive hadron production. On the other hand, as we will see below, the $y$ (or the $\theta$) dependence will not play any direct role in our analysis.

In general, the cross sections can be written in terms of convolutions of two generic TMD fragmentation functions \cite{DAlesio:2021dcx, Boer:1997mf}, defined as follows
\begin{equation}
    \mathcal{F}[\omega  D \bar{D}] = \sum_q e^2_q \int d^2\bm{k}_T d^2\bm{p}_T \, \delta^{(2)}(\bm{k}_T + \bm{p}_T - \bm{q}_T) \,\omega(\bm{k}_T, \bm{p}_T) D(z_1,\bm{k}_{\bot}) \bar{D}(z_2,\bm{p}_{\bot}) \,,
\label{eq:conv_general}
\end{equation}
where $\omega$ is a suitable weight-factor depending on the two transverse momenta, $D$ and $\bar{D}$ are the TMD-FFs of the first and second hadron and $\bm{k}_{T}$ and $\bm{p}_{T}$ are the transverse momenta of the quark/antiquark with respect to the hadron $h_1$ and $h_2$. Notice that one can easily relate these transverse momenta to $\bm{k}_{\bot}$ and $\bm{p}_{\bot}$, the transverse momenta of the hadrons with respect to their own parent quarks as follows (see Appendix \ref{apx:rotations})
\begin{equation}
    \bm{k}_T = -\frac{\bm{k}_{\bot}}{z_{p_1}}; \qquad \bm{p}_T = -\frac{\bm{p}_{\bot}}{z_{p_2}}\,.
\end{equation}

\subsection{Transversely polarized hadron production}
\label{subsec:pol_definition}
We now consider specifically the associated production of a transversely polarized spin-$1/2$ hadron, $h_1$, with an unpolarized hadron, $h_2$.
If the polarization is measured only as a function of the hadron energy fractions, with the proper use of Eqs. (\ref{eq:en_fract}) and (\ref{eq:long_fract}), we can give it as the ratio of two $\bm{q}_T$-integrated convolutions
\begin{equation}
    P^{h_1}_T(z_1,z_2) = - \sin(\phi_1 - \phi_{S_1}) \frac{ \int d^2\bm{q}_T \, F^{\sin(\phi_1 - \phi_{S_1})}_{TU} }{ \int d^2\bm{q}_T \, F_{UU}} \,,
\label{pol_qt}
\end{equation}
where $\phi_{S_1}$ is the azimuthal angle of the spin of the hadron $h_1$ and where we have simplified a common factor coming from the hard partonic subprocess (see below).
The two convolutions are defined as follows:
\begin{eqnarray}
F_{UU} & = & \mathcal{F}[D_1 \bar{D}_1] \\
F^{\sin(\phi_1 - \phi_{S_1})}_{TU} &=& \mathcal{F}\bigg[\frac{\hat{\bm{h}}\cdot \bm{k}_T}{M_{1}}D^{\perp}_{1T} \bar{D}_1\bigg]\,,
\label{eq:main_pol_conv}
\end{eqnarray}
where $D_1(z,{k}_{\bot})$ is the unpolarized TMD fragmentation function and $D^{\perp}_{1T}(z,{k}_{\bot})$ is the polarizing FF, with 
$\hat{\bm{h}}= \bm{P}_{1T}/|\bm{P}_{1T}|$.
Notice that there is another common notation for the polarizing FF, related to the probability that an unpolarized quark fragments into a transversely polarized spin-1/2 hadron~\cite{Bacchetta:2004jz}:
\begin{equation}
    \Delta^{N}\! D_{h^\uparrow/q}(z,{k}_\perp) = \frac{{k}_\perp}{zM_h} D_{1T}^\perp (z,{k}_\perp)\,.
\end{equation}

When the polarization is measured perpendicularly to the production plane, that is along the unit vector $\hat{\bm{n}}$ defined as:
\begin{equation}
    \hat{\bm{n}} \equiv (\cos\phi_n, \sin\phi_n,0)= \frac{- \bm{P}_{2}  \times  \bm{P}_{1}}{|\bm{P}_{2}  \times  \bm{P}_{1}|} = -\sin\phi_1 \hat{\bm{x}}_L + \cos\phi_1 \hat{\bm{y}}_L\,,
\end{equation}
the factor entering Eq.~(\ref{pol_qt}) simplifies as~\cite{DAlesio:2021dcx}
\begin{equation}
- \sin(\phi_1 - \phi_{S_{1}}) = 1  \,.
\end{equation}
Generally, one uses the TMD fragmentation functions in the conjugate $\bm{b}_T$-space.  
More precisely, the Fourier transform of the unpolarized FF is defined as:
\begin{equation}
\begin{split}
\widetilde{D}_1(z,b_T) = & \int d^2\bm{k}_T \, e^{i\bm{b}_T \cdot \bm{k}_T } D_1(z,{k}_{\perp})
= 2 \pi \int dk_T \, k_T J_0(b_T k_T) D_1(z,k_{\perp}) \,,
\end{split}
\end{equation}

\noindent where we have used Eq.~(\ref{bessel_0}), the integral definition of  $J_0$, the Bessel function of the first kind of order zero. With the above relation, the $F_{UU}$ convolution in $\bm{b}_T $-space can be written as:
\begin{equation}
F_{UU} = \mathcal{F}[D_1 \bar{D}_1] = \mathcal{B}_0 \Big[\widetilde{D} \widetilde{\bar{D}}\Big]
    = \sum_q e^2_q \int \frac{d b_T}{(2 \pi)} \, b_T J_0(b_T\, q_T) \widetilde{D}_1(z_1,b_T) \widetilde{\bar{D}}_1(z_2,b_T)\,.
\end{equation}
Regarding the $\bm{b}_T $-space convolution of $F^{\sin(\phi_1 - \phi_{S})}_{TU}$, we first define the Fourier transform of the product of the polarizing fragmentation function with $k^i_T$, the $i$-th component of the quark transverse momentum with respect to the hadron direction, see Appendix~\ref{apx2:conv_FT}:
\begin{equation}
    \int d^2\bm{k}_T \, \frac{k^i_T}{M_{1}} e^{i\bm{b}_T \cdot \bm{k}_T} D^{\perp}_{1T}(z,k_{\perp})
    = i b^i_T M_{1} \widetilde{D}^{\perp  (1)}_{1T}(z,b_T) \,,
\end{equation}
where $\widetilde{D}^{\perp  (1)}_{1T}(z,b_T)$, the first moment of the polarizing fragmentation function in $\bm{b}_T $-space, is defined as
\begin{equation}
    \widetilde{D}^{\perp  (1)}_{1T}(z,b_T) = -\frac{2}{M^2_{1}}\pr{}{b^2_T} \widetilde{D}^{\perp}_{1T}(z,b_T) \; ,
\label{eq:bt_first_mom_1}
\end{equation}
and $ \widetilde{D}^{\perp}_{1T}(z,b_T)$ is the Fourier transform of the polarizing FF, Eq. (\ref{eq:pFF_bt}).
Notice that the first moment of the polarizing FF in $\bm{k}_T $-space, ${D}^{\perp  (1)}_{1T}(z)$, defined as
\begin{equation}
    {D}^{\perp  (1)}_{1T}(z) = \int d^2 \bm{k}_{\perp} \, \bigg(\frac{\bm{k}^2_{\perp}}{2 z^2 M^2_h} \bigg)  {D}^{\perp}_{1T}(z,{k}_{\perp}) \,,
\end{equation}
can be related to the corresponding one in $\bm{b}_T$-space, Eq.~(\ref{eq:bt_first_mom_1}), as follows:
\begin{equation}
    \lim_{b_T \to 0} \widetilde{D}^{\perp  (1)}_{1T}(z,b_T) =\frac{1}{z^2} {D}^{\perp  (1)}_{1T}(z)\; .
    \label{eq:firstmoment_z}
\end{equation}
Employing the above equations and using the integral definition of the Bessel function $J_1$, Eq.~(\ref{bessel_1}), we can find the expression of  $F^{\sin(\phi_1 - \phi_{S_1})}_{TU}$  in  $\bm{b}_T $-space:
\begin{eqnarray}
 F^{\sin(\phi_1 - \phi_{S_1})}_{TU} & = &
 \mathcal{F}\bigg[\frac{\hat{\bm{h}}\cdot \bm{k}_T}{M_{1}}D^{\perp}_{1T} \bar{D}_1\bigg] =  M_{1}  \mathcal{B}_1 \Big[\widetilde{D}^{\perp  (1)}_{1T} \widetilde{\bar{D}}_1\Big] \nonumber\\
    & = & M_{1} \sum_q e^2_q \int \frac{d b_T}{2 \pi} \,b^2_T J_1(b_T\,q_T)  \widetilde{D}^{\perp  (1)}_{1T}(z_1,b_T) \widetilde{\bar{D}}_1(z_2,b_T)\,.
\end{eqnarray}
Finally, we can express the polarization of the final hadron, Eq.~(\ref{pol_qt}), along the $\hat{\bm{n}}$ direction as the ratio of the two convolutions in $\bm{b}_T $-space:
\begin{equation}
    P^{h_1}_n(z_1,z_2) =  \frac{ \int d^2\bm{q}_T \, F^{\sin(\phi_1 - \phi_{S_1})}_{TU} }{ \int d^2\bm{q}_T \, F_{UU}} = \frac{M_{1} \int dq_T\;q_T\;d\phi_1 \, \mathcal{B}_1 \Big[\widetilde{D}^{\perp  (1)}_{1T} \widetilde{\bar{D}}_1\Big]}{ \int dq_T\;q_T\;d\phi_1 \, \mathcal{B}_0 \Big[\widetilde{D}_1 \widetilde{\bar{D}}_1\Big] }\,,
\label{pol_ratio}
\end{equation}
where
\begin{align}
    \mathcal{B}_0 \Big[\widetilde{D}_1 \widetilde{\bar{D}}_1\Big] =& \sum_q e^2_q\int \frac{d b_T}{2 \pi} \, b_T J_0(b_T\, q_T) \widetilde{D}_1(z_1,b_T) \widetilde{\bar{D}}_1(z_2,b_T)\label{eq:all_convs1} \\
    \mathcal{B}_1 \Big[\widetilde{D}^{\perp  (1)}_{1T} \widetilde{\bar{D}}_1\Big] =& \sum_q e^2_q \int \frac{d b_T}{2 \pi} \,b^2_T J_1( b_T\,q_T)  \widetilde{D}^{\perp  (1)}_{1T}(z_1,b_T) \widetilde{\bar{D}}_1(z_2,b_T)\label{eq:all_convs2}\,.
\end{align}
The last step is to integrate both convolutions on $\bm{q}_T $. %
The integration over the azimuthal angle, $\phi_1 $, is trivial, giving a factor of $2\pi$ that cancels out in the ratio. %
Moreover, since the only terms inside the convolutions depending on $q_T $ are the Bessel functions, we can separately integrate them, obtaining
\begin{equation}
    \int^{q_{T_{\text{max}}}}_0 dq_T \, q_T J_0(b_T\, q_T) = \frac{q_{T_{\text{max}}}}{b_T}J_1(b_T\, q_{T_{\rm max}}) \\
\label{int_qtmax_1}
\end{equation}
\begin{equation}
    %
    %
    \int^{q_{T_{\text{max}}}}_0 dq_T \, q_T J_1(b_T\, q_T) =\frac{\pi q_{T_{\text{max}}}}{2 b_T}
    \{J_1(b_T\, q_{T_{\text{max}}})\bm{H}_0(b_T\, q_{T_{\rm max}}) - J_0(b_T\, q_{T_{\text{max}}})\bm{H}_1(b_T\, q_{T_{\text{max}}}) \}\,,
\label{int_qtmax_2}
\end{equation}

\noindent where ${\bm{H}}_{0,1}$ are the Struve functions of order zero and one respectively. Notice that in the above integration we have introduced a maximum value $q_{T_{\text{max}}}$, 
that has to fulfil the condition $q_{T_{\text{max}}} \ll Q$, in order to respect the validity of the TMD factorization~\cite{Collins:2016hqq} . %
In the phenomenological analysis, Section~\ref{sec:analysis}, we will test different choices of the ratio $q_{T_{\text{max}}}/Q$.

\section{Double-hadron production: CSS formalism}
\label{sec:CSS_RG_eqs}

In this Section we elaborate on the convolutions, presented in Section~\ref{sec:2_h_prod}, with the proper treatment of the scale evolution within the Collins-Soper-Sterman (CSS) approach  (see Refs.~\cite{collins_2011, Collins:2014jpa, Collins:2016hqq} for more details). According to the CSS formalism, the complete expressions of the two convolutions entering the transverse polarization observable, Eq.~(\ref{pol_ratio}),
are given by:
\begin{equation}
\begin{split}
    \mathcal{B}_0 \Big[\widetilde{D}_1 \widetilde{\bar{D}}_1\Big] &=  \sum_q e^2_q \,\mathcal{H}^{(e^+e^-)}(Q) \\
    &\times\int \frac{d b_T}{(2 \pi)} \, b_T J_0(b_T\, q_T) \widetilde{D}_{1,q/h_1}(z_1,b_T;\zeta_1,\mu) \widetilde{\bar{D}}_{1,\bar{q}/h_2}(z_2,b_T;\zeta_2,\mu) \\
\end{split}
\label{B0_css}
\end{equation}
\begin{equation}
\begin{split}
    \mathcal{B}_1 \Big[\widetilde{D}^{\perp  (1)}_{1T} \widetilde{\bar{D}}_1\Big] =& \sum_q e^2_q \, \mathcal{H}^{(e^+e^-)}(Q)\\
    \times&\int \frac{d b_T}{2 \pi} \,b^2_T J_1( b_T\,q_T)  \widetilde{D}^{\perp  (1)}_{1T, q/h_1}(z_1,b_T;\zeta_1,\mu) \widetilde{\bar{D}}_{1,\bar{q}/h_2}(z_2,b_T;\zeta_2,\mu)\,, \\
\end{split}
\label{B1_css}
\end{equation}
where $\mathcal{H}^{(e^+e^-)}(Q)$ is the hard scattering part (depending also on $y$), for the massless on-shell process  $e^+e^-\to q \bar{q}$, at the center-of-mass energy $Q$.
With respect to the expressions given in the previous section, the two fragmentation functions now depend explicitly on two scale arguments: the renormalization scale $\mu$ and the $\zeta$ scale, that describes the effect of the recoil against the emission of soft gluons into  an energy range determined approximately by $\mu$ and $\zeta$. The dependence on these two scales is regulated by the CSS and Renormalization Group (RG) equations.

\subsection{Evolution equations for TMD fragmentation functions}
The CSS evolution equation for the $\zeta$ dependence of the unpolarized TMD-FF has the following form:
\begin{equation}
    \pr{\ln \widetilde{D}_1(z,b_T;\zeta,\mu)}{\ln{\sqrt{\zeta}}} = \widetilde{K}(b_T;\mu)\,,
\label{eq_rapidity}
\end{equation}
where $ \widetilde{K}$ is the CSS kernel~\cite{collins_2011}. 
It is flavour and spin independent, but different for quarks and gluons. Its RG equation is
\begin{equation}
    \prtot{\widetilde{K}(b_T;\mu)}{\ln{\mu}} = - \gamma_K(g(\mu))\,,
\label{CSS_kern}
\end{equation}
where the anomalous dimension $\gamma_K$ has no dependence on $b_T$, since the UV divergences only arise from virtual graphs \cite{collins_2011}. The corresponding RG equation for the fragmentation function is given by
\begin{equation}
    \prtot{\ln \widetilde{D}_1 (z,b_T;\zeta,\mu)}{\ln{\mu}} = \gamma_D(g(\mu);\zeta/\mu^2)\,.
\label{eq_scalemu}
\end{equation}
Since the derivatives of the FF with respect to $\mu$ and $\zeta$ commute, we can finally obtain the energy dependence of $\gamma_D$:
\begin{equation}
    \gamma_D(g(\mu);\zeta/\mu^2) = \gamma_D(g(\mu);1) - \frac{1}{2}\gamma_K(g(\mu)) \ln{\frac{\zeta}{\mu^2}}\,.
\end{equation}
In addition, the anomalous dimensions and the CSS kernel can be computed order by order perturbatively.
By solving  Eq.~(\ref{eq_rapidity}), that gives us the evolution in $\zeta$, and by using it in Eq.~(\ref{eq_scalemu}), we get the scale evolution from $\mu_0$ to $\mu$ (with $\mu_0$ large enough to start already in the perturbative region). This eventually leads to
\begin{equation}
    \begin{split}
    \widetilde{D}_1(z,b_T;\zeta,\mu)
    &=\widetilde{D}_1(z,b_T;\zeta_{0},\mu_0) \exp{\Bigg \{ \frac{1}{2}\widetilde{K}(b_T;\mu_0) \ln\frac{\zeta}{\zeta_{0}}    \Bigg\}}\\
    &\times \exp{\Bigg\{ \int^{\mu}_{\mu_0} \frac{d\mu'}{\mu'}\,\bigg[ \gamma_D(g(\mu');1) - \frac{1}{2}\gamma_K(g(\mu')) \ln{\frac{\zeta}{\mu'^2}} \bigg]\Bigg\}}  \, .
    \end{split}
\label{eq:unp_bt_space_1}
\end{equation}
The dependence of the FF on $\zeta$ involves the function $\widetilde{K}$, implying an energy dependence on the shape of the transverse momentum distribution. Moreover, the function $\widetilde{D}_1$, at its reference scales $\zeta_{0}$ and $\mu_0$, can be thought as the Fourier transform of an intrinsic transverse momentum distribution of the hadron with respect to its parent parton.

The full solution of the evolution equations in terms of the anomalous dimensions and the CSS kernel, and all the above results, can be directly extended to the $\widetilde{D}^{\perp  (1)}_{1T}$ function~\cite{collins_2011}.
\subsection{Small-$b_T$ expansion}
The first term on the right hand side of Eq.~(\ref{eq:unp_bt_space_1}) is the TMD-FF at the reference energy scale: it is related to the short distance and small-$\bm{b}_T$ behaviour of $D_1$ and therefore computable in perturbation theory. For such reason, at small-$\bm{b}_T$, the unpolarized TMD-FF can be matched onto the corresponding integrated fragmentation function $d_{j/h}(z;\mu)$ via an Operator Product Expansion (OPE):
\begin{equation}
    \widetilde{D}_{1,q/h}(z,b_T;\zeta_{0},\mu_0)
    =\sum_j \int^1_{z} \frac{d\hat{z}}{\hat{z}^{3-2\epsilon}} \widetilde{C}_{j/q}(z/\hat{z},b_T;\zeta_{0},\mu_0,g(\mu_0)) \, d_{j/h}(\hat{z};\mu_0) + \mathcal{O}[(mb_T)^p]\,,
\label{OPE_unp}
\end{equation}
where the error term is suppressed by some power of the transverse position. The sum is over all parton types  $j$, including gluons and antiquarks. When $b_T$ is small, the coefficient function $\widetilde{C}_{j/q}$ can be expanded in perturbation theory and calculated from Feynmann graphs with external on-shell partons of type $j$, with a double-counting subtraction in order to cancel all collinear contributions \cite{collins_2011}. The lowest-order coefficient is simply given as:
\begin{equation}
    \widetilde{C}_{j/q}(z/\hat{z},b_T;\zeta,\mu,g(\mu)) = \delta_{jq}\,\delta(z/\hat{z} -1) + \mathcal{O}(g^2)\,.
\label{C0_coef}
\end{equation}
An OPE of the same kind applies also to other collinear fragmentation functions, e.g.~$G_{1L}$ and $H_{1T}$, but they generally have different coefficient functions  beyond lowest-order. For the other, polarization-dependent, TMD fragmentation functions, like the Collins and the polarizing fragmentation functions, it is possible to generalize the OPE involving quantities that are associated to matrix elements of higher-twist operators~\cite{Boer:2010ya}. For the Sivers function, for instance, this would be the Qiu-Sterman function~\cite{Qiu:1991pp,Qiu:1991wg}.

\subsection{Matching the perturbative and nonperturbative $b_T$ regions}

TMD evolution follows from generalized renormalization properties of the operator definitions for TMD parton distribution and fragmentation functions. In order to combine the small-$\bm{b}_T$ dependence coming from perturbative calculations with the one from the nonperturbative part (that must be extracted from experimental data) it is necessary to introduce a matching procedure. To match the  perturbative and nonperturbative contributions, one defines large and small $\bm{b}_T$ through a function of $\bm{b}_T$ that freezes above some $b_{\text{max}}$  and equals $\bm{b}_T$ for small values.

We adopt the following standard procedure~\cite{Collins:1984kg}. First, we introduce the parameter $b_{\text{max}}$, representing the maximum distance at which perturbation theory is to be trusted, usually taken within an interval of $[0.5-1.5]\,\text{GeV}^{-1}$. Then we define a function $\bm{b}_*(\bm{b}_T)$, that almost equals $b_T$ at small $b_T$ and saturates at $b_{\text{max}}$ at large $b_T$:
\begin{equation}
    \bm{b}_* = \frac{\bm{b}_T}{\sqrt{1+b^2_T/b^2_{\text{max}}}}\,,
\label{b_star0}
\end{equation}
and re-define the CSS Kernel as:
\begin{equation}
    \widetilde{K}(b_T;\mu) = \widetilde{K}(b_*;\mu) - g_K(b_T;b_{\text{max}})\,.
\label{K_star}
\end{equation}
In this way, $\widetilde{K}(b_*;\mu)$ is computed in a region where perturbation theory is appropriate and the correction term, $g_K$, is important only at large ${b}_T$. This last term, $g_K$, to be extracted from data fits, is a function of $b_T$ and can depend explicitly or not on the parameter $b_{\text{max}}$. Since it is the difference of $ \widetilde{K}$ calculated at two values of its position argument, it is RG invariant and has to vanish as $b_T \to 0$.

If we want to match the perturbative and nonperturbative part of the unpolarized FF $\widetilde{D}_1$, we can then use $\bm{b}_*$ as defined in Eq.~(\ref{b_star0}).

Generalizing Eq.~(\ref{K_star}), it is possible to define an intrinsically nonperturbative part of the FF with the following decomposition:
\begin{eqnarray}
&&\widetilde{D}_{1, q/h}(z,b_T;\zeta,\mu) = \widetilde{D}_{1, q/h}(z,b_*;\zeta,\mu) \Bigg[\frac{\widetilde{D}_{1, q/h}(z,b_T;\zeta,\mu)}{\widetilde{D}_{1, q/h}(z,b_*;\zeta,\mu)}\Bigg]\nonumber\\
     &&= \widetilde{D}_{1, q/h}(z,b_*;\zeta,\mu)\exp\Bigg[ -g_{q/h}(z,b_T;b_{\text{max}}) - \,g_K(b_T;b_{\text{max}})\ln\frac{\sqrt{\zeta}}{\sqrt{\zeta_{0}}}   \Bigg] \nonumber\\
     &&= \widetilde{D}_{1, q/h}(z,b_*;\zeta,\mu)\exp\Bigg[ -g_{q/h}(z,b_T;b_{\text{max}}) - \,g_K(b_T;b_{\text{max}})\ln\frac{\sqrt{\zeta}\,z}{M_{h}}   \Bigg] \, ,
\label{D_star}
\end{eqnarray}
where in the second line we have introduced the perturbatively calculable $\widetilde{D}(b_*)$ and in the last line we have used the reference value $\zeta_{0} = M^2_{h}/z^2$~\cite{collins_2011}.
By employing Eq.~(\ref{eq:unp_bt_space_1}), the anomalous dimensions, $\gamma_D$ and $\gamma_K$, cancel between numerator and denominator in the square brackets (first line) and only $g_K$ survives. 
The remaining factor, $e^{-g_{q/h}}$, defined as~\cite{collins_2011, Collins:2014jpa, Collins:2016hqq}
\begin{equation}
 \exp\big[-g_{q/h}(z,b_T;b_{\text{max}})\big]    = \frac{\widetilde{D}_{1, q/h}(z,b_T;\zeta,\mu)}{\widetilde{D}_{1, q/h}(z,b_*;\zeta,\mu)} \exp\big[  g_K(b_T;b_{\text{max}})\ln\sqrt{\zeta}/\sqrt{\zeta_{0}}\big]\,,
\label{eq:interme_css_ev}
\end{equation}
can be interpreted as the nonperturbative part of the intrinsic transverse momentum distribution.

Some comments are in order: both $g_K$ and $g_{q/h}$ vanish approximately as $b^2_T$ at small $b_T$ \cite{collins_2011}, and become significant when $b_T$ approaches $b_{\text{max}}$ and beyond; they are independent of $\zeta$ and $\mu$, being invariant under the application of the CSS and RG equations, while they do depend on the choice of $b_{\text{max}}$. On the other hand, the full TMD fragmentation function and the function  $\widetilde{K}$ are independent of $b_{\text{max}}$ and the use of the $b_*$ prescription.
The flavour and $z$ dependences of $g_K$ and  $g_{q/h}$ follow from those of the corresponding parent functions, respectively  $\widetilde{K}$ and the TMD fragmentation functions \cite{Collins:2014jpa}. Since  $\widetilde{K}$ is independent of the quark's and hadron's type, polarization and fraction $z$, so is $g_K$. %
The same, of course, is not true in general for the TMD fragmentation functions and therefore for the factor $e^{-g_{q/h}}$. %

It is worth mentioning that this last term is usually written as $M_D( b_T,{z} ;b_{\text{max}})$ or $D_{NP}(b_T,{z};b_{\text{max}})$, a generic function of $b_T$: this is because it could assume also a non-exponential functional form, still preserving its properties, and the fact that at small $b_T$ it goes like 1 + ${\cal O}(b_T^2)$; it is referred to as the \emph {nonperturbative part} of the fragmentation function, and within a parton model, can be seen as the Fourier transform of the transverse momentum distribution.
Like $g_K$, the function $M_D$ can depend explicitly or not on the parameter $b_{\text{max}}$.
In a more general way, the phenomenological extraction of both nonperturbative functions is affected by the choice of the $b_{\text{max}}$ value.

To use the perturbative small-$b_T$ result from Eq.~(\ref{OPE_unp}), it is necessary to evolve the $\widetilde{D}$ term in Eq.~(\ref{D_star}), with the $b_*$ prescription, from a region where no large kinematic ratios appear in the coefficient function $\widetilde{C}$, whose logarithms could spoil the use of the perturbative approach~\cite{collins_2011}. The standard choice is to replace $\mu_0$ by:
\begin{equation}
    \mu_{b} = \frac{C_1}{b_*(b_T)}\, ,
\end{equation}
where $C_1 = 2e^{-\gamma_E}$ (with $\gamma_E$ being the Euler-Mascheroni constant), and use for the reference value $\zeta_{0}$ the same value, that is $\zeta_0=\mu^2_{b} $. Then the TMD fragmentation function can be written as:
\begin{eqnarray}
\widetilde{D}_{1, q/h}(z,b_T;\zeta,\mu) &=&
    \sum_j \int^1_{z} \frac{d\hat{z}}{\hat{z}^{3-2\epsilon}} \widetilde{C}_{j/q}(z/\hat{z},b_*;\mu^2_b,\mu_b,g(\mu_b)) \, d_{j/h}(\hat{z};\mu_b) \nonumber\\
    &\times& M_D(b_T,z;b_{\text{max}}) \exp\Bigg\{  - \,g_K(b_T;b_{\text{max}})\ln\frac{\sqrt{\zeta}\,z}{M_{h}}   \Bigg\} \label{D_full} \\
    &\times &\exp{\Bigg\{ \frac{1}{2}\widetilde{K}(b_*;\mu_b) \ln\frac{\zeta}{\mu^2_b} + \int^{\mu}_{\mu_b} \frac{d\mu'}{\mu'}\,\bigg[ \gamma_D(g(\mu');1) - \frac{1}{2}\gamma_K(g(\mu')) \ln{\frac{\zeta}{\mu'^2}} \bigg]\Bigg\}}\nonumber \,.
\end{eqnarray}
Finally, in order to properly control the low-$b_T$ region (to ensure the matching at high $k_T$), we  modify the $\bm{b}_T$ definition using~\cite{Collins:2016hqq}:
\begin{equation}
    b_c(b_T) = \sqrt{b^2_T + b^2_{\text{min}}}\,,
\end{equation}
where $b_{\text{min}} = 2e^{-\gamma_E}/Q$, decreasing like $1/Q$ in contrast to $b_{\text{max}}$, which remains fixed. This definition reduces to $b_T$ when $b_T \gg 1/Q$ but it is of order $1/Q$ when $b_T$ is small, thereby providing an effective cutoff at small $b_T$. Consistently, $b_*$ has to be replaced by:
\begin{equation}
    b_*(b_c(b_T)) =\sqrt{ \frac{b^2_T + b^2_{\text{min}}}{1+b^2_T/b^2_{\text{max}} +b^2_{\text{min}}/b^2_{\text{max}} } }\,,
\label{b_star}
\end{equation}
in order to ensure the requested behaviour simultaneously at small and large $b_T$. Indeed, we have:
\begin{equation}
 b_*(b_c(b_T)) \to
    \begin{cases}
    b_{\text{min}} \quad b_T \ll b_{\text{min}} \\
     b_T \qquad  b_{\text{min}} \ll b_T \ll b_{\text{max}}\\
     b_{\text{max}} \quad b_T \gg b_{\text{max}}
    \end{cases}\,.
\label{b_star_gen}
\end{equation}
Lastly, we also redefine $\mu_b$, replacing it by:
\begin{equation}
    \bar{\mu}_{b}(b_c(b_T)) = \frac{C_1}{ b_*(b_c(b_T))}\, ,
\end{equation}
implying a maximum cutoff on the renormalization scale equal to $\bar{\mu}_{b} \simeq C_1/b_{\text{min}}$. Recollecting all the above results we can now write the TMD fragmentation function, Eq.~(\ref{D_full}), employing the new definitions of ${b}_T$, as:
\begin{eqnarray}
&& \widetilde{D}_{1, q/h}(z,b_c(b_T);Q^2,Q) =
    \sum_j \int^1_{z} \frac{d\hat{z}}{\hat{z}^{3-2\epsilon}} \widetilde{C}_{j/q}(z/\hat{z},b_*(b_c(b_T)); \bar{\mu}^2_b, \bar{\mu}_b,g( \bar{\mu}_b)) \nonumber\\
    && \mbox{}\times  d_{j/h}(\hat{z}; \bar{\mu}_b)\, M_D(b_c(b_T),z;b_{\text{max}}) \exp\Bigg\{  - \,g_K(b_c(b_T);b_{\text{max}})\ln\frac{Q z}{M_{h}}   \Bigg\} \nonumber\\
    &&\mbox{} \times \exp{\Bigg\{ \frac{1}{2}\widetilde{K}(b_*;\bar{\mu}_{b}) \ln\frac{Q^2}{\bar{\mu}_{b}^2} + \int^{Q}_{\bar{\mu}_{b}} \frac{d\mu'}{\mu'}\,\bigg[ \gamma_D(g(\mu');1) - \frac{1}{2}\gamma_K(g(\mu')) \ln{\frac{Q^2}{\mu'^2}} \bigg]\Bigg\}}\,,
\label{D_full_bc_Q}
\end{eqnarray}
where we have adopted $\zeta = Q^2$ and $\mu = Q$.

\subsection{Convolutions}
Thanks to the evolution equations and the matching procedure, we can write the full form of the convolutions in Eqs.~(\ref{B0_css}) and (\ref{B1_css}). For the convolution $\mathcal{B}_0 $ we have:
\begin{eqnarray}
\mathcal{B}_0 \Big[\widetilde{D} \widetilde{\bar{D}}\Big] &=&  \frac{\mathcal{H}^{(e^+e^-)}(Q)}{z^2_1 z^2_2} \sum_q e^2_q\int \frac{d b_T}{(2 \pi)} \, b_T J_0(b_T\, q_T) \, d_{q/h_1}(z_1; \bar{\mu}_b) \, d_{\bar{q}/h_2}(z_2; \bar{\mu}_b)\nonumber \\
    &\times& M_{D_1}(b_c(b_T),\textcolor{black}{z_1}) \, M_{D_2}(b_c(b_T),\textcolor{black}{z_2})
    \exp\Bigg\{\!\!-g_K(b_c(b_T);b_{\text{max}})\ln{\bigg(\frac{Q^2 z_1 z_2}{M_{1} M_{2}}\bigg)}\Bigg\} \nonumber\\
    &\times& \exp{\Bigg\{ \widetilde{K}(b_*;\bar{\mu}_{b}) \ln\frac{Q^2}{\bar{\mu}_{b}^2} + \int^{Q}_{\bar{\mu}_{b}} \frac{d\mu'}{\mu'}\,\bigg[ 2\gamma_D(g(\mu');1) - \gamma_K(g(\mu')) \ln{\frac{Q^2}{\mu'^2}} \bigg]\Bigg\}}\,,
\label{B0_full}
\end{eqnarray}
where in the second line we have omitted the implicit $b_{\rm max}$ dependence in $M_D$ and used the lowest-order coefficient, Eq.~(\ref{C0_coef}), of the OPE expression for both the fragmentation functions. 
Similarly, for the convolution $\mathcal{B}_1$ we have:
\begin{eqnarray}
    \mathcal{B}_1 \Big[\widetilde{D}^{\perp  (1)}_{1T} \widetilde{\bar{D}}_1\Big]
    &=& \frac{\mathcal{H}^{(e^+e^-)}(Q)}{z^2_1 z^2_2}\sum_q e^2_q\int \frac{d b_T}{(2 \pi)} \, b^2_T J_1(b_T\, q_T) {D}^{\perp  (1)}_{1T} (z_1;\bar{\mu}_b) \, d_{\bar{q}/h_2}(z_2; \bar{\mu}_b) \nonumber\\
    &\times& M^{\perp}_{D_1}(b_c(b_T),\textcolor{black}{z_1}) \, M_{D_2}(b_c(b_T),\textcolor{black}{z_2})\exp\Bigg\{-g_K(b_c(b_T);b_{\text{max}})\ln{\bigg(\frac{Q^2 z_1 z_2}{M_{1} M_{2}}\bigg)}\Bigg\} \nonumber\\
    &\times& \exp{\Bigg\{ \widetilde{K}(b_*;\bar{\mu}_{b}) \ln\frac{Q^2}{\bar{\mu}_{b}^2} + \int^{Q}_{\bar{\mu}_{b}} \frac{d\mu'}{\mu'}\,\bigg[ 2\gamma_D(g(\mu');1) - \gamma_K(g(\mu')) \ln{\frac{Q^2}{\mu'^2}} \bigg]\Bigg\}}\,,\nonumber\\
\label{B1_full}
\end{eqnarray}
where again we have used the lowest-order coefficient for the OPEs and $M^{\perp}_{D_1}$ as the nonperturbative part for the polarizing fragmentation function for the hadron $h_1$. %

The last lines of Eqs.~(\ref{B0_full}) and (\ref{B1_full}) are usually referred to as perturbative Sudakov factors and, as explained above, they can be computed analytically. The anomalous dimension of the fragmentation functions at order $\alpha_s(\mu)$
is~\cite{Aidala:2014hva}:
\begin{equation}
    \gamma_D(\alpha_s(\mu);\zeta/\mu^2) = 4 C_F \bigg( \frac{3}{2} - \ln\frac{\zeta}{\mu^2}    \bigg) \bigg(\frac{\alpha_s(\mu) }{4\pi} \bigg) + \mathcal{O}(\alpha^2_s(\mu) )\,,
\label{gamma_D}
\end{equation}
where $C_F = 4/3$. Meanwhile the anomalous dimension of the CSS Kernel $\widetilde{K}$ at one-loop order is:
\begin{equation}
\gamma_K(\alpha_s(\mu)) = 8 C_F\bigg(\frac{\alpha_s(\mu) }{4\pi} \bigg) + \mathcal{O}(\alpha^2_s(\mu) )\,,
\label{gamma_K}
\end{equation}
with $\widetilde{K} = 0$ at first order. For the running coupling \cite{Aidala:2014hva}  we use the form:
\begin{equation}
    \alpha_s(\mu) = \frac{A}{2 \ln(\mu/\Lambda_{\rm QCD})}\,,
\end{equation}
with
\begin{equation}
    A = \frac{1}{\beta_0} = \frac{12 \pi}{33 - 2 n_f}\,,
\end{equation}
where $n_f$ is the number of active flavours.\footnote{Actually we consistently use $n_f=3$ with $\Lambda_{\rm QCD} =$  0.2123~GeV~\cite{Aidala:2014hva}.} We can then get, by analytical integration, the following expression for the exponent in the perturbative Sudakov factor
\begin{eqnarray}
&& \widetilde{K}(b_*;\bar{\mu}_{b}) \ln\frac{Q^2}{\bar{\mu}_{b}^2} + \int^{Q}_{\bar{\mu}_{b}} \frac{d\mu'}{\mu'}\,\bigg[ 2\gamma_D(g(\mu');1) - \gamma_K(g(\mu')) \ln{\frac{Q^2}{\mu'^2}} \bigg] \\
     &&= \frac{2A}{\pi}\Bigg[ \ln\bigg( \frac{\ln(Q/\Lambda_{\rm QCD})}{\ln(\bar{\mu}_{b}/\Lambda_{\rm QCD})} \bigg) - \frac{4}{3} \ln(Q/\Lambda_{\rm QCD}) \ln\bigg( \frac{\ln(Q/\Lambda_{\rm QCD})}{\ln(\bar{\mu}_{b}/\Lambda_{\rm QCD})} \bigg) +  \frac{4}{3}\ln(Q/\bar{\mu}_{b})\Bigg] \nonumber\,.
\end{eqnarray}


\subsection{Nonperturbative parts}

We discuss now the nonperturbative contributions, entering the above convolution formulas. We will give some details on the corresponding parametrizations available in the literature, focusing on those we will use directly in our phenomenological analysis: namely $g_K$ and $M_D$. For the moment, we will leave apart the other fundamental quantity, $M_D^\perp$, one of the main focus of our study, that will be properly addressed when we discuss our fitting procedure.

Notice that, while for $g_K$ we will use functional forms depending explicitly on $b_{\rm max}$, for $M_D$ this dependence enters only implicitly (as discussed previously) and, for the sake of notation, we will drop it in the sequel.

\subsubsection{$g_K(b_T;b_{\rm max})$}

We start considering $g_k$, an intrinsically nonperturbative quantity, that cannot be computed from first principles.  Nevertheless, as shown in Ref.~\cite{Collins:2014jpa}, it is possible to extract some of its properties from perturbative calculations. Indeed, the lowest-order formula for $\widetilde{K}$ gives:
\begin{equation}
    g_K(b_T,b_{\text{max}}) = \frac{\alpha_s(C_1/b_*) C_F}{\pi} \ln(1+ b^2_T/b^2_{\text{max}})\,,
\label{ll_lgm}
\end{equation}
that behaves like $b^2_T$ at small $b_T$, but shows a slower logarithmic rise above $b_{\text{max}}$. A similar functional form has been adopted in Refs.~\cite{Aidala:2014hva, Sun:2014dqm} (AFGR/SIYY in the following)  with the following expression
\begin{equation}
    g_K(b_T;b_{\text{max}})= g_2 \ln\bigg(\frac{b_T}{b_*}\bigg) \,,
\label{}
\end{equation}
with an extracted value of $g_2 = 0.84$ .

For large-$b_T$ values, Eq.~(\ref{ll_lgm}) is not expected to be an accurate parametrization. Indeed, it is an extrapolation of a lowest-order perturbative calculation and it depends strongly on $b_{\text{max}}$ at large $b_T$.  The $b^2_T$ behaviour at small $b_T$ can be found expanding Eq.~(\ref{ll_lgm}) in powers of $b_T$, obtaining:
\begin{equation}
    g_K(b_T;b_{\text{max}})= \frac{C_F}{\pi} \frac{b^2_T}{b^2_{\text{max}}}\alpha_s(\mu_{b})\,,
\label{}
\end{equation}
with an explicit quadratic form of the $g_K$ function. This justifies the use of the following expression:
\begin{equation}
     g_K(b_T;b_{\text{max}}) = \frac{g_2  b^2_T}{2}\,,
\end{equation}
employed and fitted to data in Ref.~\cite{Landry:2002ix} (BLNY in the following) and in Ref.~\cite{Konychev:2005iy} (KN) where they found, respectively, a value of $g_2 = 0.68\,\text{GeV}^2$, with $b_{\text{max}} = 0.5 \,\text{GeV}^{-1}$, and a value of $g_2 = 0.18\,\text{GeV}^2$, with $b_{\text{max}} = 1.5 \,\text{GeV}^{-1}$. Notice that such a large value of $b_{\rm max}$, as we will discuss in the next section, implies a too small renormalization scale, preventing its use in our calculation.

Since the real nonperturbative physics is at larger $b_T$ values, one wants to extract $g_K(b_T;b_{\text{max}})$ with a more general parametrization and be sure that the data used to extract it are sensitive to high values of $b_T$. Moreover, the complete TMD factorization formalism is $b_{\text{max}}$ independent and, in principle, optimized fits should not depend on its choice. %

For a more exhaustive comparison, we will also consider another set of nonperturbative functions, as those extracted from fits on SIDIS, Drell-Yan and $Z$-boson production and discussed in Ref.~\cite{Bacchetta:2017gcc}. Concerning $g_K$, this has the same functional form as BLNY, but with a different value for $g_2$, namely $g_2 = 0.13\,\text{GeV}^2$ (PV17). This implies a softer behaviour in $b_T$. 
Moreover, this has been obtained with an \emph{ad hoc} $b_*$ prescription
\begin{equation}
  b_* \equiv b_*(b_T;b_{\rm min},b_{\rm max}) = b_{\rm max}\bigg(\frac{1-e^{-b_T^4/b^4_{\rm max}}}{1-e^{-b_T^4/b^4_{\rm min}}} \bigg)^{1/4}
\end{equation}
and the corresponding parametrization of $M_D$ (see below).
Quite recently a new and refined global analysis, at N$^3$LL accuracy, has been performed by the Pavia group~\cite{Bacchetta:2022awv}. For consistency we will not adopt it in our study, where we are using the anomalous dimensions only at one-loop order.

\begin{figure}[!bth]
\centerline{\includegraphics[trim =  50 30 30 0,width=9cm]{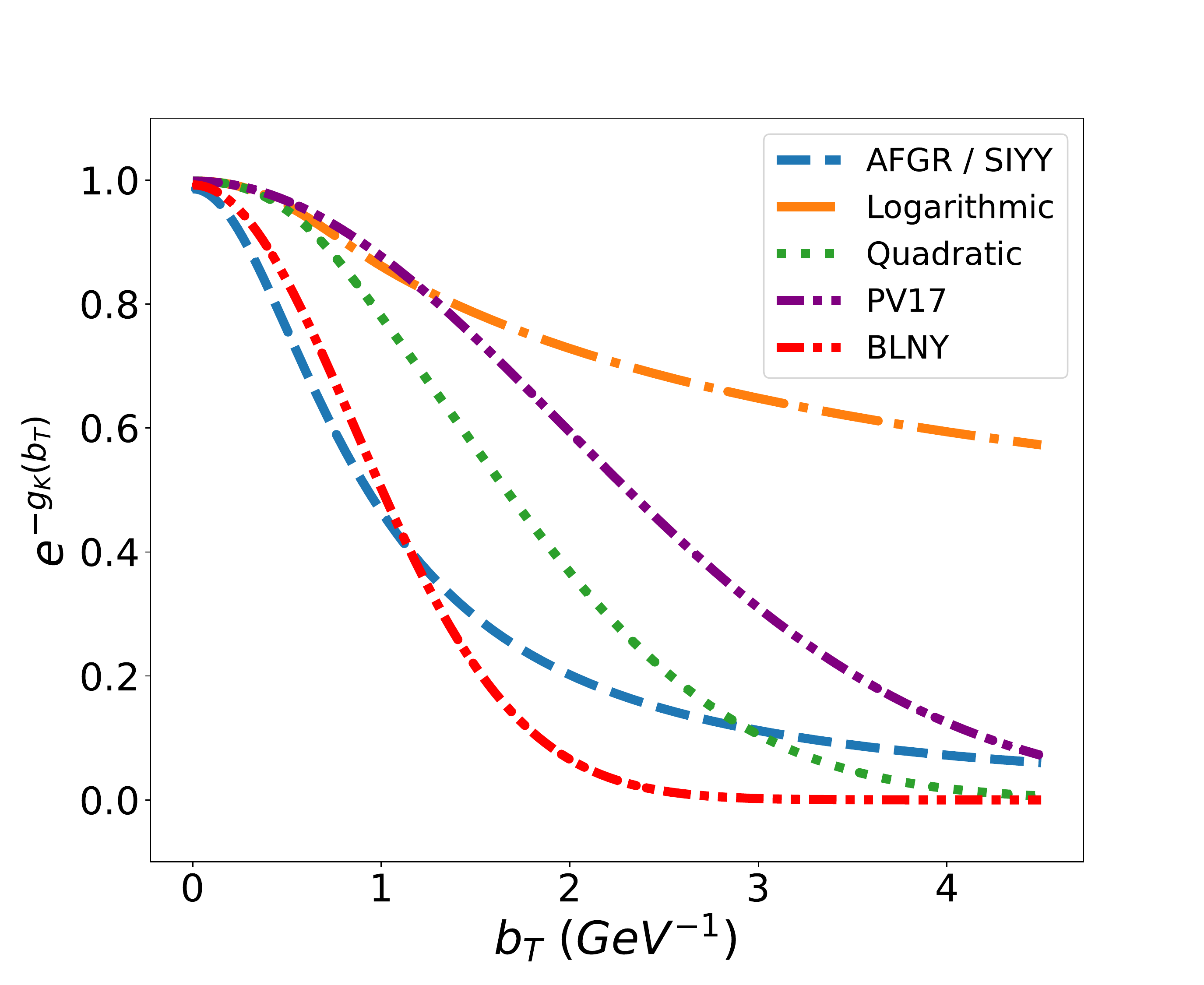}}
    \caption{Graphical representation of the different forms of the nonperturbative function $\exp(-g_K)$, listed in Eq.~(\ref{eq:all_gk}), with $b_{\rm max} = 0.6$~GeV$^{-1}$.}
\label{fig:all_gk}
\end{figure}

Summarizing, in the phenomenological analysis, Section~\ref{sec:analysis}, we will employ the following functional forms of the nonperturbative function, $g_K$, adopting $b_{\rm max} = 0.6$~GeV$^{-1}$, see Fig.~\ref{fig:all_gk}:
\begin{equation}
    \begin{split}
        g_K(b_T;b_{\text{max}}) &= \frac{g_2  b^2_T}{2}\,; \quad g_2 = 0.68\,\text{GeV}^2 \qquad \qquad\;\,  \text{BLNY}   \\
        g_K(b_T;b_{\text{max}}) &= \frac{C_F}{\pi} \frac{b^2_T}{b^2_{\text{max}}}\alpha_s(\mu_{b_*}) \qquad\qquad\qquad\qquad \text{Quadratic}\\
        g_K(b_T;b_{\text{max}}) &= \frac{\alpha_s(C_1/b_*) C_F}{\pi} \ln(1+ b^2_T/b^2_{\text{max}})  \qquad \text{Logarithmic}\\
        g_K(b_T;b_{\text{max}}) &= g_2 \ln\bigg(\frac{b_T}{b_*}\bigg) \quad g_2 = 0.84  \qquad\qquad\quad \text{AFGR/SIYY} \\
        g_K(b_T;b_{\text{max}}) &= \frac{g_2  b^2_T}{2}\,; \quad g_2 = 0.13\,\text{GeV}^2 \qquad \qquad\;\,\,  \text{PV17}  \,. \\
    \end{split}
\label{eq:all_gk}
\end{equation}
Thanks to their universality, all of them can be used to predict observables or be supportive in the extraction of other nonperturbative functions, in processes like $e^+e^-$ collisions, Semi-inclusive DIS and Drell-Yan processes.


\subsubsection{$M_D(b_T,z)$}

\noindent
The other relevant nonperturbative function entering our convolutions is $M_D(b_T,z)$.
In Ref.~\cite{Collins:2014jpa} it has been shown that the arguments for the approximately quadratic behaviour of $g_K(b_T)$ at small $b_T$ are also valid for the function $g_{q/h}(b_T)$, and this corresponds, after an exponentiation, to a Gaussian model for the TMD-FF:
\begin{equation}
    M_D(b_T,z)
    = \exp{\bigg(-\frac{a b^2_T}{2}\bigg)}\,.
\label{gauss_mod}
\end{equation}
This justifies the use of the following parameterization:
\begin{equation}
    M_D(b_T,z)
    = \exp{\bigg(-\frac{\langle p_\perp^2 \rangle b^2_T}{4 z^2}\bigg)}\,,
\label{eq:gaussian_np}
\end{equation}
corresponding in the conjugate   $\bm{p}_{\perp}$-space\footnote{This space is equivalent to the $\bm{k}_{\perp}$-space used in previous sections.} to
\begin{equation}
     \widetilde{M}_D(p_{\perp}) =\frac{e^{- p^2_{\perp}/\langle p_\perp^2 \rangle}}{\pi \langle p_\perp^2 \rangle}\,,
     \label{eq:gaussian_pperp}
\end{equation}
where $\langle p_\perp^2 \rangle$ is the usual transverse momentum Gaussian width. Notice that, assuming it as a constant, in $p_\perp$-space there is no explicit $z$ dependence.

The commonly assumed quadratic behaviour of $g_K(b_T)$ and the Gaussian behaviour of the TMD fragmentation function can only be a valid approximation, at best, for moderate $b_T$.
Appropriate parametrizations for the nonperturbative large-$b_T$ behaviour of the TMD-FFs and of the CSS kernel need to be inferred from general principles of quantum field theory~\cite{Collins:2014jpa}, that suggest an exponentially decaying behaviour at large $b_T$. From several one-loop calculations of TMD quantities, a typical integral giving the proper $b_T$ dependence is of the form
\begin{equation}
    \int d^2\bm{p}_T \frac{e^{i\bm{p}_T\cdot \bm{b}_T}}{p^2_T + m^2}\,.
\end{equation}
One possible functional parametrization that generalizes, in $\bm{b}_T$-space, the Fourier transform of the previous equation, and preserves the quadratic behaviour at small $b_T$, used in Refs.~\cite{Boglione:2017jlh,Boglione:2020auc,Boglione:2022nzq}, is the following:
\begin{equation}
    M_D(b_T,z,p,m) = \frac{2^{2-p}}{\Gamma(p-1)}\,(b_T m/z)^{p-1}{K}_{p-1}(b_T m/z)\,,
\label{eq:pwrlw_mod_np}
\end{equation}
where $K_{p-1}$ is a Bessel function of the second kind, with the condition $p>1$. Its Fourier transform in $\bm{p}_{\perp}$-space is given by:
\begin{equation}
    \widetilde{M}_D(p_{\perp}) = \frac{\Gamma(p)}{\pi \Gamma(p-1)}\frac{m^{2(p-1)}}{(p^2_{\perp} + m^2)^p}\,.
    \label{eq:pwelw_md_pperp}
\end{equation}
We will refer to this as the Power-law model.

Notice that for a massive hadron, on the right-hand side of Eq.~(\ref{eq:gaussian_np}) one has to replace $z_p$ with $z$, in order to properly get Eq.~(\ref{eq:gaussian_pperp}). The same happens for going from Eq.~(\ref{eq:pwrlw_mod_np}) to Eq.~(\ref{eq:pwelw_md_pperp}).

\begin{figure}[!ht]
\centerline{\includegraphics[trim =  50 30 30 0,width=14cm]{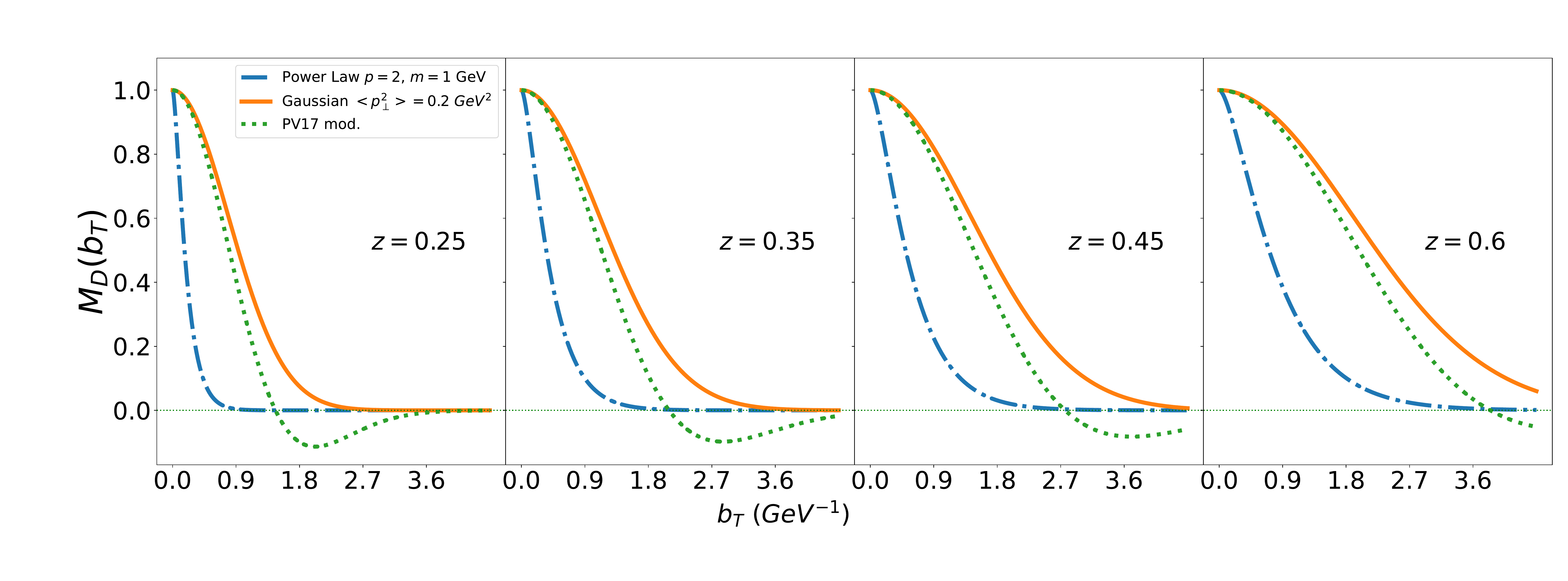}}
    \caption{Representation of nonperturbative hadronic models for $M_D(b_T,z)$ as a function of $b_T$ at different values of $z$: Gaussian (orange solid line), Power-Law (blue dot-dashed line) and PV17 (green dotted line) model.}
\label{fig:np_models}
\end{figure}
%
As mentioned above, the Pavia group~\cite{Bacchetta:2017gcc} provided a complete set of nonperturbative functions, for pions and kaons: for $M_D$, to be used together with its corresponding $g_K$ (last line of Eq.~(\ref{eq:all_gk})), they propose:
\begin{equation}
    M_D(b_T,z) =\frac{g_3\, e^{-b^2_T\frac{g_3}{4z^2}} +\frac{\lambda_F}{z^2}g^2_4(1 -g_4\frac{b_T^2}{4z^2} )e^{-b^2_T\frac{g_4}{4z^2}} }{g_3 + \frac{\lambda_F}{z^2}g^2_4 }\,,
\label{PV17_md}
\end{equation}
where
\begin{equation}
    g_{3,4} = N_{3,4}\frac{(z^{\beta}+\delta)(1-z)^{\gamma} }{(\hat{z}^{\beta}+\delta)(1-\hat{z})^{\gamma}} \quad %
\end{equation}
\begin{equation}
\begin{split}
    \hat{z}&= 0.5; \quad N_3= 0.21 \,\text{GeV}^2; \quad N_4 = 0.13 \,\text{GeV}^2\\
    \beta &= 1.65; \quad \delta = 2.28; \quad \gamma=0.14; \quad \lambda_F= 5.50 \,\text{GeV}^{-2}\,.
\end{split}
\end{equation}
The three parametrizations, discussed above, are shown in Fig.~\ref{fig:np_models}.

Finally, for the nonperturbative component of the polarizing FF, $M_D^\perp$, in the subsequent phenomenological analysis we will adopt the same functional forms as those in Eqs.~(\ref{eq:gaussian_np}) and (\ref{eq:pwrlw_mod_np}), extracting the corresponding parameters from the fit.

\section{Single-inclusive hadron production}
\label{sec:1-h_prod}
As already mentioned, there is another interesting and related process relevant in this context: namely the single-inclusive production of (un)polarized spin-$1/2$ hadrons in $e^+e^-$ annihilation processes. In Ref.~\cite{DAlesio:2020wjq} a first attempt to consider this case, within a simplified phenomenological TMD scheme, was discussed.
As we will see, this case is more subtle and deserves a proper and dedicated treatment within the TMD factorization scheme.
We will present here only some relevant formulas, summarizing the kinematics and giving the expression of the cross section. We refer the reader to Refs.~\cite{Kang:2020yqw,Gamberg:2021iat}, where a complete TMD formulation of the process $e^+e^-\to h(S_{h})+X$ is discussed. It is worth noticing that the issues of proper factorization and universality for such a process have been formally addressed in a series of recent papers~\cite{Makris:2020ltr,Boglione:2020auc, Boglione:2020cwn, Boglione:2021wov}, where a detailed, and somehow complementary, approach can be found.

\begin{figure}[!th]
\centering
\includegraphics[width=12cm]{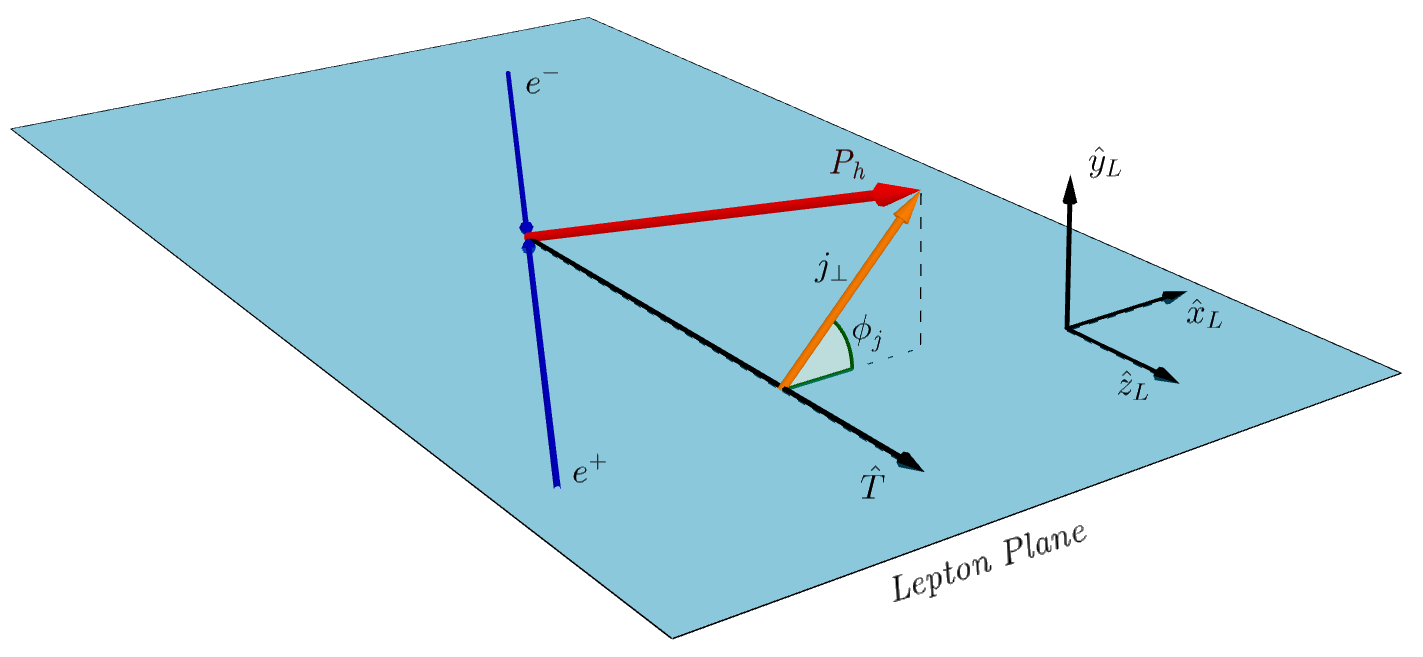}
    \caption{Kinematics for the process $e^+e^-\to h+X$ in the \emph{thrust-frame} configuration.}
    \label{fig:kin_1_h}
\end{figure}

In this configuration, as shown in Fig.~\ref{fig:kin_1_h},
the hadron is produced with a transverse momentum $\bm{j}_{\perp}$ with respect to the thrust axis $\hat{\bm{T}}$, defined as the vector, $\hat{\bm{T}}$, which maximizes the thrust variable $T$
\begin{equation}
    T = \frac{\sum_i |\bm{p}_i \cdot \hat{\bm{T}}|}{\sum_i |\bm{p}_i|}\,,
\end{equation}
where $\bm{p}_i$ represent the three-momenta of the measured final-state particles. This is referred to as the \emph{thrust frame} configuration, where one plane is defined by the lepton direction and the thrust vector (Lepton plane) and the second plane by the thrust axis and the hadron momentum.
Moreover, the full phase space is divided into two hemispheres by the plane perpendicular to the thrust axis at the $e^+e^-$ interaction point.
Similarly to the previous case, the hadron has a certain energy fraction
\begin{equation}
    z_{h} = \frac{2 P_{h} \cdot q}{Q^2}\,,
\end{equation}
where once again $q$ is the virtual photon four-momentum and $Q^2 = q^2 = s$.

The most important aspect is that, for this process, one hemisphere is fully inclusive, while
the single-inclusive measurement is carried out in the hemisphere that contains the thrust axis. Thus, only soft radiation which is emitted into this hemisphere will contribute to $\bm{j}_{\perp}$.
Indeed the factorized expression used in Ref.~\cite{Gamberg:2021iat} and given at next-to-leading logarithm accuracy (NLL) in \cite{Kang:2020yqw} introduces the hemisphere soft function $S_{\text{hemi}}$, that is different from the typical soft function $S$ usually defined to describe the almost back-to-back double hadron production in $e^+e^-$ collisions.
On the other hand, as demonstrated in \cite{Kang:2020yqw}, since $S_{\text{hemi}}$ at one-loop order accuracy equals $\sqrt{S}$, both the unpolarized and polarizing FFs, in the single-inclusive process, are the same FFs appearing in the double-hadron production process.
The cross section for the unpolarized hadron production, for $j_T\ll Q$, is then given by:
\begin{eqnarray}
    \frac{d \sigma}{dz d^2 \bm{j}_{\perp}} &=& \frac{\sigma_0}{z^2} \sum_q e^2_q\int \frac{d b_T}{(2 \pi)} \, b_T J_0(b_T\, q_T)\,d_{q/h}(z; \bar{\mu}_b) \, U_{NG}(\bar{\mu}_b,Q) \nonumber\\
    &\times & M_{D}(b_c(b_T),z) \, \exp\Bigg\{-g_K(b_c(b_T);b_{\text{max}})\ln{\bigg(\frac{Q z}{M_{h}}\bigg)}\Bigg\} \nonumber\\
    &\times& \exp\Bigg\{ \Tilde{K}(b_*;\bar{\mu}_b)\ln{\frac{Q}{\bar{\mu}_b}} + \int^Q_{\bar{\mu}_b} \frac{d\mu'}{\mu'} \Bigg[ \gamma_D(g(\mu'),1) - \gamma_K(g(\mu'))\ln{\frac{Q}{{\mu'}}} \Bigg] \Bigg\}\,,
\label{eq:1-h_unp}
\end{eqnarray}
where $z$ is the hadron light-cone momentum fraction, related to $z_{h}$ as shown in Eq.~(\ref{eq:en_fract}), and
\begin{equation}
    \sigma_0 = \frac{4 N_c \pi \alpha^2_{em}}{3 Q^2}\,.
\end{equation}
Here we find the same elements already discussed in Section~\ref{sec:CSS_RG_eqs}: $d_{q/h}$ is the integrated unpolarized FF, $M_D$ and $g_K$ are the nonperturbative model functions.
Similarly, the cross section for transversely polarized hadron production has the following form:
\begin{eqnarray}
    \frac{d \Delta\sigma}{dz d^2\bm{j}_{\perp}}& =&\sin(\phi_{S_{h}} - \phi_j)\,\frac{\sigma_0}{z^2} \sum_q e^2_q\int \frac{d b_T}{(2 \pi)} \, b^2_T J_1(b_T\, q_T) {D}^{\perp (1)}_{1T}(z,\bar{\mu}_b) \, U_{NG}(\bar{\mu}_b,Q) \nonumber\\
    &\times& M^{\perp}_{D}(b_c(b_T),z)  \, \exp\Bigg\{-g_K(b_c(b_T);b_{\text{max}})\ln{\bigg(\frac{Q z}{M_{h}}\bigg)}\Bigg\} \nonumber\\
    &\times &\exp\Bigg\{ \Tilde{K}(b_*;\bar{\mu}_b)\ln{\frac{Q}{\bar{\mu}_b}} + \int^Q_{\bar{\mu}_b} \frac{d\mu'}{\mu'} \Bigg[ \gamma_D(g(\mu'),1) - \gamma_K(g(\mu'))\ln{\frac{Q}{{\mu'}}} \Bigg] \Bigg\}\,,
\label{eq:1-h_pol}
\end{eqnarray}
where ${D}^{\perp (1)}_{1T}$, see Eq.~(\ref{eq:firstmoment_z}), is the small-${b}_T$ limit of the first moment of the polarizing fragmentation function.

Since soft radiation is restricted to only one hemisphere, the cross section is a non-global observable. The factorization formulas for this kind of observables have been derived within an effective field theory framework  \cite{Becher:2015hka,Becher:2016mmh,Becher:2016omr,Becher:2017nof}, where a multi-Wilson-line structure \cite{Caron-Huot:2015bja,Nagy:2016pwq,Nagy:2017ggp} is the key ingredient to capture the non-linear QCD evolution effects from the so-called \emph{non-global logarithms}. %
For this reason in both cross sections, Eqs.~(\ref{eq:1-h_unp}) and (\ref{eq:1-h_pol}), we have introduced the function $ U_{NG}$ (see Ref. \cite{Kang:2020yqw}), which accounts for the effects of such non-global effects.

In the following we will use the parametrization given in Ref.~\cite{Dasgupta:2001sh}
\begin{equation}
    U_{NG}(\bar{\mu}_b,Q) = \exp{\bigg[ -C_A C_F \frac{\pi^2}{3} u^2 \frac{1+ (au)^2}{1+(bu^c)} \bigg]}\,,
\end{equation}
with $a = 0.85 \, C_A$, $b = 0.86 \, C_A$, $c = 1.33$ and
\begin{equation}
    u = \frac{1}{4 \pi \beta_0} \ln{\bigg[\frac{\alpha_s(\bar{\mu}_b)}{\alpha_s(Q)} \bigg]}\,,
\end{equation}
where $\beta_0 = (11 C_A - 4 T_F n_f)/12 \pi$, with $T_F = 1/2$, $C_A=3$, $C_F=4/3$ and $n_f$ is the number of the active flavours.
In addition, when the polarization is measured along the axis $\hat{\bm{n}}= \hat{\bm{T}} \times \hat{\bm{P}}_{h}$, the spin and transverse momentum azimuthal angles are such that $\sin(\phi_{S_{h}} - \phi_j) = 1$. Finally, the expression of the transverse polarization can be given as:

\begin{equation}
    \mathcal{P}(z,j_{\perp}) = \frac{{d\Delta \sigma}/{dz d^2\bm{j}_{\perp}}}{{d \sigma}/{dz d^2\bm{j}_{\perp}}}\,,
\label{eq:1_h_polarization}
\end{equation}
that will be used to fit the Belle data. This will allows us to extract the first moment of the polarizing fragmentation function and its nonperturbative model function.

\section{Phenomenological analysis}
\label{sec:analysis}
We can now proceed with the analysis of Belle data \cite{Guan:2018ckx} for the transverse $\Lambda$ polarization measured in $e^+e^-$ collisions, employing the approach presented in the previous sections. Two data sets are available: one where the $\Lambda$ particle is produced almost back-to-back with respect to a light unpolarized hadron, that we will refer to as double-hadron production (2-h) data set, and one where the $\Lambda$ transverse momentum is measured with respect to the thrust axis, the single-inclusive production (1-h) data set.
We will start considering only the double-hadron production data set and present the corresponding results. In a second phase, we will include in the study also the single-inclusive hadron production case.

\subsection{Double-hadron production data fit}

In this section, by employing Eqs.~(\ref{pol_ratio}),~(\ref{int_qtmax_1}),~(\ref{int_qtmax_2}),~(\ref{B0_full}) and~(\ref{B1_full}), we present the analysis of the polarization of  $\Lambda$/$\bar{\Lambda}$ hyperons produced with a light hadron, $\pi^{\pm}$ or $K^{\pm}$, measured at $\sqrt{s} = 10.58\,\text{GeV}$. The 128 data points are given as a function of $z_{\Lambda}$ and $z_{\pi/K}$, the energy fractions of the $\Lambda$/$\bar{\Lambda}$ and $\pi/K$ particles. For the current analysis we start imposing a cut on large values of the light-hadron energy fractions, $z_{\pi/K}<0.5$, keeping only 96 data points. We will come back to this point in the following.

Notice that here we consider the transverse polarization for inclusive $\Lambda$ particles, namely those directly produced from $q\bar{q}$ fragmentation and those indirectly produced from strong decays of heavier strange baryons.
The purpose of this analysis is to extract ${D}^{\perp  (1)}_{1T,\, \Lambda/q}$ and $M^{\bot}_{D, \Lambda}$, the first moment and the nonperturbative component of the $\Lambda$ polarizing FF.

We will use the following expression to parametrize the $z$ dependence of ${D}^{\perp  (1)}_{1T,\, \Lambda/q}$:
\begin{equation}
     {D}^{\perp  (1)}_{1T,\, \Lambda/q}(z;\mu_b)=\mathcal{N}^p_q(z)\,d_{q/\Lambda}(z;\mu_b)\,,
\label{eq:first_mom1}
\end{equation}
with, as adopted and motivated in Ref.~\cite{DAlesio:2020wjq},  $q = u,\, d,\, s$ and sea, and where
$\mathcal{N}^{\rm p}_q(z)$ (the apex refers to the polarizing FF)  is parametrized as:
\begin{equation}
    \mathcal{N}^{\rm p}_q(z) = N_q z^{a_q}(1-z)^{b_q}\frac{(a_q +b_q )^{(a_q +b_q )}}{a_q^{a_q}b_q^{b_q}}\,.
\label{eq:first_mom2}
\end{equation}
$d_{q/\Lambda}$ is the collinear unpolarized $\Lambda$ fragmentation function
for which we employ 
the AKK08 set~\cite{Albino:2008fy}. This parametrization is given for $\Lambda + \bar{\Lambda}$ and adopts the longitudinal momentum fraction, $z_p$, as scaling variable. In order to separate the two contributions we assume
\begin{equation}
    d_{q/\bar{\Lambda}}(z_p) = d_{\bar{q}/\Lambda}(z_p) = (1- z_p)\, d_{q/\Lambda}(z_p) \,.
\end{equation}
This is a common way to take into account the expected difference between the quark and antiquark FF with a suppressed sea at large $z_p$ with respect to the valence component. Other similar choices have a very little impact on the fit.

Concerning the nonperturbative function $M^{\bot}_{D, \Lambda}$ we employ two different functional forms. The first one is the Gaussian model, analogous to Eq.~(\ref{eq:gaussian_np}):
\begin{equation}
M^{\bot}_{D, \Lambda}(b_T,z) = \exp{\bigg(-\frac{\langle p_\perp^2 \rangle_\text{p} b^2_T}{4 z^2_p}\bigg)}\,,
\end{equation}
where
$\langle p_\perp^2 \rangle_{\text{p}} $ is the Gaussian width, a free parameter that we extract from the fit. The second one is the Power-Law model, Eq.~(\ref{eq:pwrlw_mod_np}):
\begin{equation}
M^{\bot}_{D, \Lambda}(b_T,z) = \frac{2^{2-p}}{\Gamma(p-1)}\,(b_T m/z_p)^{p-1}{K}_{p-1}(b_T m/z_p)\,,
\end{equation}
where we will extract the values of $p$ and $m$ (with the condition $p>1$).

Regarding the collinear FFs of the unpolarized light hadrons, $\pi$ and $K$, we adopt the DSS07 set~\cite{deFlorian:2007aj}, while for $M_D$ we assume a Gaussian model,  compatible with previous extractions, with $\langle p_\perp^2 \rangle = 0.2 \, \text{GeV}^2$ \cite{Anselmino:2005nn}.
We will also consider the PV17 model with its proper nonperturbative functions.

For what concerns the $\Lambda$ unpolarized FFs, for $M_D$ we use either a Gaussian model, with the same width as for the light hadrons, or a Power-Law model, Eq.~(\ref{eq:pwrlw_mod_np}), with the parameters values $p=2$ and $m=1\,\text{GeV}$. These are represented in Fig.~\ref{fig:np_models}.
Notice that all the conversions among the different scaling variables $(z, z_h, z_p)$ involved, Eqs.~(\ref{eq:light_con_def}),~(\ref{eq:en_fract}) and~(\ref{eq:long_fract}), are properly taken into account.
For the $g_K$ function, we use the expressions presented in Section \ref{sec:CSS_RG_eqs}, see Eq. (\ref{eq:all_gk}) and Fig. \ref{fig:all_gk}.
Concerning the $\bm{b}_*$-prescription, we use:
\begin{equation}
    b_{\text{min}} = 2e^{-\gamma_E}/Q \,; \qquad  b_{\text{max}} = 0.6\,\text{GeV}^{-1}\,,
\nonumber
\end{equation}
with $Q = 10.58 \,\text{GeV}$. %
Bearing in mind that the larger is the value of $b_{\text{max}}$, the smaller is the value assumed by $\mu_b$, we chose the value of $b_{\text{max}}$ to be as large as possible, taking into account
at the same time that the AKK set is defined for scales $\ge$ 1 GeV. %

Lastly, for the integration in Eqs.~(\ref{int_qtmax_1}) and~(\ref{int_qtmax_2}), we use $q_{T_{\text{max}}} = 0.25\,Q $, exploring also the impact of different values of $q_{T_{\text{max}}}/Q$.
To perform the phenomenological analysis we use \emph{iMINUIT}~\cite{iminuit} as a minimizer for the $\chi^2$ function, and for the Fourier transforms we employ the \emph{Fast Bessel Transform} algorithm presented in \cite{DBLP:journals/cphysics/KangPST21}.

\subsection{Fit results}
Concerning the first moment of the polarizing FF, Eqs.~(\ref{eq:first_mom1}) and (\ref{eq:first_mom2}), we adopt the same parameter choice considered in Ref.~\cite{DAlesio:2020wjq}, that is:

\begin{equation}
    N_u,\quad N_d,\quad N_s,\quad N_{\text{sea}},\quad a_s,\quad b_u,\quad b_{\text{sea}} \,,
\end{equation}
with all other $a$ and $b$ parameters set to zero. This indeed ensures, once again, a good quality of the fit, keeping the number of parameters under control.
Regarding the nonperturbative functions, we have explored various combinations of them, for a total of 36 fits, plus the PV17 set. %
We have also considered different initial values of the $p$ parameter
of the Power-Law model, noticing that this leads to different chi-square minimum values. This means that we have 8 or 9 free parameters depending on whether we use the Gaussian or the Power-Law model for the polarizing nonperturbative function.
The best results for the double-hadron production fit are reported in Tab.~\ref{tab:best_fit}, adopting different combinations of the $g_K$, $M_D$ and $M_D^\perp$ parametrizations.

\begin{table}[h!]
\centering
\begin{tabular}{|c c c c c|}
 \hline
 $M_D^\perp$ & $M_D$ & $g_K$ & $\chi^2_{\text{dof}}\,$(2-h) & $\chi^2_{\text{dof}}\,$(2-h + 1-h) \\ [.7ex]
 \hline
 Gaussian & Power-Law & Logarithmic & 1.192 & 2.9 \\ [.7ex]
 Power-Law & Power-Law & Logarithmic & 1.21 & 2.43 \\ [.7ex]
 \hline
 Gaussian & Power-Law & PV17 & 1.198 & 3.159 \\ [.7ex]
 \hline
\end{tabular}
\caption{Values of the $\chi^2_{\text{dof}}$ obtained fitting the double hadron production data set only (column ``2-h''), and those obtained for the combined fit (column ``2-h + 1-h''). }
\label{tab:best_fit}
\end{table}

\begin{figure}[!h]
\centerline{\includegraphics[trim =  50 30 30 0,width=7cm]{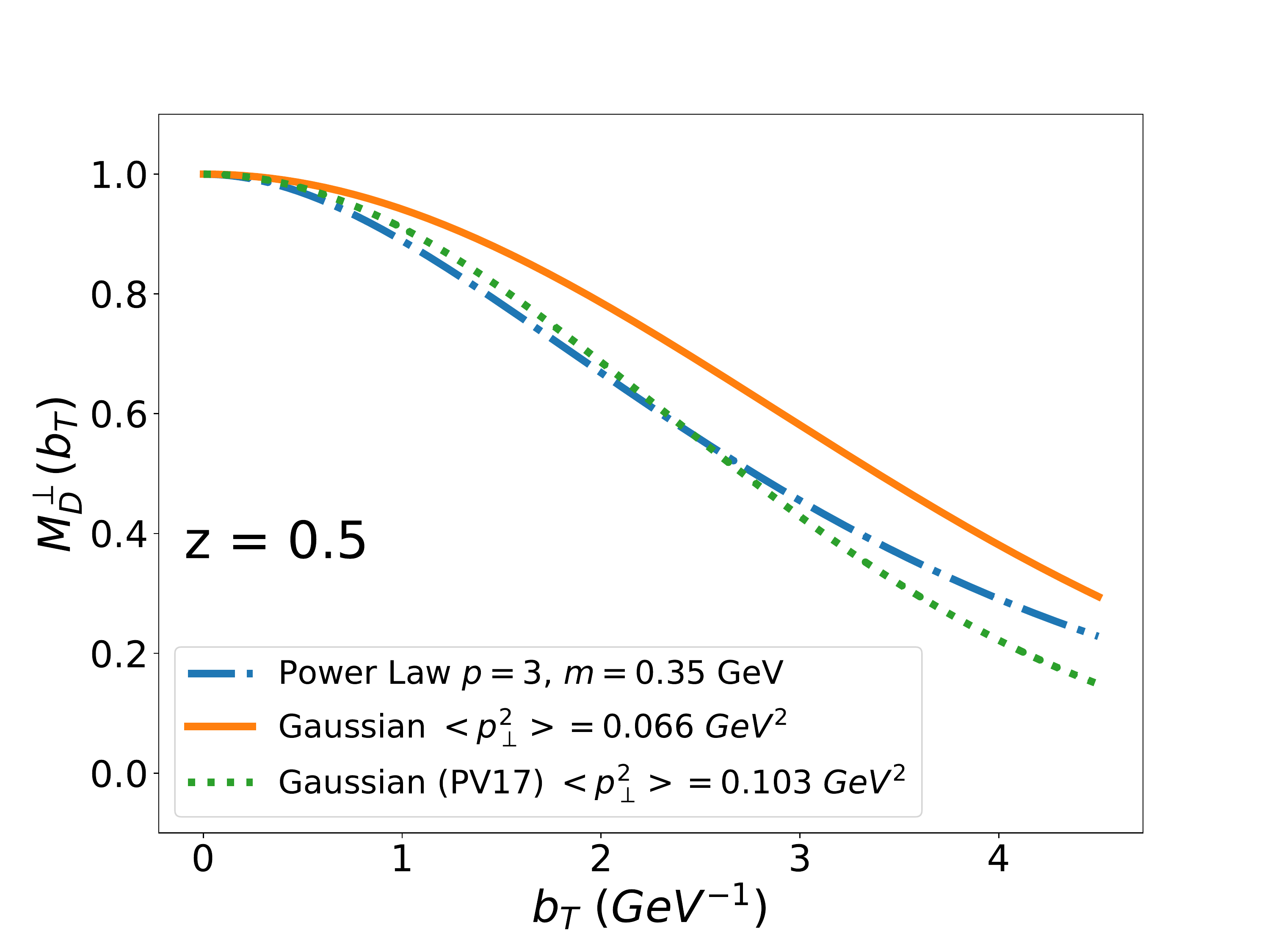}}
    \caption{Nonperturbative functions extracted from the double-hadron data fit: Gaussian (orange solid line), Power-Law (blue dash-dotted line) and PV17 Gaussian (green dotted line) model.}
\label{fig:two_pol_model}
\end{figure}

As reported in Tab.~\ref{tab:parameters}, the parameters values extracted employing the Gaussian or Power-Law models are totally consistent and, similarly, the two polarizing nonperturbative models $M^{\perp}_D$ are compatible, as shown in Fig.~\ref{fig:two_pol_model}. As in the case of the previous extraction~\cite{DAlesio:2020wjq}, we find that only the first moment of the up quark is positive, confirming, moreover, that the contributions to the $\Lambda$ transverse polarization given by the up and the down quarks are opposite in sign. %
\begin{table}[h!]
    \centering
    \begin{tabular}{|ccc|c|}
    \hline
    Parameters & Gaussian & Power-Law & Gaussian (PV17) \\ [.7ex]
    \hline
    $N_u$ & $0.093^{-0.052}_{+0.092}$ & $0.100^{-0.054}_{+0.095}$ & $0.168^{-0.007}_{+0.008}$ \\ [.7ex]
    $N_d$ & $-0.100^{-0.036}_{+0.035}$ & $-0.107^{-0.041}_{+0.036}$  & $-0.138^{-0.011}_{+0.012}$ \\ [.7ex]
    $N_s$ & $-0.117^{-0.09}_{+0.059}$ & $-0.115^{-0.089}_{+0.057}$& $-0.161^{-0.03}_{+0.033}$ \\ [.7ex]
    $N_{sea}$ & $-0.055^{-0.058}_{+0.033}$ & $-0.058^{-0.062}_{+0.034}$& $-0.104^{-0.008}_{+0.008}$ \\ [.7ex]
    $a_s$ & $2.19^{-0.83}_{+1.07}$ & $2.12^{-1.0}_{+1.5}$ & $2.19^{-0.32}_{+0.28}$\\ [.7ex]
    $b_u$ & $3.5^{-2.2}_{+2.8}$ & $3.5^{-1.9}_{+2.8}$ & $4.02^{-0.26}_{+0.28}$ \\ [.7ex]
    $b_{sea}$ & $2.3^{-1.8}_{+2.5}$ & $2.3^{-1.9}_{+2.7}$ & $2.91^{-0.16}_{+0.18}$ \\ [.7ex]
    $\langle p_\perp^2 \rangle_p$ & $0.066^{-0.031}_{+0.039}$ && $0.103^{-0.014}_{+0.015}$  \\ [.7ex]
    $p$ &  & $3.0^{-1.4}_{+2.5}$ &\\ [.7ex]
    $m$ &  & $0.35^{-0.22}_{+0.3}$& \\ [.7ex]
    \hline
\end{tabular}
\caption{Best parameter values for the first moment of the polarizing FF and for the two nonperturbative functions employed to fit the double-hadron data set.}
\label{tab:parameters}
\end{table}

\begin{figure}[t]
{\includegraphics[trim =  0 50 0 140,clip,width=7.5cm]{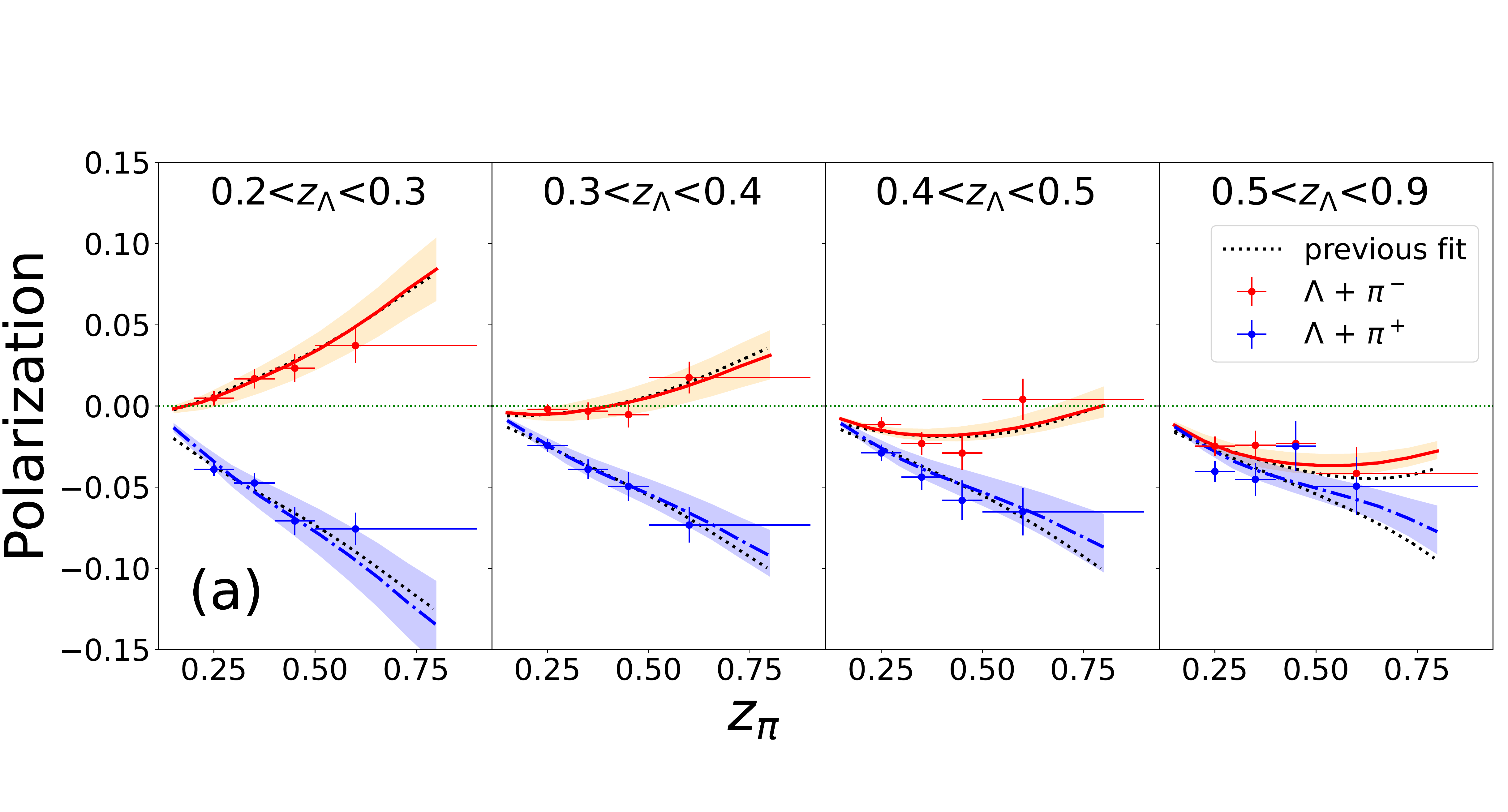}}
{\includegraphics[trim =  0 50 0 140,clip,width=7.5cm]{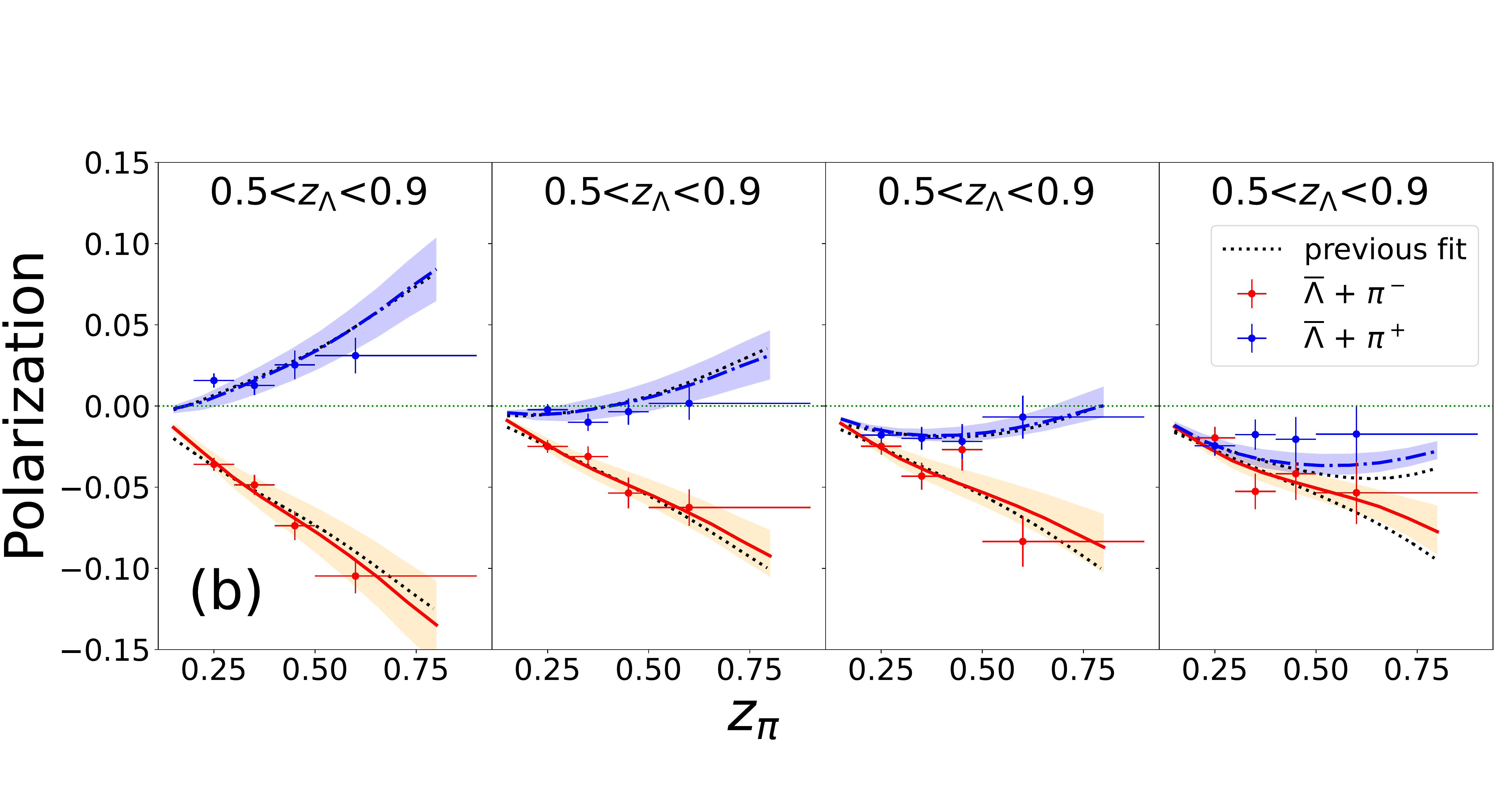}}
\newline
{\includegraphics[trim =  0 50 0 140,clip,width=7.5cm]{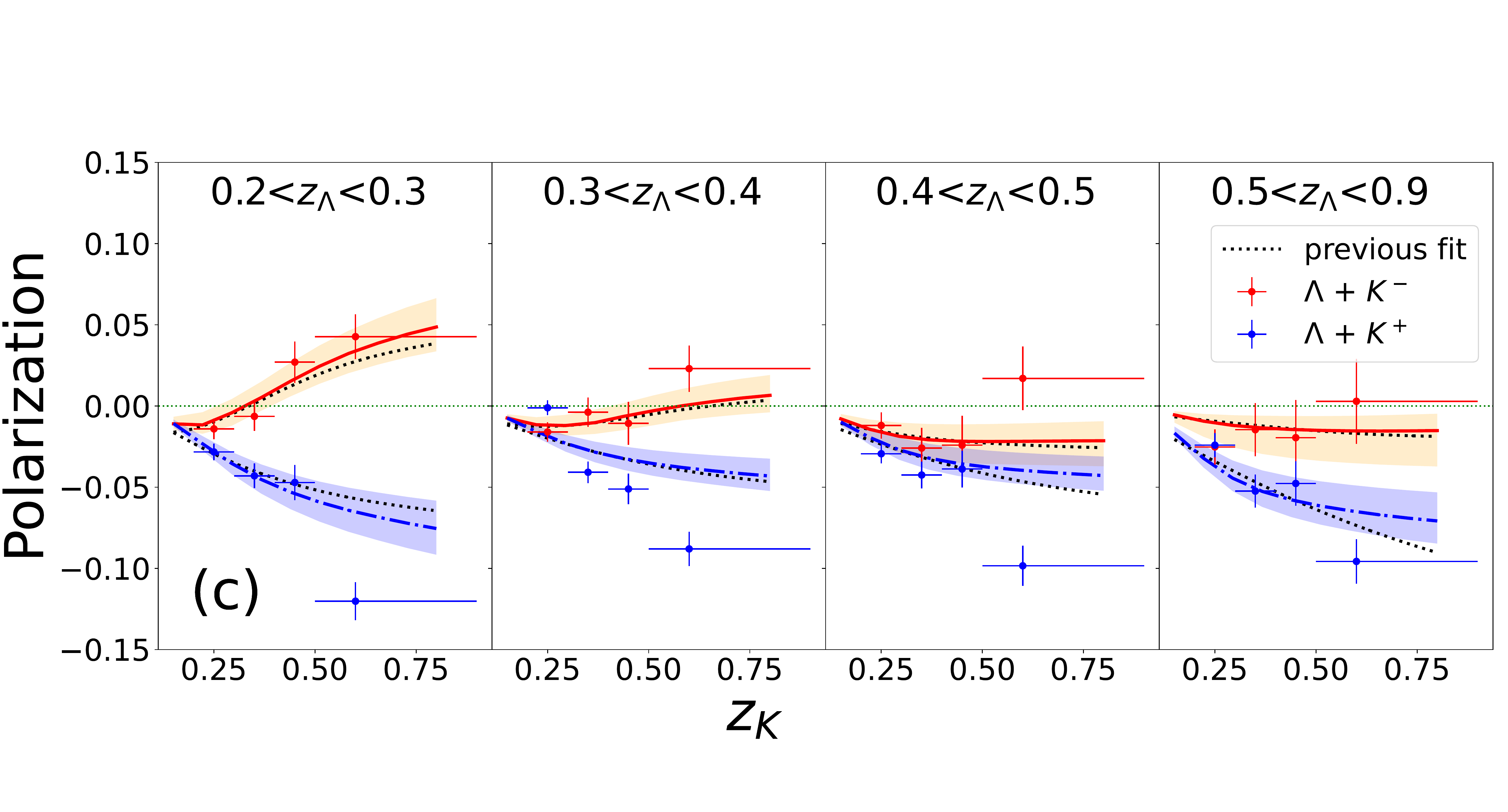}}
{\includegraphics[trim =  0 50 0 140,clip,width=7.5cm]{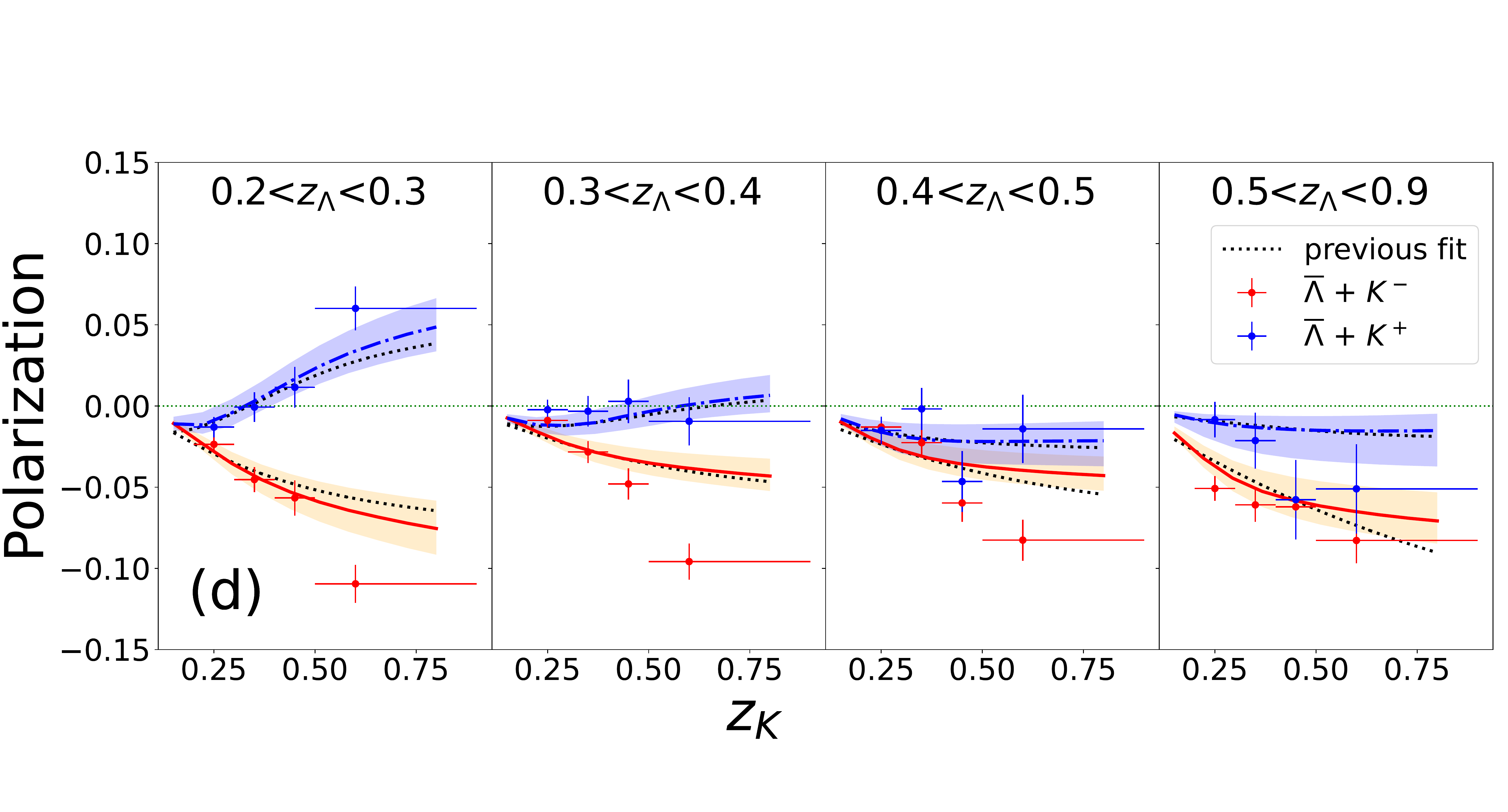}}
\caption{Best-fit estimates (Gaussian model with the parameters of  Tab.~\ref{tab:parameters}) of the transverse polarization for $\Lambda$, $\bar{\Lambda}$ in $e^+e^-\to \Lambda (\bar{\Lambda}) h +X $, for $\Lambda \pi^{\pm}$ (a), $\bar{\Lambda} \pi^{\pm}$ (b), $\Lambda K^{\pm}$ (c), $\bar{\Lambda} K^{\pm}$ (d), as a function of $z_h$ ($h= \pi, K$) for different $z_{\Lambda}$ bins. Data are from Belle \cite{Guan:2018ckx}. The statistical uncertainty bands, at $2\sigma$ level, are also shown. Data for $z_{\pi, K} > 0.5$ are not included in the fit. The results of the fit of Ref.~\cite{DAlesio:2020wjq} (dotted line) are also shown.}
\label{fig:Lh_gauss}
\end{figure}

In Fig.~\ref{fig:Lh_gauss} we show the estimates of the transverse $\Lambda$ polarization, produced in association with a light-hadron, compared against Belle data~\cite{Guan:2018ckx}, adopting the parameters extracted with the Gaussian model. The shaded areas, corresponding to a $2\sigma$ uncertainty, are computed according to the procedure explained in the Appendix of Ref.~\cite{Anselmino:2008sga}.

As in the previous fit~\cite{DAlesio:2020wjq}, where the full TMD machinery was not employed, $\Lambda K^\pm$ data with $z_{K}>0.5$ cannot be described. This can due to different reasons, like the large uncertainty on the last $z_{\pi,K}$ bin and/or the uncertainty affecting the unpolarized FFs in this region.

In Tab.~\ref{tab:chi_range} we report the $\chi^2_{\text{dof}}$ ranges for different fits.
In this comparison we combine different nonperturbative functions to fit data. At fixed $g_K$ function and adopting four combinations of the polarizing and unpolarized FFs with Gaussian and Power-law model, the $\chi^2$ goes from a minimum to a maximum value as reported in Tab~\ref{tab:chi_range}.
It is worth noticing that the extractions are consistent and stable when we employ the same $g_K$.

Moreover, we see that the best fits are found when we make use of the \emph{Logarithmic} $g_K$ function (quite similar to the PV17 model), while the \emph{Quadratic} and \emph{AFGR} functional forms give similar results with worse $\chi^2_{\text{dof}}$. Finally, the worst $\chi^2_{\text{dof}}$s are obtained with the \emph{BLNY} functional form.
\begin{table}[ht!]
    \centering
\begin{tabular}{|c c| }
 \hline
 $g_K$ & $\chi^2_{\text{dof}}$ range  \\ [.7ex]
 \hline

 Logarithmic & 1.192 - 1.287  \\ [.7ex]
 Quadratic & 1.4 - 1.472  \\[.7ex]
 AFGR & 1.474 - 1.514  \\[.7ex]
 BLNY & 1.67 - 1.783  \\[.7ex]
 PV17 & 1.198 - 1.524 \\[.7ex]
 \hline
\end{tabular}
\caption{Range of the $\chi^2_{\text{dof}}$ values for different nonperturbative $g_K$ functions.}
\label{tab:chi_range}
\end{table}

\begin{figure}[!t]
\centerline{\includegraphics[trim =  50 30 30 0,width=12cm]{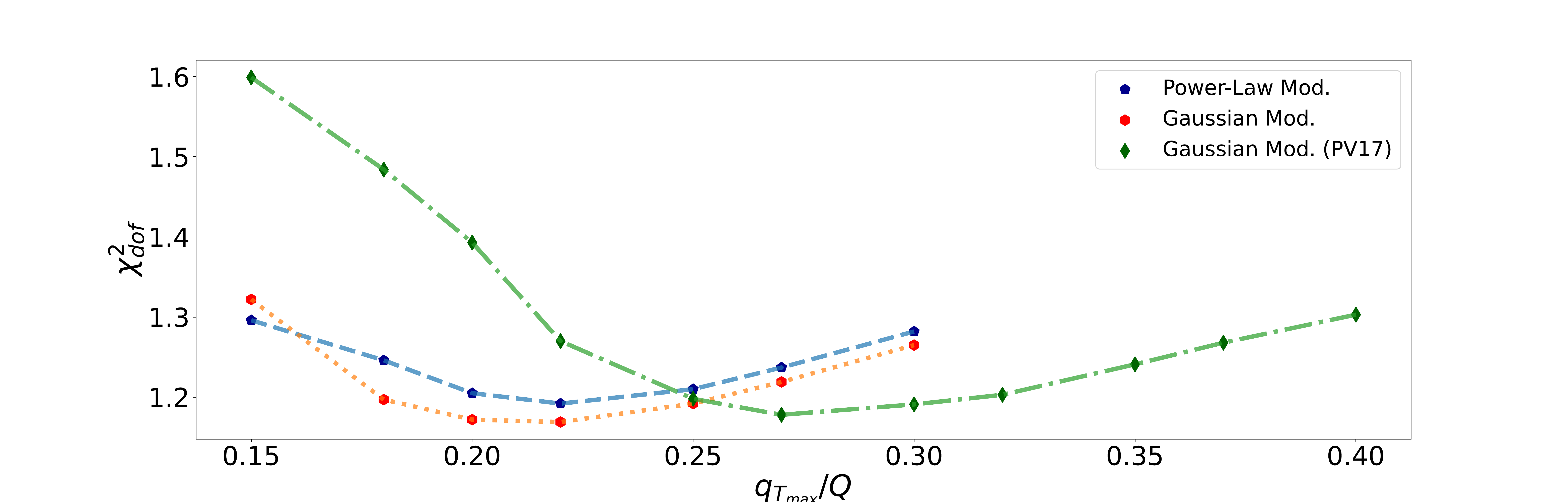}}
    \caption{$\chi^2_{\text{dof}}$ values for the fits obtained adopting the Power-Law  (blue short-dashed line), the Gaussian (red dotted line), both with the \emph{Logarithmic} $g_K$ function, and the PV17 model (green long-dashed line) with the PV17 $g_K$ functional form, as a function of $q_{T_{\text{max}}}/Q$.}
\label{fig:qt_impact}
\end{figure}

In Fig.~\ref{fig:qt_impact} we show the impact of choosing different values of $q_{T_{\text{max}}}/Q$ on the quality of the fit, obtained using the Power-Law and the Gaussian models. In general, the Gaussian model gives smaller $\chi^2_{\text{dof}}$ values. Both models reach their minimum $\chi^2_{\text{dof}}$ value around  $q_{T_{\text{max}}}/Q = 0.22$, while  the Gaussian model with the PV17 parametrization reaches its minimum $\chi^2_{\text{dof}}$ value at $q_{T_{\text{max}}}/Q \simeq 0.27$. %
The growth of the $\chi^2_{\text{dof}}$, as $q_{T_{\text{max}}}/Q$ increases, can be explained considering that we are gradually going out of the validity region of the TMD factorization. Meanwhile, the growth for lower values of $q_{T_{\text{max}}}/Q$ can be attributed to the fact that the smaller is $q_{T_{\text{max}}}/Q$, the larger is the distance between the nodes of the Bessel functions. Hence, the \emph{Fast Bessel Transform} algorithm~\cite{DBLP:journals/cphysics/KangPST21} is not anymore able to sample sufficiently well the integrand, whose Fourier transform is to be computed.

\subsection{Combined fit: double- and single-inclusive hadron production data}
We now discuss the role of single-inclusive polarization data in extracting the polarizing FF, in view of a combined fit of both data sets.
We start checking whether by adopting the results from the 2-h fit one is able to describe the single-inclusive hadron data. This data set is given as a function of $p_{\perp}$, the transverse momentum of the $\Lambda/\bar{\Lambda}$ particle with respect to the thrust axis, that coincides with $j_{\perp}$ in Eqs.~(\ref{eq:1-h_unp}) and (\ref{eq:1-h_pol}), for different bins of the energy fraction $z_{\Lambda}$~\cite{Guan:2018ckx}.
\begin{figure}[h!]
\begin{center}
\includegraphics[trim =  0 50 0 140,clip,width=12.5cm]{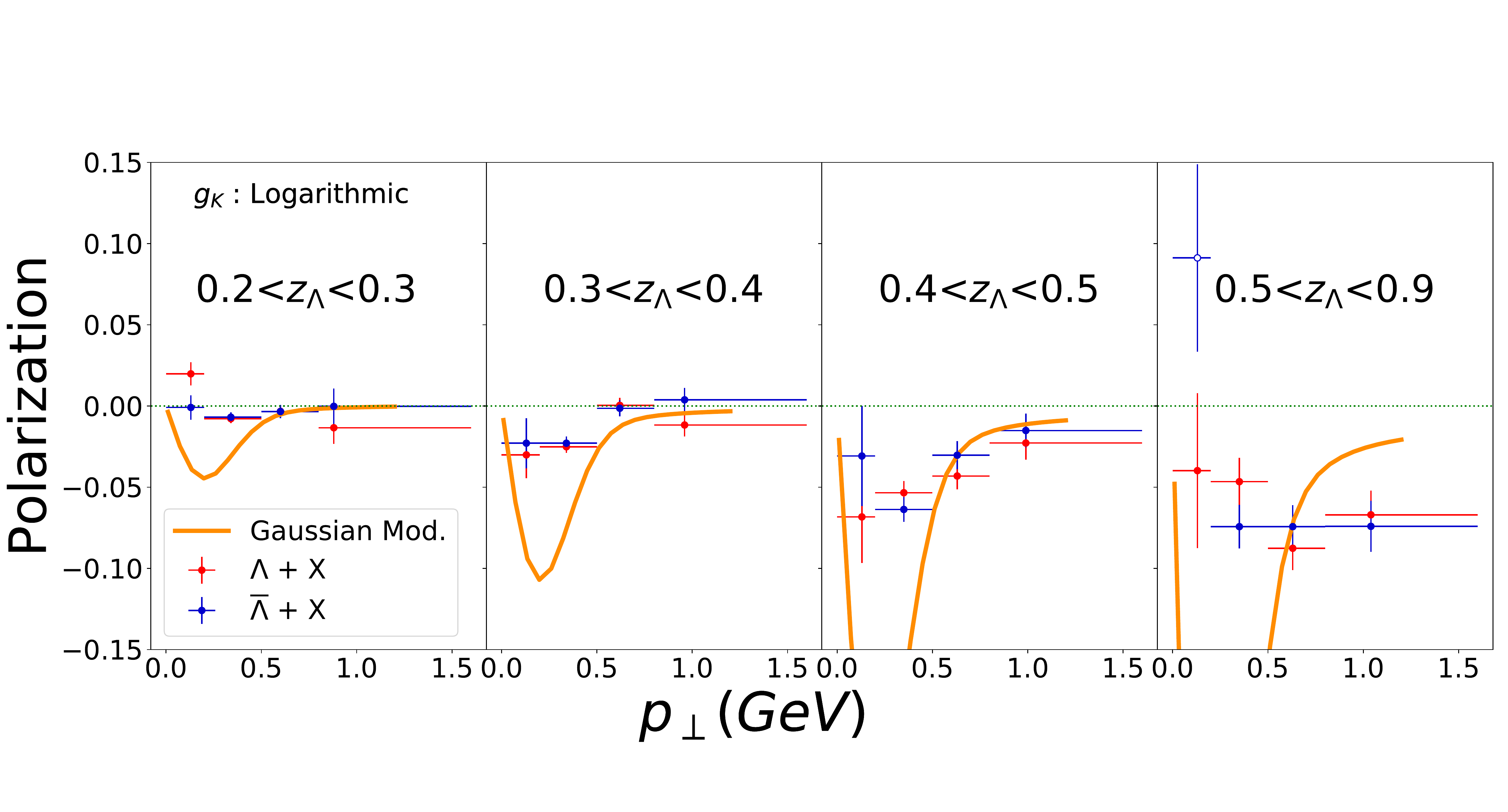}
\end{center}
\caption{Estimates of the transverse polarization for single-inclusive $\Lambda/\bar\Lambda$ production compared with Belle data. The results are obtained using Eqs.~(\ref{eq:1-h_unp}),~(\ref{eq:1-h_pol}) and~(\ref{eq:1_h_polarization}) and the parameter values of Tab.~\ref{tab:parameters}, for the Gaussian model (with the logarithmic functional form for $g_K$)  and the 2-h fit.}
\label{fig:pred_1h}
\end{figure}

As shown in Fig.~\ref{fig:pred_1h} the Gaussian model from the 2-h fit (but the same happens also for the Power-Law model) cannot describe the pattern and the size of the polarization.

As a further step, we perform a combined fit including  {\it both} the single-inclusive (1-h) and the 2-h hadron data sets. In such a case we obtain a large $\chi^2_{\text{dof}}$, as shown in the last column of Tab.~\ref{tab:best_fit}.  The main outcome is that while the single-inclusive data can be described better than in the previous case,
the agreement for the associated light-hadron production case (in particular for pions) is spoiled. This is the main reason for the increasing of the $\chi^2$ (see also Tab.~\ref{tab:chi_square_point}, where we show the $\chi^2$ for data points).
Even exploring all other different combinations of nonperturbative models, as we have done in the double-hadron production section, we keep getting $ \chi^2_{\text{dof}}$ values ranging from $2.4$ to $5.4$.

Since the first moment of the polarizing FF is a collinear quantity, it is expected to be the same in both the double-hadron and the single-inclusive cross sections. Therefore, the fact that the two data sets cannot be fitted simultaneously could suggest that these processes cannot be described within the same factorization theorem and/or by the same  nonperturbative function $M^{\perp}_D$ (see also Refs.~\cite{Boglione:2020cwn,Boglione:2020auc,Boglione:2021wov}). An attempt to explore this issue will be discussed in the following section.

\subsubsection{Different nonperturbative functions $M_D^\perp$ for the 2-h and 1-h cases }
In order to investigate the possible reasons why the combined fit is not satisfactory and why the parameters extracted in the double-hadron fit cannot describe the single-inclusive polarization data, we try to fit both data sets using the same parametrizations for the first moment of the polarizing FF and the same functional form for $M^{\perp}_D$, but with two different sets of parameters for the latter.
Notice that this has to be considered as an attempt to explore the compatibility of the two data sets with the use of a unique and universal nonperturbative function.

More precisely, for the Gaussian model we fit two different Gaussian widths, while in the case of the Power-Law model we fit two different  $(p, m)$ parameter pairs, one for the 2-h data set and one for the 1-h data set. Concerning $g_K$ and the unpolarized nonperturbative functions we use the same functional forms as in Tab.~\ref{tab:best_fit}. As reported in   Tab.~\ref{tab:double_mod_prm}, with this approach we can find much better $\chi^2_{\text{dof}}$'s with respect to those reported in the last column of Tab.~\ref{tab:best_fit}. Indeed, we obtain a $\chi^2_{\text{dof}}$ = 1.801 and 1.565 respectively for the Gaussian and the Power-Law models.
\begin{table}[!h]
\centering
\begin{tabular}{|c c c | c c c|}
 \hline
 \multicolumn{3}{|c|}{Gaussian} & \multicolumn{3}{c|}{Power-Law} \\ [.7ex]
 \hline
 \multicolumn{3}{|c|}{$\chi^2_{\text{dof}} = 1.801$} & \multicolumn{3}{c|}{$\chi^2_{\text{dof}} = 1.565$} \\ [1.ex]
 \hline
  & 2-h & 1-h &   & 2-h & 1-h \\ [.7ex]
  \hline
\multirow{2}{*}{$\langle p_\perp^2 \rangle_p$}  & \multirow{2}{*}{$0.04^{-0.02}_{+0.03}$} & \multirow{2}{*}{$0.2^{-0.01}_{+0}$} & $p$ & $1.352^{-0.055}_{+0.068}$ & $1.623^{-0.011}_{+0.011}$ \\ [1.2ex]
  &   &   & $m$ & $0.151^{-0.024}_{+0.026}$ & $0.48^{-0.005}_{+0.005}$ \\ [.9ex]
 \hline
\end{tabular}
\caption{Values of the best fit parameters for the nonperturbative function, $M_D^\perp$, using two independent sets of parameters for the 2-h and 1-h data sets, for the Gaussian and Power-Law models.}
\label{tab:double_mod_prm}
\end{table}
\begin{figure}[!h]
\begin{center}
\hspace{0.cm}
\includegraphics[width=12cm]{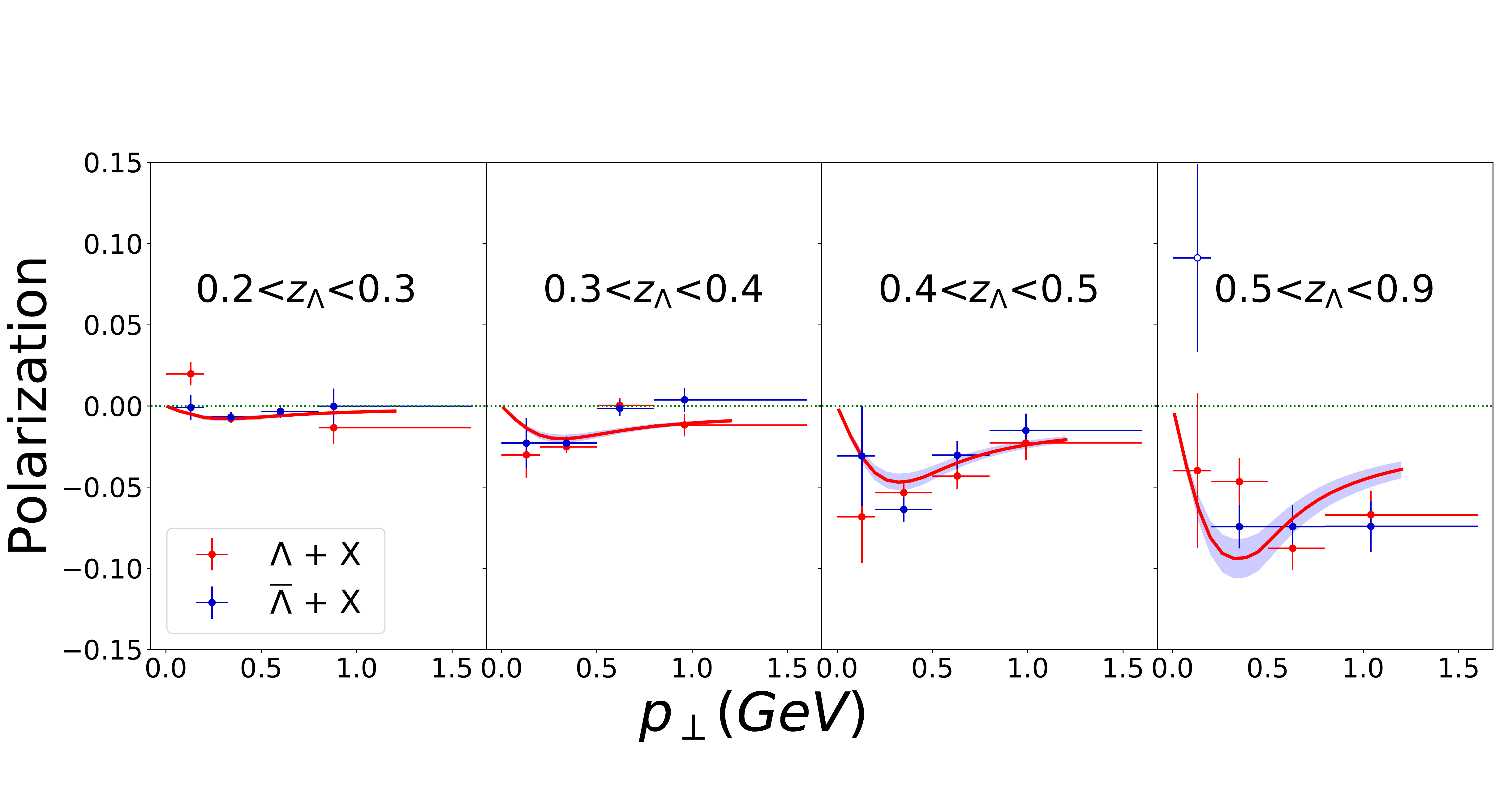}
\caption{Comparison of our fit estimates for the $\Lambda/\bar\Lambda$ single-inclusive polarization against Belle data, adopting the double parametrization for the Power-Law model.}
\label{fig:L1h_pwrlw}
\end{center}
\end{figure}

\begin{center}
\begin{figure}[!th]
\includegraphics[trim =  0 0 0 0,clip,width=9cm]{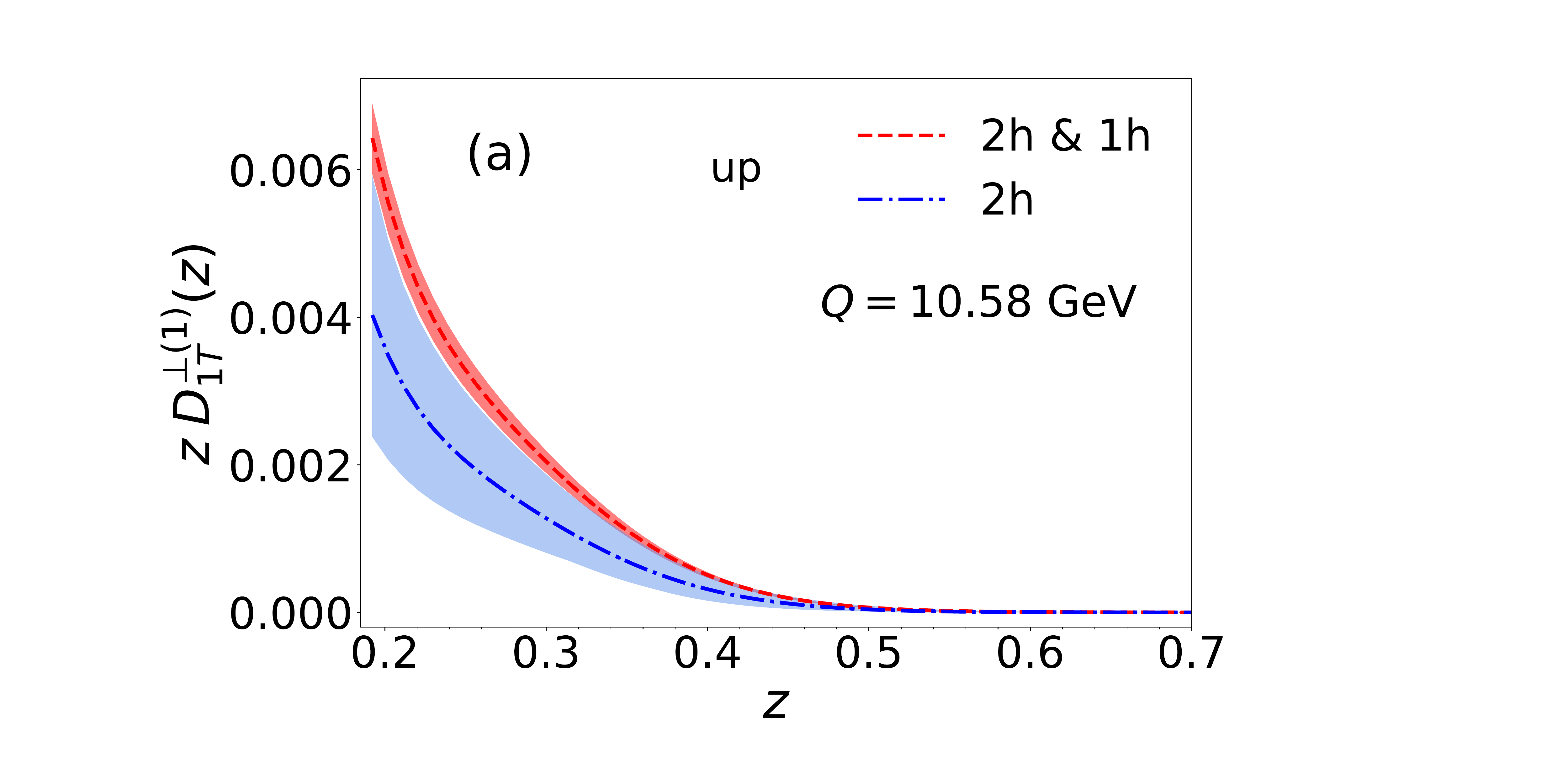}\hspace{-2.7cm}
\includegraphics[trim =  0 0 0 0,clip,width=9cm]{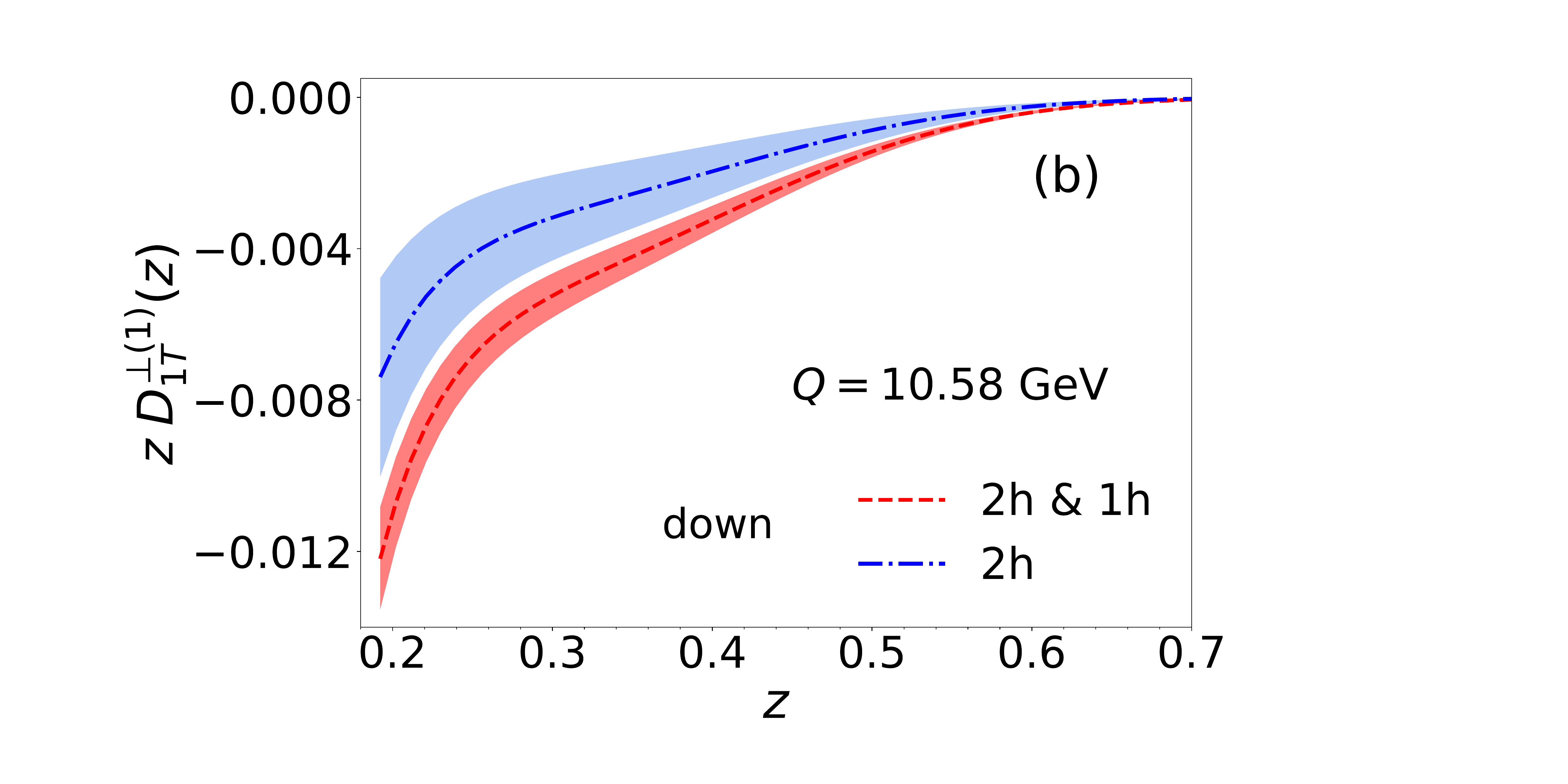}
\includegraphics[trim =  0 0 0 40,clip,width=9cm]{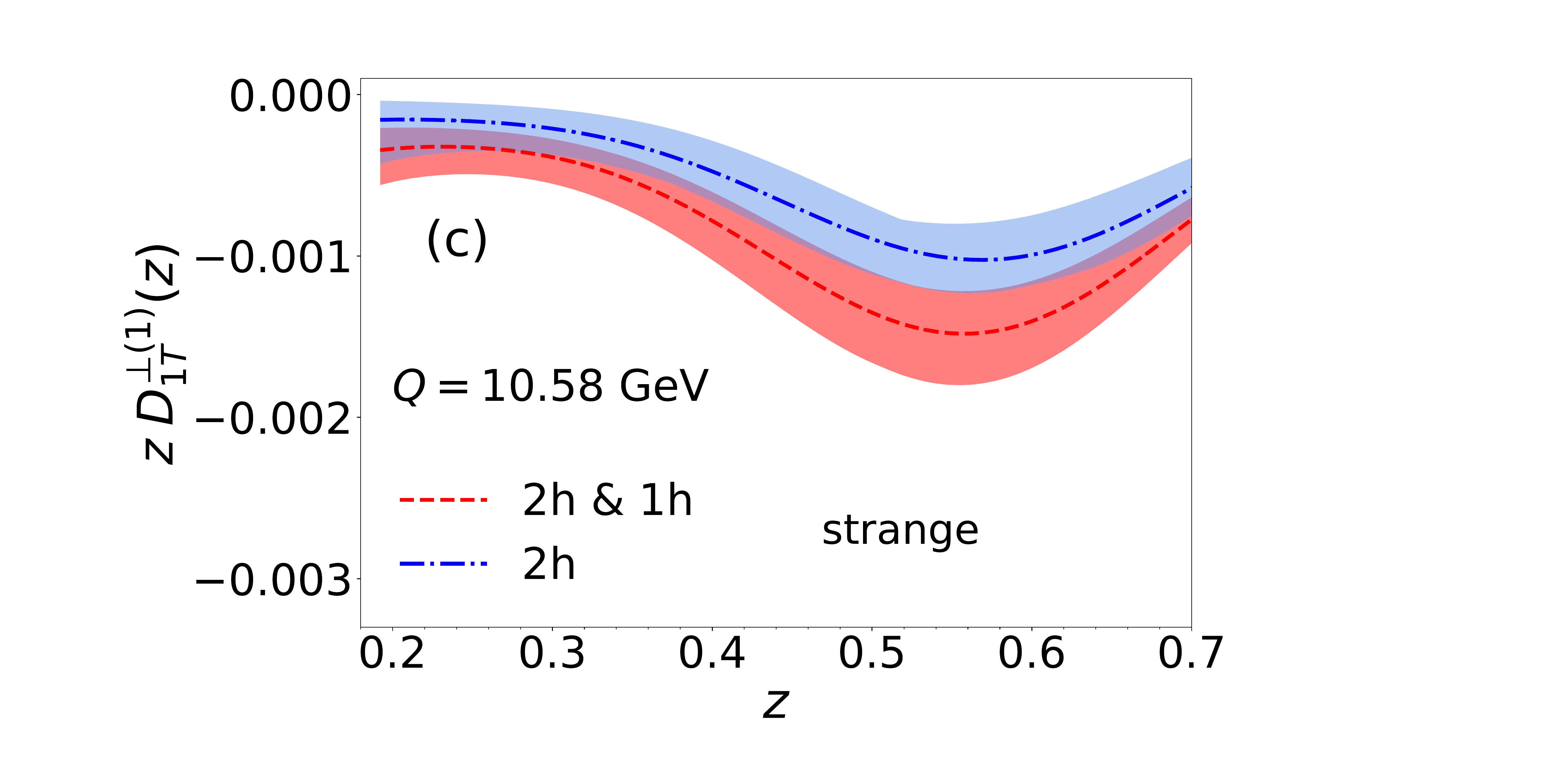}\hspace{-2.5cm}
\includegraphics[trim =  0 0 0 40,clip,width=9cm]{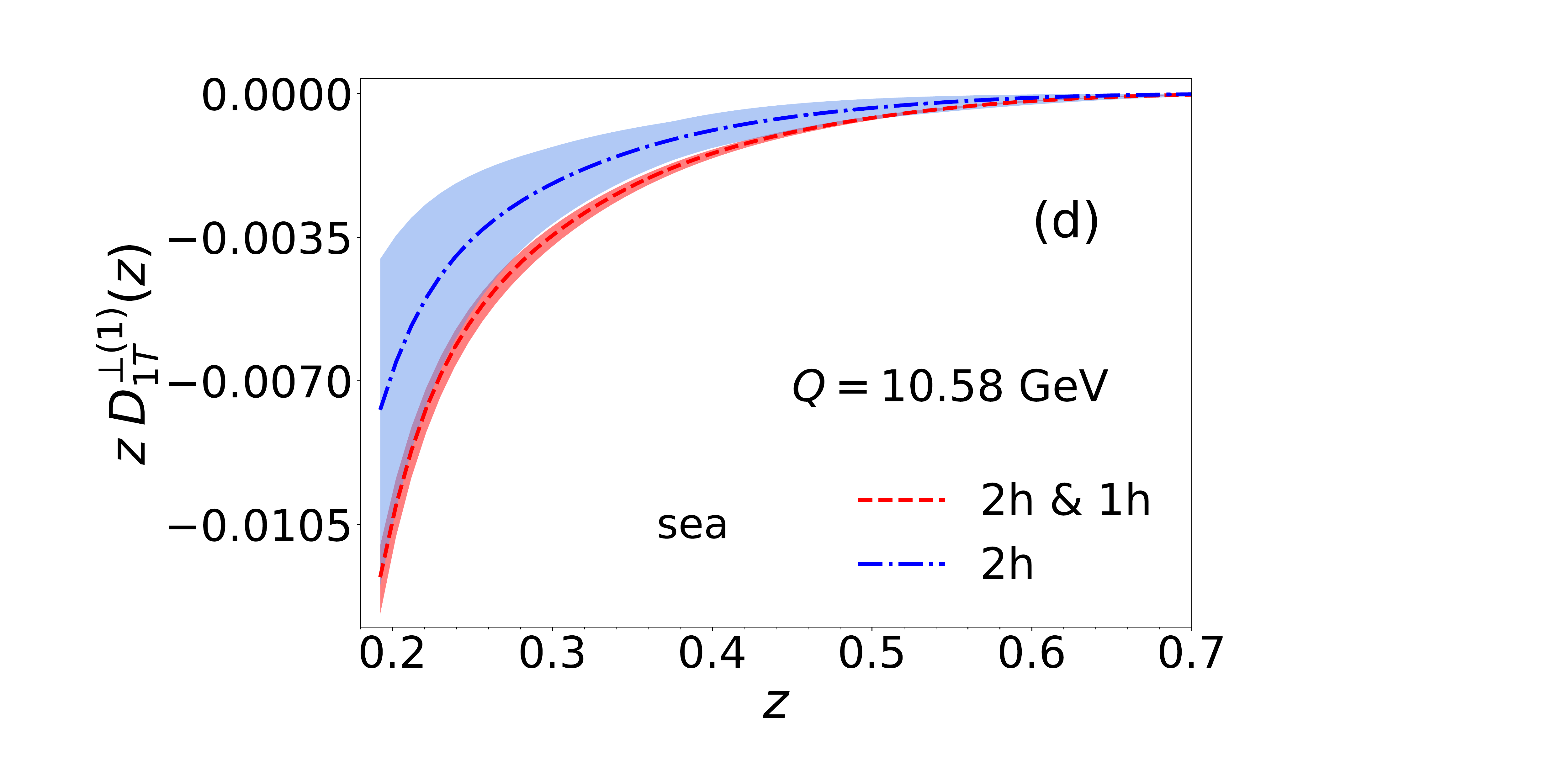}
\caption{First moments of the polarizing FFs, for the up (a), down (b), strange (c) and sea (d) quarks, as obtained from the combined fit (red dashed lines) and the 2-h fit (blue dot-dashed lines). The corresponding statistical uncertainty bands are also shown. }
\label{fig:first_mom}
\end{figure}
\end{center}

Focusing on the results obtained with the Power-Law model, that gives a better $\chi^2_{\text{dof}}$, in Fig.~\ref{fig:L1h_pwrlw} we can see how the estimates for the single-inclusive polarization  describe the experimental data much better than those in Fig.~\ref{fig:pred_1h} (without spoiling the agreement with the 2-h data set, see below). The two different pairs of $(p, m)$ are given in Tab.~\ref{tab:double_mod_prm}.
In Fig.~\ref{fig:first_mom} we also show a comparison of the first moments of the polarizing FFs as extracted in the 2-h fit and in the combined fit, adopting the double model parametrization for $M_D^\perp$. As one can see the two extractions seem not to be compatible, at least  within the uncertainty bands (the full theoretical uncertainty bands, very difficult to estimate, might be larger than the statistical ones). On the other hand, as already stressed, the combined fit requires further insight for what concerns the single-inclusive hadron production. We also notice that the qualitative behaviour and the size of the first moments are comparable with those extracted in Ref.~\cite{DAlesio:2020wjq} .

In Tab.~\ref{tab:chi_square_point} we try to summarize the main findings of our phenomenological analysis, by reporting the $\chi^2$ per data points for each case considered, and still focusing on the Power-Law model.
Starting with the 2-h fit (second column), we see that, besides a small tension in the $\Lambda K$ data set, the overall $\chi^2_{\rm point}$ is extremely good (as already discussed above).
Moving to the results for the combined fit (2-h+1-h), adopting a unique parametrization for $M_D^\perp$ (third column), we see that the description of the $\Lambda\pi$ data set is completely spoiled. Moreover, the $\chi^2_{\rm point}$ for the inclusive data is also extremely high and  the overall $\chi^2_{\rm point}$ doubles its value with respect to the 2-h fit. Finally, the combined fit, but with two separate parametrizations for $M_D^\perp$ (last column), shows that for this scenario the 2-h data sets can be described at the same level of accuracy as in the 2-h fit. More important, the $\chi^2_{\rm point}$ for the inclusive data set reduces significantly, leading to an improvement in the  overall description ($\chi^2_{\rm point}=1.43$)

\begin{table}[!h]
\centering
\begin{tabular}{| c || c c || c  c|| c c|}
 \hline
 &\multicolumn{2}{|c||}{2-h} & \multicolumn{2}{c||}{2-h+1-h (one param.)} & \multicolumn{2}{c|}{2-h+1-h (two param.)} \\ [.7ex]
 \hline
 &  Data &$\chi^2_{\rm point}$ &  Data &$\chi^2_{\rm point}$ &  Data &$\chi^2_{\rm point}$  \\ [1.ex]
 $\Lambda \pi$ &  48 & 0.79 &  48  & 2.2 &  48  & 0.8 \\ [1.ex]
$\Lambda K$ &  48 & 1.4 &  48  & 1.7 &  48  & 1.4 \\ [1.ex]
 $\Lambda $&   &   &  31 & 3.1 &  31 & 2.4  \\ [1.ex]
 TOT. &  96 & 1.1 &  127 & 2.3 &  127 & 1.43 \\ [1.ex]
\hline
\end{tabular}
\caption{$\chi^2$ per data points for the 2-h, 2-h+1-h (single parametrization) and 2-h+1-h (double parametrization) fits, adopting the Power-Law model.}
\label{tab:chi_square_point}
\end{table}

\begin{figure}[!th]
\begin{center}
\hspace{0.cm}
\includegraphics[trim =  20 40 0 100,clip,width=12cm]{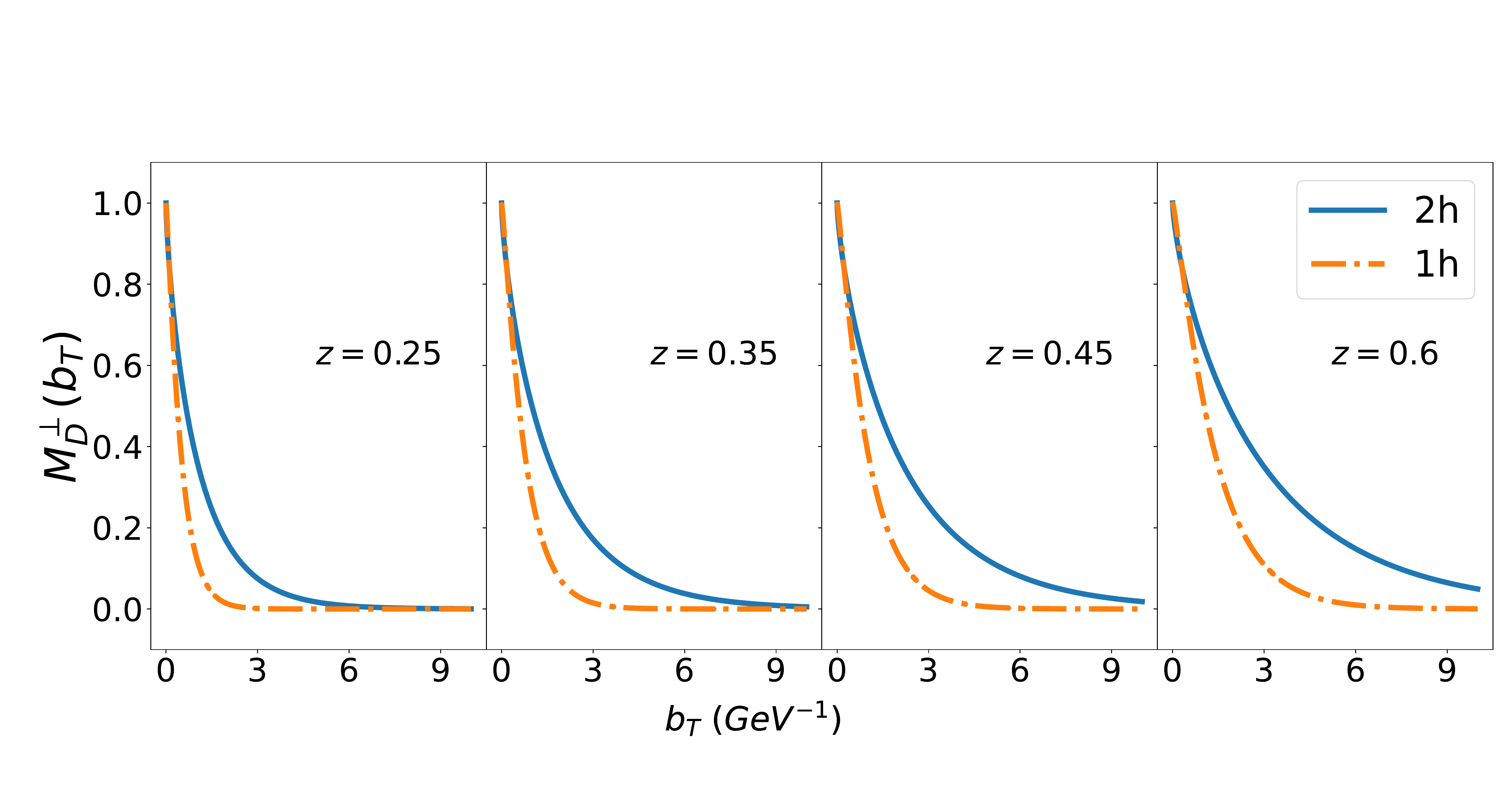}
\caption{Results within the Power-Law model fit for the two $M_D^\perp$ parametrizations in $b_T$-space: 2-h (blue solid line) and 1-h  (orange dash-dotted line) case. }
\label{fig:double_mod_comparison}
\end{center}
\end{figure}

\begin{figure}[!th]
\centerline{\includegraphics[trim =  0 0 0 0,width=8cm]{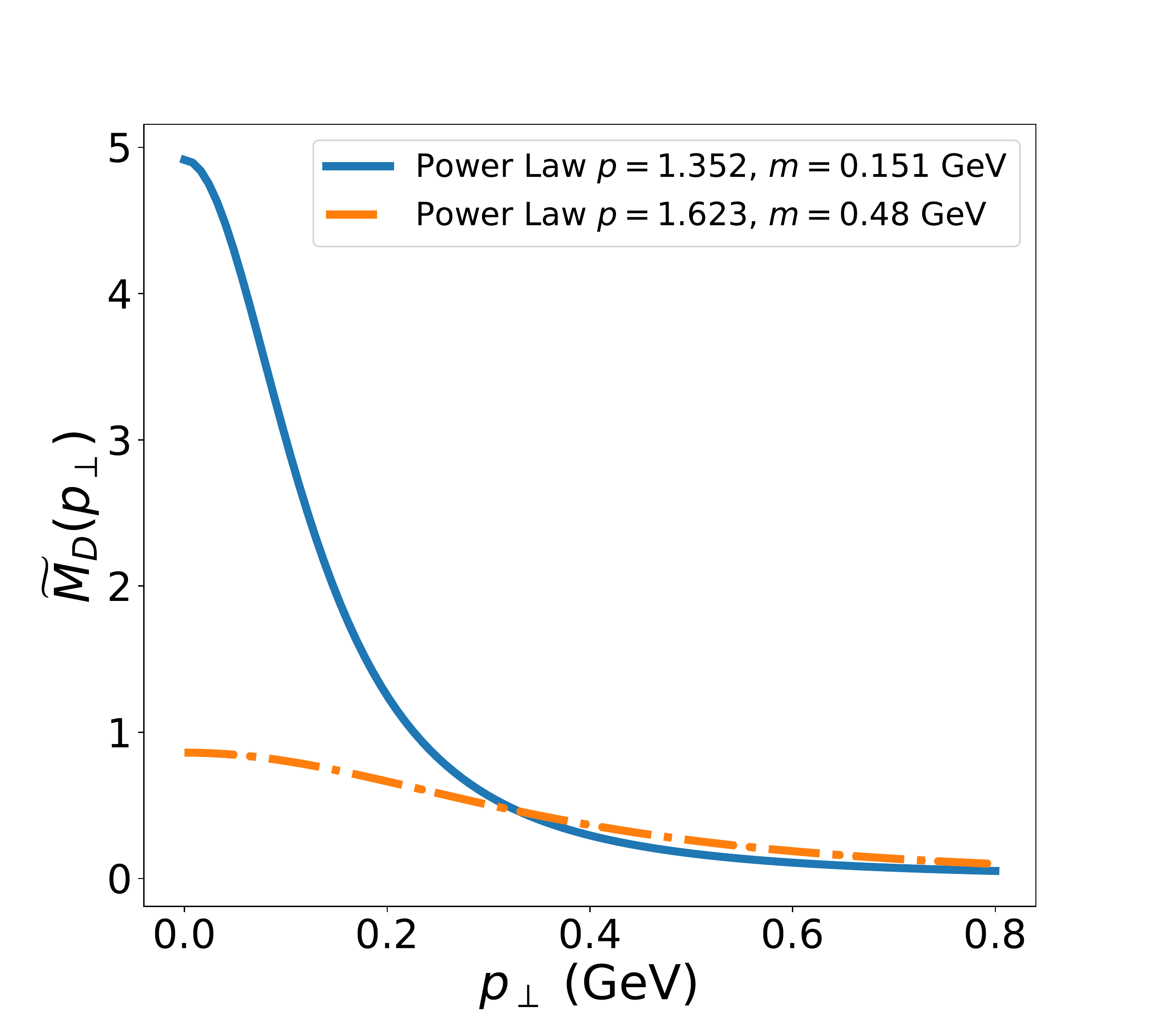}}
    \caption{Results within the Power-Law model fit for the two $M_D^\perp$ parametrizations in $p_{\perp}$-space: 2-h (blue solid line) and 1-h  (orange dash-dotted line) case. The functional form is given in Eq.~(\ref{eq:powerlaw_apx_p}).}
\label{fig:double_mod_comp_pt}
\end{figure}

Let us move to the results obtained for the soft nonperturbative functions, $M_D^\perp$.
Comparing the two Power-Law models as extracted from the 2-h and 1-h data fits in Fig.~\ref{fig:double_mod_comparison}, we can see that they have almost the same behaviour and size at small $b_T$, as expected since the two fragmentation functions should be the same in the collinear limit, but they differ significantly as $b_T$ increases: the model related to the 2-h data set, blue solid line, is wider than the model related to the 1-h data set, orange dash-dotted line.
This corresponds in $\bm{p}_{\perp}$-space to a similar behaviour at large $p_{\perp}$ values, Fig.~\ref{fig:double_mod_comp_pt}, and a sizeably different one at small $p_{\perp}$.

These findings suggest that the two models (same functional form but different parameters) could represent effective convolutions involving two distinct soft factors and that the effects of the recoil against the emission of soft gluons might be different in the two cross sections~\cite{Boglione:2021wov}.

Notice that the fragmentation functions of the single-inclusive hadron production, in the factorization scheme presented in Ref.~\cite{Kang:2020yqw}, coincide with those in the double-hadron production at one-loop order. Indeed, only in this case the hemisphere soft factor, $S_{\text{hemi}}$, corresponds to the soft factor, $\sqrt{S}$, convoluted with each one of the fragmentation functions in the double-hadron production cross sections.

Hence in future analyses, in order to have a more consistent and robust combined fit without using, as an ansatz, two different sets of parameters, one should try to calculate and employ the soft factor $S_{\text{hemi}}$ beyond the one-loop order. An alternative strategy would be to exploit the cross sections as formulated in Ref.~\cite{Makris:2020ltr} within an effective theory framework, or in Refs.~\cite{Boglione:2020cwn,Boglione:2021wov}, where the derivation is based on a CSS approach.

\subsection{Predictions}

We can now try to check explicitly the consistency of the whole approach as well as the role of the TMD evolution by considering another set of data, namely those from the OPAL Collaboration~\cite{OPAL:1997oem}, collected at much larger energy, $\sqrt s =~M_Z$. This data set refers to the single-inclusive $\Lambda$ production, integrated over almost the entire $z_h$ range [0.15-1]. Even if they have large errors, this is the only other available data set for this observable and it is therefore worth studying the impact on our findings.

\begin{figure}[!bt]
\centerline{
\includegraphics[trim =  50 30 30 0,width=7cm]{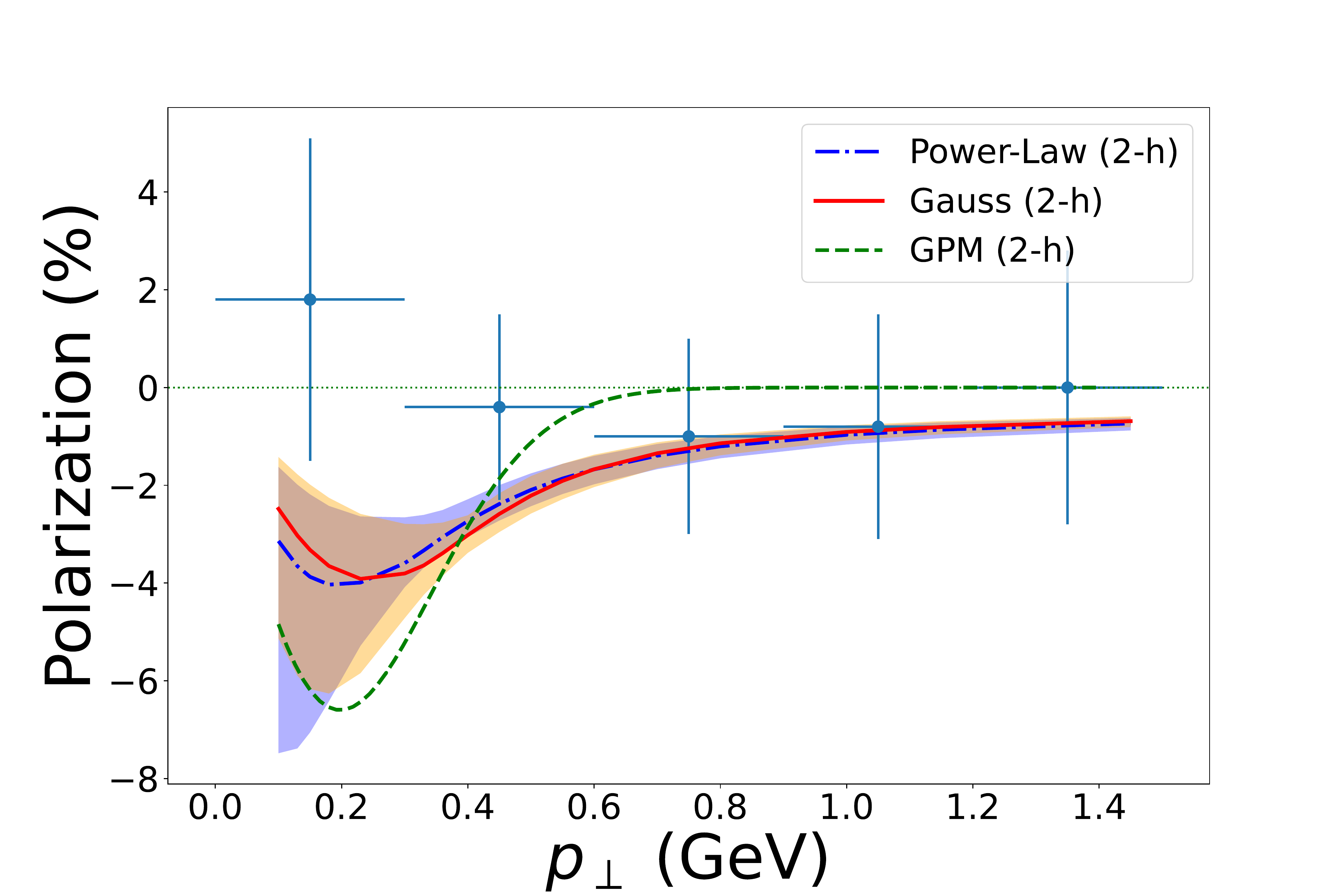}
\includegraphics[trim =  50 30 30 0,width=7cm]{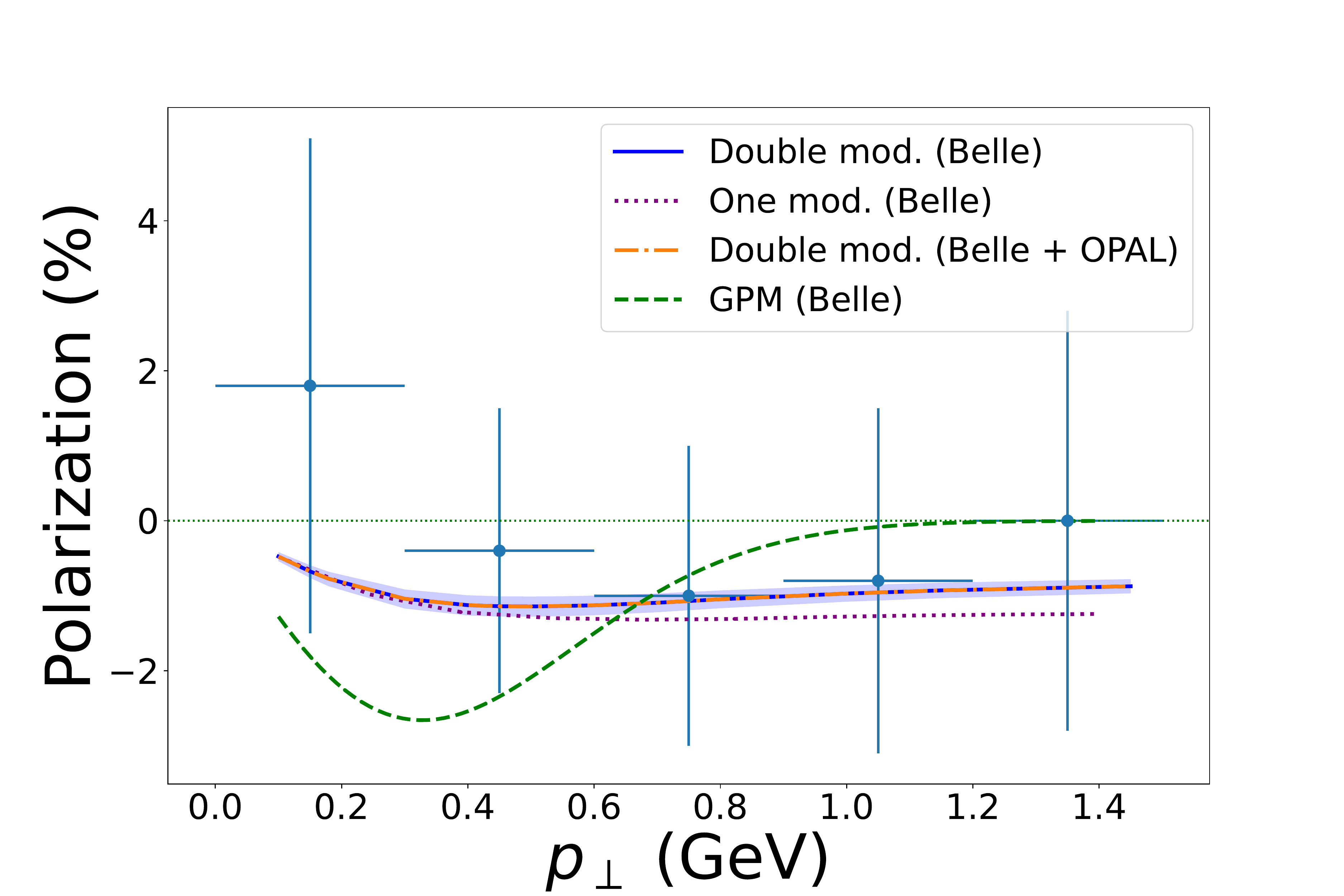}
}
\caption{Predictions for the transverse $\Lambda$ polarization in $e^+e^-\to \Lambda\, X$ as a function of $p_\perp$ at $\sqrt s = M_Z$, integrated over $z_h$ in the range  $[0.15,1]$, against OPAL data~\cite{OPAL:1997oem}. Left panel, 2-h fit:  Power-Law (blu dot-dashed line), Gaussian (red solid line) model and GPM results~\cite{DAlesio:2020wjq} (dashed green line) from double-hadron fits. Right panel, combined fits within the Power-Law parametrization: double-model fit (blue solid line), one-model fit (violet dotted line) and global analysis (Belle + OPAL data) within the double-model fit (orange dot-dashed line). We also show the results within the GPM approach~\cite{DAlesio:2020wjq} (green dashed line). For some cases uncertainty bands are shown.}
\label{fig:opal}
\end{figure}

In Fig.~\ref{fig:opal} we show a series of predictions obtained adopting the different models discussed in the previous sections: all of them are almost compatible with data, within the large error bars, with some differences that deserve several comments.

Let us start from the 2-h fit extraction (left panel). Both the Power-law (blue dot-dashed line) and the Gaussian (red solid line) models are not able to describe the lowest $p_\perp$ data, while they work pretty well above 0.4~GeV. This behaviour shows the same features of the description of the single-inclusive Belle data adopting the 2-h parametrization, see Fig.~\ref{fig:pred_1h}. On the other hand, OPAL data are integrated over a single, much larger $z_h$ bin and this somehow reduces the discrepancies with the theoretical estimates.
For completeness we also show the predictions from the analysis performed in Ref.~\cite{DAlesio:2020wjq}, where a TMD scheme with a simple Gaussian parametrization at fixed scale was adopted (green dashed line, labelled GPM for simplicity). The main difference with respect to the results obtained in the present analysis is the strong suppression, starting already at $p_\perp \simeq 0.5$~GeV. We will come back to this point below.

In the right panel we present the corresponding predictions obtained from the combined fit. In such a case we focus on the Power-Law model that gives the best $\chi^2$ values.
Both extractions, the one based on the single-model parametrization (red dotted line) and the one coming from the double-parametrization fit, with the parameter set for $M_D^\perp$ from the single-inclusive hadron production (blue solid line), are able to describe the data in size and sign pretty well.
On the other hand, one has to recall, see Tab.~\ref{tab:chi_square_point}, that the single-model parametrization for the combined fit gives very large $\chi^2$ for the associated pion production data set.

Even if with some caution, we could observe how the flattening behaviour in $p_\perp$ of these predictions reproduce the \emph{puzzling} plateau of the transverse $\Lambda$ polarization observed in unpolarized hadron-hadron collisions as a function of the transverse momentum of the $\Lambda$ with respect to the direction of the incoming hadrons.

The corresponding GPM results (recall that for the single-inclusive case a simplified phenomenological TMD scheme was adopted), while qualitatively good, show once again their distinguishing Gaussian suppression at large $p_\perp$. The main difference with the corresponding GPM curve shown in the left panel is that the combined fit, even in the GPM approach, gives a much larger Gaussian width and the suppression is somehow shifted at larger $p_\perp$.

It is worth noticing that the overall common good agreement, within the large error bars, of the two approaches, CSS framework (this analysis) and GPM results from Ref.~\cite{DAlesio:2020wjq}, is due to the fact that both of them are controlled by the collinear DGLAP evolution. In the first case from the formal matching onto the collinear scale-dependent FFs, in the second one by construction. What makes the difference, and improves somehow the description in the present study, is the indirect scale dependence of the widths, through the CSS evolution~\cite{collins_2011}, not included in the simplified TMD scheme.

For the sake of completeness, we have also tried a sort of global fit, including the OPAL data set in our analysis. By  adopting the double-model parametrization and the Power-Law model we get an overall $\chi^2_{\rm dof}$ =~1.52. We have to notice that this somehow very low value (at least for the combined fit) is affected by the fact that we have included further data points with large errors. The resulting estimate, shown in the right panel of Fig.~\ref{fig:opal} (orange dot-dashed line) almost coincides with the prediction obtained within the same scenario without including the OPAL data.
Obviously, the large error bars prevent to draw any definite conclusion. Nonetheless, the agreement is quite promising.

More precise data as well as more data at larger $p_\perp$ would be extremely useful to test the model predictions.

\subsection{$SU(2)$ symmetry and charm contribution}

We conclude our phenomenological analysis by discussing here some important (and somehow related)
issues, as we will see below. We stress that what follows has to be considered as a preliminary and exploratory study, to be addressed more carefully in future work.

The first aspect we would like to address is the $SU(2)$ isospin symmetry, advocated for instance in Ref.~\cite{Chen:2021hdn}. In our phenomenological analysis, presented in the previous sections, we have not imposed any constraint on the polarizing FFs for up and down quarks. As already found in the first extraction of the $\Lambda$ polarizing FFs~\cite{DAlesio:2020wjq,Callos:2020qtu}, the use of a unique FF for up and down quarks (within a three-flavour parametrization), even adopting the full TMD formalism, would lead to a very poor quality of the fit with a much larger $\chi^2_{\rm dof}$ (around 2, reaching up to 2.5 if no large-$z_h$ cuts are imposed). When the normalization of the up and down quark polarizing FFs are left free the $\chi^2$ minimization naturally leads to different sizes and, more importantly, to opposite signs.
Therefore, within a three-flavour fit and with present data $SU(2)$ symmetry seems to be ruled out. In this respect nothing changes adopting the proper TMD framework.

Another important issue is the relevance of the charm contribution, as explicitly discussed in the Belle experimental analysis~\cite{Guan:2018ckx}. As we will show below, this could play a role also concerning the isospin symmetry issue, and, quite interestingly, in the description of the large-$z_h$ $\Lambda K$ data. These indeed have not been included in the fit, since they would spoil its quality and, at variance with the large-$z_\pi$ $\Lambda\pi$ data, are very difficult to describe. We have also to notice that the Kaon FFs, especially at large $z$, are affected by large uncertainties and this has to be considered in conjunction with the large last experimental $z$ bin.

\begin{figure}[!ht]
\begin{center}
\hspace{0.cm}
\includegraphics[width=15cm]{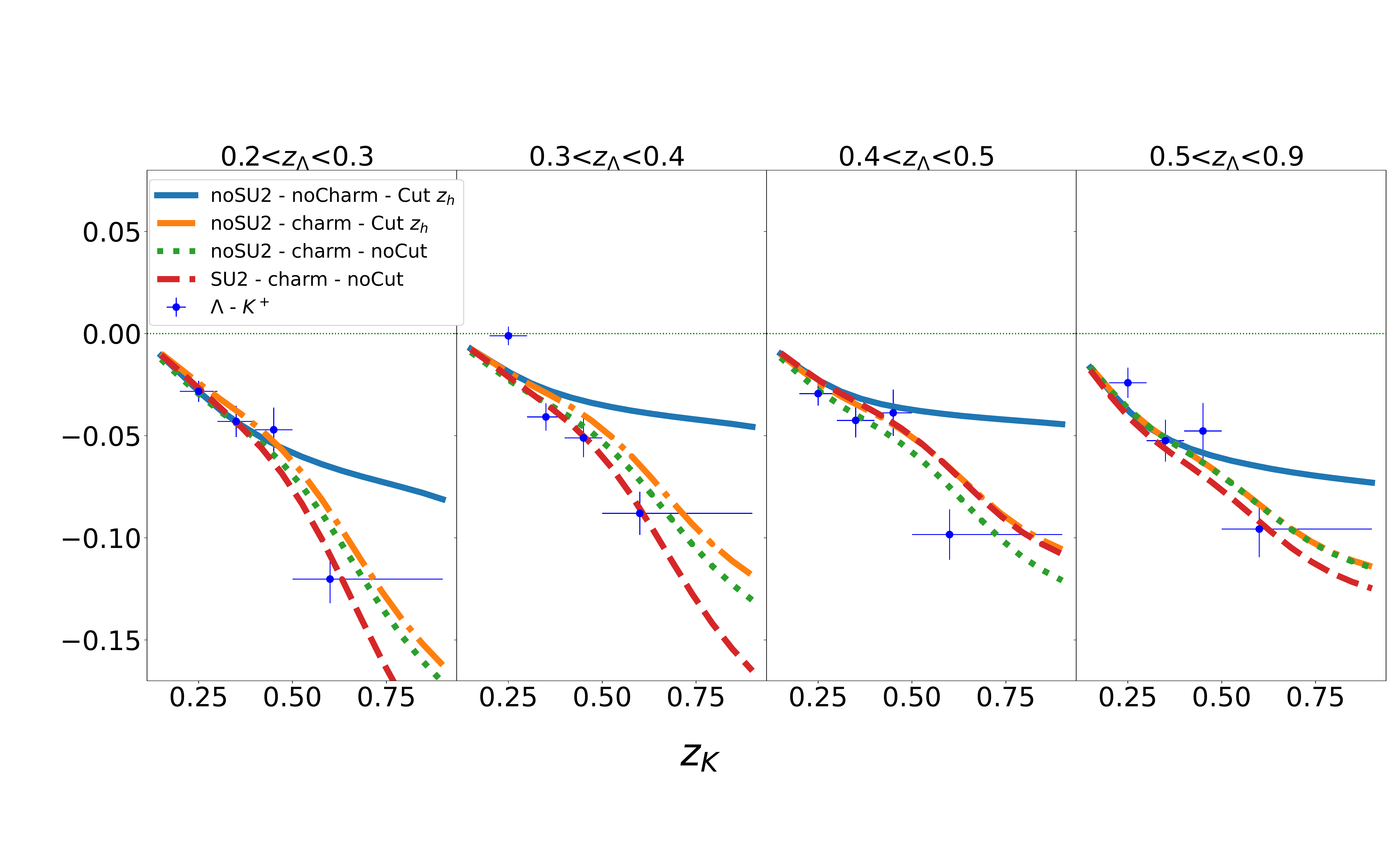}
\caption{Impact of $SU(2)$ symmetry, charm contribution and large $z_K$-cut in the description of $\Lambda K^+$ data: without $SU(2)$ sym., no charm and $z_K$ cut (blue solid line), without $SU(2)$ sym., with charm and $z_K$ cut (orange dot-dashed line), without $SU(2)$ sym., with charm and without $z_K$ cut (green dotted line) and imposing $SU(2)$ sym., with charm and without  $z_K$ cut (red dashed line).}
\label{fig:charm2}
\end{center}
\end{figure}

At this stage we have tried to see what happens by including the charm contribution only in the unpolarized cross section, the denominator of the transverse polarization. While still obtaining a similar $\chi^2_{\rm dof}$ around 1.2 with no $SU(2)$ symmetry, now one can obtain a $\chi^2_{\rm dof}$ around 1.45 when imposing it. Quite interestingly also the description of the large $z_h$ bins for $\Lambda K$ data, even if not included in the fit, improves a lot, as one can see in Fig.~\ref{fig:charm2} for the $\Lambda K^+$ data set. Notice that the agreement for $\Lambda\pi$ data is preserved in all cases considered.

Concerning the polarizing FFs 
we obtain the following results: the inclusion of the charm contribution leads to an increase, in size, of all polarizing FFs; this effect is mainly driven by the increase of the unpolarized cross section in the denominator of the transverse polarization. When we impose also $SU(2)$ symmetry, both the down- and  up-quark polarizing FFs come out positive (it is worth recalling that without $SU(2)$ symmetry the down-quark polarizing FF was negative) and there is a further general increase of all polarizing FFs.

As a general conclusion we can say that by including also the charm contribution (at least in the unpolarized cross section) one can obtain similar good fits imposing or not $SU(2)$ symmetry and/or imposing or not the cut on the large-$z_h$ bin. On the other hand these choices could affect in a different way the corresponding predictions for the transverse $\Lambda$ polarization in other processes, like in SIDIS. We have indeed carried out a preliminary study confirming this hypothesis. A detailed analysis is in progress and will be presented in a future paper.

For its relevance, we also checked the impact of the above assumptions in the combined fit. Once again the one-model fit would give very large $\chi^2_{\rm dof}$. In the double-model fit, focusing on the Power-law parametrization, we obtain similar results as those discussed for the 2-h fit. In other words, imposing $SU(2)$ symmetry once again leads to a very large $\chi^2$.
On the other hand, if one includes the charm contribution, without imposing $SU(2)$ symmetry the $\chi^2$ does not change, while imposing it the $\chi^2$ increases up to 1.7. In both cases the agreement with the large $z_h$ bins for $\Lambda$-$K$ data, even if not included in the fit, improves significantly. 
\section{Conclusions}
\label{sec:conclusions}
In this paper we have carried out a reanalysis of Belle data for the transverse $\Lambda$ polarization in $e^+e^-$ annihilation processes by employing proper TMD factorization theorems and QCD evolution equations within the CSS approach.
This is indeed an important observable in the context of hadron physics and the fragmentation process, in particular to reach deeper insights on the transverse polarization mechanism of $\Lambda$ hyperons. Moreover, it is well defined in terms of a TMD approach, allowing to extract the still poorly known polarizing fragmentation function.

One of our main findings is that this study confirms, in many aspects, the results previously obtained within a former, simplified, extraction of the $\Lambda$ polarizing FF, allowing, at the same time, to have a more robust framework for future studies at different energies and/or for different processes.
A common feature of the two analyses is that for the extracted polarizing FFs a clear separation in flavours can be achieved. Within a three flavour scenario, the best description is obtained with three different valence polarizing FFs (breaking the $SU(2)$ isospin symmetry), with their relative
sign and size determined quite accurately. The need of a sea contribution is also well supported.

Two data sets have been considered with their proper features, employing the corresponding formal description and paying special attention to the nonperturbative functions: the associated production case with a light unpolarized hadron and the inclusive case, where one has to reconstruct the thrust axis direction.
Focusing on the double-hadron production data, we have shown how they can be described extremely well through different combinations of nonperturbative functions. Moreover, the $M^{\perp}_D$ functions so extracted, within the Gaussian and the Power-Law model, are totally compatible. However, the other relevant nonperturbative function, $g_K$, plays a more distinctive role. Indeed, the smallest  $\chi^2_{\text{dof}}$s are found by employing a specific functional form:  the \emph{Logarithmic} one.
One main remark is that none of the models extracted from the double-hadron production data can describe either the size or the pattern of the transverse $\Lambda$  polarization data in the single-inclusive production.

Another striking feature is that the $\Lambda K^+$ and $\bar{\Lambda} K^-$ polarization data, with $z_{K}>0.5$, cannot be described even employing the CSS evolution equations.

As a further attempt we have performed a combined fit of the double and single-inclusive hadron production data sets, by extracting a single or two different sets of parameters for the $M^{\perp}_D$ function. In the first case, the results are still unsatisfactory, while the double-model fit, to be considered as a phenomenological attempt, allows for an overall quite good description.
These findings raise the issue that the two data sets could hardly analysed within the same factorization scheme. For the same reason, the discrepancy between the two models, extracted independently from the two data sets,
could highlight that there are different effects of recoil against emission of soft gluons, and distinct polarization mechanisms for $\Lambda$ hyperons between double and single-inclusive hadron production processes. This deserves a more detailed analysis.

For completeness, we have also considered another data set, the one from the OPAL Collaboration, for the inclusive production case, but at much larger energies. We have checked our predictions against the data as well tried to include them in a global fit. As for the Belle data, the predictions from the associated production data fit are not able to describe the OPAL data, while the combined fit works pretty well. Including the OPAL data set in the fit confirms this finding, with an even smaller $\chi^2$.

Finally, we have explored the role of $SU(2)$ symmetry and of the charm contribution (limiting it to the unpolarized cross section).
Even without carrying out a detailed phenomenological analysis, we can say that  imposing $SU(2)$ symmetry alone spoils the description of the data. On the other hand, the inclusion of the charm contribution (with or without $SU(2)$ symmetry) allows still to get a good fit, improving also the description of the $\Lambda K$ data at large $z_K$. Also in this case further work seems necessary and is underway.

Future experimental analyses and further theoretical developments will certainly help in  understanding and possibly unveiling the hadronization mechanism involved in the transverse $\Lambda$ polarization observed in the two processes considered.
This issue together with the universality properties of the polarizing FF and its TMD evolution, as well as the role of $SU(2)$ symmetry and heavy flavour contributions, could  eventually be explored and clarified in other processes, like SIDIS (in particular at the Electron Ion Collider) and inclusive hadron production in $pp$ collisions.

\acknowledgments
We thank Z.-B.~Kang and A.~Simonelli for useful comments and discussions.
This project has received funding from the European Union’s Horizon 2020 research and innovation programme under grant agreement N.~824093 (STRONG-2020). U.D. and M.Z.~also acknowledge financial support by Fondazione di Sardegna under the project Proton tomography at the LHC, project number F72F20000220007 (University of Cagliari).
L.G. acknowledges support from the US Department of Energy under contract No. DE-FG02-07ER41460.

\appendix

\section{Transverse momenta in different frames and kinematic corrections}
\label{apx:rotations}

Here we give the relations among the different transverse momenta, presented in Section~\ref{sec:2_h_prod}, in the hadron frame. Even if already discussed in the literature, it is worth presenting them once again, taking into account also mass effects.

More precisely, we work in a frame where the two final-state hadrons are exactly back-to-back along the $\pm \hat{\bm{z}}$ direction, the fragmenting quarks have a transverse momentum with respect to the direction of their parent hadrons and the photon has a transverse momentum as well. We label the momenta as follows:
\begin{itemize}
    \item $k$ is the four-momentum of the quark fragmenting into the hadron $h_1$, with four-momentum $P_1$ and moving  along $-\hat{\bm{z}}$;
    \item $p$ is the four-momentum of the anti-quark fragmenting into the hadron $h_2$, with four-momentum $P_2$ and moving along $+\hat{\bm{z}}$;
    \item $\bm{k}_T$ and $\bm{p}_T$  are respectively the transverse momenta of the quark and the anti-quark with respect to their associated hadron momenta, while  $\bm{k}_{\perp}$ and $\bm{p}_{\perp}$ are respectively the transverse momenta of the two hadrons with respect to their parent quark directions of motion;
    \item $q$ is the four-momentum of the virtual photon and $\bm{q}_T$ its transverse component.
\end{itemize}
By adopting standard light-cone coordinates ($a=(a^+,a^-, \bm{a}_T)$),
we define the quarks and photon momenta as follows~\cite{Boer:1997mf}:
\begin{align}
        k &=\Bigg( \frac{k^2_T}{2 k^-},k^-,\bm{k}_T  \Bigg) \quad \text{where} \quad k_T=\abs{\bm{k}_T}\\
        p &= \Bigg( p^+,\frac{p^2_T}{2 p^+}, \bm{p}_T  \Bigg) \quad \text{where} \quad p_T=\abs{\bm{p}_T}\\
        q &= \Bigg( \frac{\widetilde{Q}}{\sqrt{2}},\frac{\widetilde{Q}}{\sqrt{2}},\bm{q}_T  \Bigg) \,,
\end{align}
with $\widetilde{Q}^2 = Q^2 + q^2_T$, where $q_T=\abs{\bm{q}_T}$. Since $ k +p = q$, we have directly
\begin{align}
        k^- =  \frac{{Q}}{\sqrt{2}} +\mathcal{O}{(q^2_T/Q^2)}&; \qquad k^+ = \frac{k^2_T}{\sqrt{2}Q} +\mathcal{O}{(q^2_T/Q^2)} \\
        k^0 = \frac{Q}{2} +\mathcal{O}{(k^2_T/Q^2)}&; \qquad  k^3 = - \frac{Q}{2} +\mathcal{O}{(k^2_T/Q^2)}\\
        p^+ =  \frac{{Q}}{\sqrt{2}} +\mathcal{O}{(q^2_T/Q^2)}&; \qquad p^- = \frac{p^2_T}{\sqrt{2}Q} +\mathcal{O}{(q^2_T/Q^2)}\,,
\end{align}
with $\bm{k}_T + \bm{p}_T = \bm{q}_T$.
Concerning the hadron momenta, we have:
\begin{equation}
    \begin{split}
        P_1 &= \Bigg(\frac{M^2_1}{z_1 {Q}\sqrt{2}}, \frac{z_1 {Q}}{\sqrt{2}},\bm{0}_{T}  \Bigg)  \\
        P_2 &= \Bigg( \frac{z_2 {Q}}{\sqrt{2}} ,\frac{M^2_2}{z_2 {Q}\sqrt{2}} ,\bm{0}_{T}  \Bigg) \,, \\
    \end{split}
\end{equation}
where we have used the light-cone momentum fractions:
\begin{equation}
    z_1 = \frac{P_1^-}{k^-}; \qquad z_2 = \frac{P^+_2}{p^+}\,.
\end{equation}

\subsection{$k_T$ vs.~$k_{\perp}$ }
We now derive the relation between the transverse momenta $\bm{k}_T$ and $\bm{k}_{\perp}$.
Firstly, without loss of generality, we consider the hadron $h_1$ and the fragmenting quark momenta sitting on the $\widehat{xz}$ plane, namely
\begin{equation}
    P_1=(P_1^0,0,0,P_1^3)\qquad k= (k^0, k_{T}, 0, k^3)\,,
\end{equation}
with the quark transverse momentum, $\bm{k}_T$, along the positive $\hat{\bm{x}}$-axis. Then, we move to a frame where the quark has zero transverse momentum component by employing the following rotation:
\begin{equation}
        R_{y}(\xi, \lambda) = \begin{pmatrix}
                    1 & 0 & 0 & 0 \\
                    0 & \xi & 0 & -\lambda \\
                    0 & 0 & 1 & 0 \\
                    0 & \lambda & 0 & \xi \\
                    \end{pmatrix}     \,,
\end{equation}
where  $\xi = - k^3/k^0 \,$ and $\lambda =- k_T/k^0 $, with $\xi^2 + \lambda^2 =1$. The transformed quark four-momentum $k'$ is then given as:
\begin{equation}
    k'=R_{y}(\xi, \lambda) k = (k^0,0,0,k'^3)\,.
\end{equation}
 By applying the above rotation to the hadron four-momentum $P_1$, we get
\begin{equation}
    P'_{1}= R_{y}(\xi, \lambda)\,P_{1}  = (P_1^0, -\lambda{P}_1^3 ,0,{P'}_1^3)\,,
\end{equation}
with
\begin{align}
    P_1^0 &= \frac{z_1 Q}{2} \Bigg( 1 + \frac{M^2_1}{ z^2_1 {Q}^2}\Bigg) \\
    {P'}_1^3 &=- \xi \frac{z_1 Q}{2}\Bigg( 1 - \frac{M^2_1}{ z^2_1 {Q}^2}\Bigg)\,\\
    -\lambda {P}_1^3 &=-k_T z_1 \Bigg( 1 - \frac{M^2_1}{ z^2_1 {Q}^2}\Bigg)\,.
\end{align}
The hadron transverse momentum with respect to the quark direction, usually called $\bm{k}_{\perp}$, is now along the negative $\hat{\bm{x}}$-axis.
Therefore, from the above relations, we can see that $\bm{k}_{T}$ and $\bm{k}_{\perp}$ are anti-parallel (neglecting $k_T^2/Q^2$ corrections) and, using Eq.~(\ref{eq:long_fract}), that $k_{\perp} = k_T\, z_{p_1}$ (this indeed can be derived also as an exact relation). This eventually leads to:
\begin{align}
        \bm{k}_{\perp} &= -\bm{k}_T z_{p_1} \,,
\end{align}
that, for a massless hadron, reduces to the usual $\bm{k}_T = -\bm{k}_{\perp}/z_{1}$ relation.  The same is valid for the second hadron, that is $\bm{p}_{\perp} = -\bm{p}_T z_{p_2}$.

\subsection{$P_{1T}$ vs.~$q_{T}$ }

In order to find the relation between $\bm{P}_{1T}$, the transverse momentum of $h_1$ with respect to $h_2$ in the hadron frame, and $\bm{q}_T$, the transverse momentum of the photon in the new frame, where the two hadrons are back-to-back, we can use more directly the tensor $g^{\mu \nu}_{\perp}$ \cite{Boer:1997mf}:
\begin{equation}
  g^{\mu \nu}_{\perp} = g^{\mu \nu} -\hat{t}^{\mu} \hat{t}^{\nu} + \hat{z}^{\mu} \hat{z}^{\nu} \quad\quad {\rm with}\quad\quad
        \hat{t}^{\mu} = \frac{q^{\mu}}{Q}  \quad\quad
        \hat{z}^{\mu} = 2 \frac{P^{\mu}_2}{z_{h_2}Q} - \hat{t}^{\mu} \,.
\end{equation}
By contracting the above tensor with the four-momentum of hadron $h_1$,  we find
\begin{equation}
   P^{\mu}_{1T} = g^{\mu \nu}_{\perp} P_{1 \nu} = P^{\mu}_1 + P^{\mu}_2 \bigg(2 \frac{z_1}{z_{h_2}} - \frac{z_{h_1}}{z_{h_2}}  \bigg) - z_1 q^{\mu} \,,
\end{equation}
where we have neglected the mass of the second hadron, $M_{2} =0$. By explicit calculation one can then see that all components but the transverse one are zero, that is
\begin{equation}
    \bm{P}_{1T} = - z_1 \bm{q}_T\,,
\end{equation}
where $z_1$ is the light-cone momentum fraction. 

\section{Convolutions and Fourier transforms}
\label{apx2:conv_FT}
In order to exploit the CSS evolution equations, in Section~\ref{sec:2_h_prod} we showed how the convolutions, in $\bm{k}_T$-space, can be written in the conjugate $\bm{b}_T$-space through Fourier transforms of the TMD-FFs. We give here some details about this procedure. The first convolution we consider is $F_{UU}$, defined according to Eq.~(\ref{eq:conv_general}),
\begin{equation}
\begin{split}
    \mathcal{F}[\omega  D \bar{D}] = \sum_q e^2_q \int d^2\bm{k}_T d^2\bm{p}_T \, \delta^{(2)}(\bm{k}_T + \bm{p}_T - \bm{q}_T) \,\omega(\bm{k}_T, \bm{p}_T) D(z_1,\bm{k}_{\bot}) \bar{D}(z_2,\bm{p}_{\bot}) \,,
\end{split}
\end{equation}
as follows:
\begin{equation}
\begin{split}
    F_{UU} =& \mathcal{F}[D_1 \bar{D}_1]\,.
\end{split}
\end{equation}
It is trivial to convert this convolution in $\bm{b}_T$-space by employing the Fourier transform definition of the TMD unpolarized FF:
\begin{equation}
\widetilde{D}_1(z,b_T) =  \int d^2\bm{k}_T \, e^{i\bm{b}_T \cdot \bm{k}_T } D_1(z,{k}_{\perp})
= 2 \pi \int dk_T \, k_T J_0(b_T k_T) D_1(z,k_{\perp})\,,
\end{equation}
where we have used the integral representation of $J_0$, the Bessel function of the first kind
\begin{equation}
    \int^{2\pi}_0 d\theta \, e^{i b_T k_T \cos\theta} = 2\pi J_0(b_T k_T)  \,.
\label{bessel_0}
\end{equation}
With the above relations, the $F_{UU}$ convolution in $\bm{b}_T $-space can be written as:
\begin{equation}
\begin{split}
     F_{UU} =& \mathcal{F}[D_1 \bar{D}_1]
    =  \sum_q e^2_q \int \frac{d^2 \bm{b}_T}{(2 \pi)^2} \, e^{-i\bm{b}_T \cdot \bm{q}_T} \widetilde{D}_1(z_1,b_T) \widetilde{\bar{D}}_1(z_2,b_T) \\
    =& \sum_q e^2_q \int \frac{d b_T}{(2 \pi)} \, b_T J_0(b_T\, q_T) \widetilde{D}_1(z_1,b_T) \widetilde{\bar{D}}_1(z_2,b_T)
    = \mathcal{B}_0 \Big[\widetilde{D}_1 \widetilde{\bar{D}}_1\Big] \,.
\end{split}
\end{equation}
The second convolution we consider is $F^{\sin(\phi_1 - \phi_{S_1})}_{TU}$, defined as follows:
\begin{equation}
\begin{split}
    F^{\sin(\phi_1 - \phi_{S_1})}_{TU} =& \mathcal{F}\bigg[\frac{\hat{\bm{h}}\cdot \bm{k}_T}{M_{1}}D^{\perp}_{1T} \bar{D}_1\bigg]\,. \\
\end{split}
\end{equation}
To work out this convolution we need the Fourier transform of the unpolarized fragmentation function, presented above, and the one of the polarizing fragmentation function multiplied by $k^i_T$, the $i$-th component of the quark  transverse momentum with respect to the hadron direction:
\begin{eqnarray}
\int d^2\bm{k}_T \, \frac{k^i_T}{M_{h}} e^{i\bm{b}_T \cdot \bm{k}_T} D^{\perp}_{1T}(z,k_{\perp})
    &=& \frac{-i}{M_{h}} \pr{}{b^i_T}  \int  d^2\bm{k}_T \, e^{i\bm{b}_T \cdot \bm{k}_T} D^{\perp}_{1T}(z,k_{\perp})  \\
    &=& -\frac{i}{M_{h}} \pr{}{b^i_T} \widetilde{D}^{\perp}_{1T}(z,b_T) \\
    %
    %
    &=& - 2\frac{i}{M_{h}} b^i_T \pr{}{b^2_T} \widetilde{D}^{\perp}_{1T}(z,b_T) \\
    &=& i b^i_T M_{h} \widetilde{D}^{\perp  (1)}_{1T}(z,b_T)\,,
\end{eqnarray}
where we have used the definition of the Fourier transform of the polarizing FF
\begin{equation}
\widetilde{D}^{\perp}_{1T}(z,b_T) = \int d^2\bm{k}_T \, e^{i\bm{b}_T \cdot \bm{k}_T } D^{\perp}_{1T}(z,\bm{k}_{\perp}) \,.
\label{eq:pFF_bt}
\end{equation} 
In the very last step we have also introduced
the first moment of the polarizing FF in $\bm{b}_T $-space:
\begin{equation}
    \widetilde{D}^{\perp  (1)}_{1T}(z,b_T) = \bigg(-\frac{2}{M^2_{h}}\pr{}{b^2_T} \widetilde{D}^{\perp}_{1T}(z,b_T) \bigg)\; .
\label{eq:apx_first_mom_bt}
\end{equation}
This term can be related to the first moment in $\bm{k}_T $-space, defined as:
\begin{equation}
    {D}^{\perp  (1)}_{1T}(z) = \int d^2 \bm{k}_{\perp} \, \bigg(\frac{\bm{k}^2_{\perp}}{2 z^2 M^2_{h}} \bigg)  {D}^{\perp}_{1T}(z,{k}_{\perp})\,,
\label{eq:first_mom_apx}
\end{equation}
by using the relation:
\begin{equation}
    \lim_{b_T \to 0} \widetilde{D}^{\perp  (1)}_{1T}(z,b_T) =\frac{1}{z^2} {D}^{\perp  (1)}_{1T}(z)\; .
\label{eq:first_mom_kt_bt}
\end{equation}
We have indeed that:
\begin{equation}
    \begin{split}
        \widetilde{D}^{\perp  (1)}_{1T}(z,b_T) &= -\frac{2}{M^2_{h}}\pr{}{b^2_T} \widetilde{D}^{\perp}_{1T}(z,b_T) \\
        &= -\frac{2 \pi}{b_T M^2_{h}}\pr{}{b_T}\int \frac{dk_{\perp} \, k_{\perp}}{z^2} J_0\bigg(\frac{b_T k_{\perp}}{z}\bigg) D^{\perp  }_{1T}(z,k_{\perp }) \\
        &=\frac{2 \pi}{M^2_{h} z^3}  \int dk_{\perp} \frac{k^2_{\perp}}{b_T} J_1\bigg(\frac{b_T k_{\perp}}{z}\bigg)D^{\perp }_{1T}(z,k_{\perp })\,,
    \end{split}
\end{equation}
where we have used 
\begin{equation}
        \pr{}{b_T} J_0(a b_T) = - a J_1(ab_T)\,.
\end{equation}
Then by employing the additional relation
\begin{equation}
    \lim_{b_T\to0} \frac{J_1(a b_T)}{b_T} = \frac{a}{2}\,,
\end{equation}
we are able to proof Eq.~(\ref{eq:first_mom_kt_bt}):
\begin{equation}
    \begin{split}
         \lim_{b_T\to0} \widetilde{D}^{\perp  (1)}_{1T}(z,b_T) = \frac{1}{z^2}\int d^2 \bm{k}_{\perp} \, \bigg(\frac{\bm{k}^2_{\perp}}{2 z^2 M^2_h} \bigg)  {D}^{\perp}_{1T}(z,{k}_{\perp})\,.
    \end{split}
\end{equation}
\noindent Finally, with the above results we can write the $F^{\sin(\phi_1 - \phi_{S_1})}_{TU}$ convolution in $\bm{b}_T $-space:
\begin{equation}
\begin{split}
    & F^{\sin(\phi_1 - \phi_{S_1})}_{TU} = \mathcal{F}\bigg[\frac{\hat{h}\cdot \bm{k}_T}{M_{1}}D^{\perp}_{1T} \bar{D}_1\bigg] \\
     =&  \sum_q e^2_q \int d^2\bm{k}_T d^2\bm{p}_T \, \delta^{(2)}(\bm{k}_T + \bm{p}_T - \bm{q}_T) \,\frac{\hat{h}\cdot \bm{k}_T}{M_{1}} D^{\perp}_{1T}(z_1,\bm{k}_T) \bar{D}_1(z_2,\bm{p}_T) \\
    =& M_{1} \sum_q e^2_q \int \frac{d^2 \bm{b}_T}{(2 \pi)^2} \, e^{-i\bm{b}_T \cdot \bm{q}_T} (i \hat{h}\cdot \bm{b}_T) \widetilde{D}^{\perp  (1)}_{1T}(z_1,b_T) \widetilde{\bar{D}}_1(z_2,b_T) \\
    =& M_{1} \sum_q e^2_q \int \frac{d b_T}{2 \pi} \,b^2_T J_1(q_T\, b_T)  \widetilde{D}^{\perp  (1)}_{1T}(z_1,b_T) \widetilde{\bar{D}}_1(z_2,b_T) \\
    =& M_{1}  \mathcal{B}_1 \Big[\widetilde{D}^{\perp  (1)}_{1T} \widetilde{\bar{D}}_1\Big]\,,
\end{split}
\end{equation}
where we have used the integral definition of the $J_1$ Bessel function:
\begin{equation}
    \int^{2\pi}_0 d\theta \, e^{i b_T k_T \cos\theta}\,\cos\theta = (2\pi i) J_1(b_T k_T) \,.
\label{bessel_1}
\end{equation}
%
\subsection{Gaussian model}
As a first example of parametrization for the transverse momentum dependence of the FFs we present a simple Gaussian model, deriving its Fourier transform and first moment. We use the following Gaussian model for a generic TMD fragmentation function:
\begin{equation}
    {D}(z,k_{\perp}) = {D}(z,0) \,\frac{e^{-k^2_{\perp }/\langle k_\perp^2 \rangle_D }}{\pi \langle k_\perp^2 \rangle} .
\end{equation}
where $\langle k_\perp^2 \rangle$ is the Gaussian width of the unpolarized TMD-FF. 
According to Eq.~(\ref{eq:first_mom_apx}), its first moment is:
\begin{equation}
        D^{(1)}(z) = \int d^2 \bm{k}_{\perp} \, \bigg(\frac{\bm{k}^2_{\perp}}{2 z^2 M^2_h} \bigg)  {D}(z,{k}_{\perp})
        = {D}(z,0)\, \frac{\langle k_\perp^2 \rangle^2_D}{2 z^2 M^2_h \langle k_\perp^2 \rangle}
\end{equation}
and the $\bm{b}_T$-space fragmentation function is
\begin{eqnarray}
\widetilde{D}(z,b_T) & = & \int d^2\bm{k}_T \, e^{i\bm{b}_T \cdot \bm{k}_T } D(z,k_{\perp}) = 2 \pi \int dk_T \, k_T J_0(b_T k_T) D(z,k_{\perp})\nonumber \\
&=& 2 \pi \int \frac{dk_{\perp } \, k_{\perp}}{z^2} J_0\bigg(\frac{b_T k_{\perp}}{z}\bigg) D(z,k_{\perp})
= \frac{{D}(z,0) }{z^2} \frac{\langle k_\perp^2 \rangle_D}{\langle k_\perp^2 \rangle} e^{-b^2_T \langle k_\perp^2 \rangle_D/(4 z^2)}\,.\nonumber\\
\end{eqnarray}
We notice that, when $\langle k_\perp^2 \rangle_D = \langle k_\perp^2 \rangle$, as in the case of the unpolarized FF,
\begin{equation}
    \lim_{b_T\to 0}\widetilde{D}_1(z,b_T) = \frac{1}{z^2}{D}_1(z,0) \,,
\end{equation}
where ${D}_1 $ coincides with $d_{j/h}$ in the OPE in Eq.~(\ref{OPE_unp}).
Employing the first-moment definition in $\bm{b}_T$-space, Eq.~(\ref{eq:apx_first_mom_bt}), we have:
\begin{equation}
    \widetilde{D}^{ (1)}(z,b_T) = \bigg(-\frac{2}{M^2_{h}}\pr{}{b^2_T} \widetilde{D}(z,b_T) \bigg)
    %
    = {D}(z,0) \frac{1}{2 z^4 M^2_h} \frac{\langle k_\perp^2 \rangle^2_D}{\langle k_\perp^2 \rangle} \, e^{-b^2_T \langle k_\perp^2 \rangle_D/(4 z^2)} \,.
\end{equation}
Then, by using Eq.~(\ref{eq:first_mom_kt_bt}), we find:
\begin{equation}
        \lim_{b_T\to 0}\widetilde{D}^{ (1)}(z,b_T) = \frac{1}{z^2} \bigg[ \frac{1}{2 M^2_h z^2} \frac{\langle k_\perp^2 \rangle^2_D}{\langle k_\perp^2 \rangle} {D}(z,0) \bigg]
        = \frac{1}{z^2} D^{(1)}(z) \, .
\end{equation}

\subsection{Power-Law model}
The second model used to parametrize the transverse momentum dependence of FFs, of which we calculate here its Fourier transform and first moment, is the the Power-Law model. This is defined as follows:
\begin{equation}
    D(z,k_{\perp}) = D(z,0) \frac{\Gamma(p)}{\pi \Gamma(p-1)}\frac{m^{2(p-1)}}{(k^2_{\perp} + m^2)^p} \,.
\label{eq:powerlaw_apx_p}
\end{equation}
Its Fourier transform is:
\begin{equation}
   \widetilde{D}(z,b_T) = D(z,0)\frac{2^{2-p}}{\Gamma(p-1)}\,(b_T m/z)^{p-1}{K}_{p-1}(b_T m/z)\,.
\end{equation}
The integrated first moment and the first moment in $\bm{b}_T$ space are:
\begin{equation}
    D^{(1)}(z) = D(z,0) \frac{1}{M^2_h z^2} \frac{m^2}{2(p-2)}
\end{equation}

\begin{equation}
    \widetilde{D}^{(1)}(z,b_T) =D(z,0) \frac{2^{2-p}}{\Gamma(p-1)} \frac{ m^2}{M^2_h z^2} (b_T m/z)^{p-2}{K}_{p-2}(b_T m/z)\,.
\end{equation}



\begin{thebibliography}{10}

\bibitem{Dick:1975ty}
L.~Dick et~al., \emph{{Spin Effects in the Inclusive Reactions $\pi^\pm \,
  p(\uparrow) \to \pi^\pm$ Anything at 8~GeV/c}},
  \href{https://doi.org/10.1016/0370-2693(75)90252-X}{\emph{Phys. Lett. B}
  {\bfseries 57} (1975) 93}.

\bibitem{Klem:1976ui}
R.D.~Klem, J.E.~Bowers, H.W.~Courant, H.~Kagan, M.L.~Marshak, E.A.~Peterson
  et~al., \emph{{Measurement of Asymmetries of Inclusive Pion Production in
  Proton Proton Interactions at 6~GeV/c and 11.8~GeV/c}},
  \href{https://doi.org/10.1103/PhysRevLett.36.929}{\emph{Phys. Rev. Lett.}
  {\bfseries 36} (1976) 929}.

\bibitem{Dragoset:1978gg}
W.H.~Dragoset, J.B.~Roberts, J.E.~Bowers, H.W.~Courant, H.~Kagan, M.L.~Marshak
  et~al., \emph{{Asymmetries in Inclusive Proton-Nucleon Scattering at
  11.75~GeV/c}}, \href{https://doi.org/10.1103/PhysRevD.18.3939}{\emph{Phys.
  Rev. D} {\bfseries 18} (1978) 3939}.

\bibitem{Bunce:1976yb}
G.~Bunce et~al., \emph{{$\Lambda^0$ Hyperon Polarization in Inclusive
  Production by 300~GeV Protons on Beryllium}},
  \href{https://doi.org/10.1103/PhysRevLett.36.1113}{\emph{Phys. Rev. Lett.}
  {\bfseries 36} (1976) 1113}.

\bibitem{Schachinger:1978qs}
L.~Schachinger et~al., \emph{{A Precise Measurement of the $\Lambda^0$ Magnetic
  Moment}}, \href{https://doi.org/10.1103/PhysRevLett.41.1348}{\emph{Phys. Rev.
  Lett.} {\bfseries 41} (1978) 1348}.

\bibitem{Heller:1978ty}
K.J.~Heller et~al., \emph{{Polarization of $\Lambda$'s and $\bar\Lambda$'s
  Produced by 400~GeV Protons}},
  \href{https://doi.org/10.1103/PhysRevLett.41.607}{\emph{Phys. Rev. Lett.}
  {\bfseries 41} (1978) 607}.

\bibitem{Adams:1991cs}
{\scshape FNAL-E704} collaboration, \emph{{Analyzing power in inclusive $\pi^+$
  and $\pi^-$ production at high $x_F$ with a 200~GeV polarized proton beam}},
  \href{https://doi.org/10.1016/0370-2693(91)90378-4}{\emph{Phys. Lett. B}
  {\bfseries 264} (1991) 462}.

\bibitem{Adams:1991ru}
{\scshape E581} collaboration, \emph{{Large $x_F$ spin asymmetry in $\pi^0$
  production by 200~GeV polarized protons}},
  \href{https://doi.org/10.1007/BF01555512}{\emph{Z. Phys. C} {\bfseries 56}
  (1992) 181}.

\bibitem{Adams:1991rx}
{\scshape E581} collaboration, \emph{{First results for the two spin parameter
  $A_{LL}$ in $\pi^0$ production by 200~GeV polarized protons and
  anti-protons}},
  \href{https://doi.org/10.1016/0370-2693(91)91350-5}{\emph{Phys. Lett.}
  {\bfseries B261} (1991) 197}.

\bibitem{Bravar:1996ki}
{\scshape Fermilab E704} collaboration, \emph{{Single-spin asymmetries in
  inclusive charged pion production by transversely polarized antiprotons}},
  \href{https://doi.org/10.1103/PhysRevLett.77.2626}{\emph{Phys. Rev. Lett.}
  {\bfseries 77} (1996) 2626}.

\bibitem{Adams:2003fx}
{\scshape STAR} collaboration, \emph{{Cross sections and transverse single-spin
  asymmetries in forward neutral pion production from proton collisions at
  $\sqrt s = 200$~GeV}},
  \href{https://doi.org/10.1103/PhysRevLett.92.171801}{\emph{Phys. Rev. Lett.}
  {\bfseries 92} (2004) 171801}
  [\href{https://arxiv.org/abs/hep-ex/0310058}{{\ttfamily hep-ex/0310058}}].

\bibitem{Adams:2006uz}
{\scshape STAR} collaboration, \emph{{Forward neutral pion production in p+p
  and d+Au collisions at $\sqrt{s_{NN}} = 200$~GeV}},
  \href{https://doi.org/10.1103/PhysRevLett.97.152302}{\emph{Phys. Rev. Lett.}
  {\bfseries 97} (2006) 152302}
  [\href{https://arxiv.org/abs/nucl-ex/0602011}{{\ttfamily nucl-ex/0602011}}].

\bibitem{PHENIX:2003qdw}
{\scshape PHENIX} collaboration, \emph{{Absence of suppression in particle
  production at large transverse momentum in $\sqrt{s_{NN}} = 200$~GeV d + Au
  collisions}},
  \href{https://doi.org/10.1103/PhysRevLett.91.072303}{\emph{Phys. Rev. Lett.}
  {\bfseries 91} (2003) 072303}
  [\href{https://arxiv.org/abs/nucl-ex/0306021}{{\ttfamily nucl-ex/0306021}}].

\bibitem{Adler:2005in}
{\scshape PHENIX} collaboration, \emph{{Measurement of transverse single-spin
  asymmetries for mid-rapidity production of neutral pions and charged hadrons
  in polarized $p+p$ collisions at $\sqrt{s} = 200$~GeV}},
  \href{https://doi.org/10.1103/PhysRevLett.95.202001}{\emph{Phys. Rev. Lett.}
  {\bfseries 95} (2005) 202001}
  [\href{https://arxiv.org/abs/hep-ex/0507073}{{\ttfamily hep-ex/0507073}}].

\bibitem{:2008mi}
{\scshape BRAHMS} collaboration, \emph{{Single Transverse Spin Asymmetries of
  Identified Charged Hadrons in Polarized p+p Collisions at $\sqrt{s}$ = 62.4
  GeV}}, \href{https://doi.org/10.1103/PhysRevLett.101.042001}{\emph{Phys. Rev.
  Lett.} {\bfseries 101} (2008) 042001}
  [\href{https://arxiv.org/abs/0801.1078}{{\ttfamily 0801.1078}}].

\bibitem{BRAHMS:2007tyt}
{\scshape BRAHMS} collaboration, \emph{{Production of mesons and baryons at
  high rapidity and high $p_T$ in proton-proton collisions at $\sqrt s =
  200$~GeV}}, \href{https://doi.org/10.1103/PhysRevLett.98.252001}{\emph{Phys.
  Rev. Lett.} {\bfseries 98} (2007) 252001}
  [\href{https://arxiv.org/abs/hep-ex/0701041}{{\ttfamily hep-ex/0701041}}].

\bibitem{Lundberg:1989hw}
B.~Lundberg et~al., \emph{{Polarization in Inclusive $\Lambda$ and
  $\bar{\Lambda}$ Production at Large $p_T$}},
  \href{https://doi.org/10.1103/PhysRevD.40.3557}{\emph{Phys. Rev. D}
  {\bfseries 40} (1989) 3557}.

\bibitem{Ramberg:1994tk}
E.J.~Ramberg et~al., \emph{{Polarization of $\Lambda$ and $\bar\Lambda$
  produced by 800~GeV protons}},
  \href{https://doi.org/10.1016/0370-2693(94)91397-8}{\emph{Phys. Lett. B}
  {\bfseries 338} (1994) 403}.

\bibitem{Fanti:1998px}
V.~Fanti et~al., \emph{{A Measurement of the transverse polarization of
  $\Lambda$ hyperons produced in inelastic $p N$ reactions at 450~GeV proton
  energy}}, \href{https://doi.org/10.1007/s100520050337}{\emph{Eur. Phys. J. C}
  {\bfseries 6} (1999) 265}.

\bibitem{Abt:2006da}
{\scshape HERA-B} collaboration, \emph{{Polarization of $\Lambda$ and
  $\bar\Lambda$ in 920~GeV fixed-target proton-nucleus collisions}},
  \href{https://doi.org/10.1016/j.physletb.2006.05.040}{\emph{Phys. Lett. B}
  {\bfseries 638} (2006) 415}
  [\href{https://arxiv.org/abs/hep-ex/0603047}{{\ttfamily hep-ex/0603047}}].

\bibitem{Erhan:1979xm}
S.~Erhan et~al., \emph{{$\Lambda^0$ Polarization in Proton Proton Interactions
  at $\sqrt{s}=53$~GeV and 62~{GeV}}},
  \href{https://doi.org/10.1016/0370-2693(79)90761-5}{\emph{Phys. Lett. B}
  {\bfseries 82} (1979) 301}.

\bibitem{Kane:1978nd}
G.L.~Kane, J.~Pumplin and W.~Repko, \emph{{Transverse Quark Polarization in
  Large $p_T$ Reactions, $e^+ e^-$ Jets, and Leptoproduction: A Test of QCD}},
  \href{https://doi.org/10.1103/PhysRevLett.41.1689}{\emph{Phys. Rev. Lett.}
  {\bfseries 41} (1978) 1689}.

\bibitem{HERMES:2006lro}
{\scshape HERMES} collaboration, \emph{{Longitudinal Spin Transfer to the
  $\Lambda$ Hyperon in Semi-Inclusive Deep-Inelastic Scattering}},
  \href{https://doi.org/10.1103/PhysRevD.74.072004}{\emph{Phys. Rev. D}
  {\bfseries 74} (2006) 072004}
  [\href{https://arxiv.org/abs/hep-ex/0607004}{{\ttfamily hep-ex/0607004}}].

\bibitem{HERMES:2007fpi}
{\scshape HERMES} collaboration, \emph{{Transverse Polarization of $\Lambda$
  and $\bar\Lambda$ Hyperons in Quasireal Photoproduction}},
  \href{https://doi.org/10.1103/PhysRevD.76.092008}{\emph{Phys. Rev. D}
  {\bfseries 76} (2007) 092008}
  [\href{https://arxiv.org/abs/0704.3133}{{\ttfamily 0704.3133}}].

\bibitem{HERMES:2014fmx}
{\scshape HERMES} collaboration, \emph{{Transverse polarization of $\Lambda$
  hyperons from quasireal photoproduction on nuclei}},
  \href{https://doi.org/10.1103/PhysRevD.90.072007}{\emph{Phys. Rev. D}
  {\bfseries 90} (2014) 072007}
  [\href{https://arxiv.org/abs/1406.3236}{{\ttfamily 1406.3236}}].

\bibitem{NOMAD:2000wdf}
{\scshape NOMAD} collaboration, \emph{{Measurement of the $\Lambda$
  polarization in $\nu_\mu$ charged current interactions in the NOMAD
  experiment}},
  \href{https://doi.org/10.1016/S0550-3213(00)00503-4}{\emph{Nucl. Phys. B}
  {\bfseries 588} (2000) 3}.

\bibitem{NOMAD:2001iup}
{\scshape NOMAD} collaboration, \emph{{Measurement of the $\bar\Lambda$
  polarization in $\nu_\mu$ charged current interactions in the NOMAD
  experiment}},
  \href{https://doi.org/10.1016/S0550-3213(01)00181-X}{\emph{Nucl. Phys. B}
  {\bfseries 605} (2001) 3}
  [\href{https://arxiv.org/abs/hep-ex/0103047}{{\ttfamily hep-ex/0103047}}].

\bibitem{Guan:2018ckx}
{\scshape Belle} collaboration, \emph{{Observation of Transverse
  $\Lambda/\bar{\Lambda}$ Hyperon Polarization in $e^+e^-$ Annihilation at
  Belle}}, \href{https://doi.org/10.1103/PhysRevLett.122.042001}{\emph{Phys.
  Rev. Lett.} {\bfseries 122} (2019) 042001}
  [\href{https://arxiv.org/abs/1808.05000}{{\ttfamily 1808.05000}}].

\bibitem{OPAL:1997oem}
{\scshape OPAL} collaboration, \emph{{Polarization and forward - backward
  asymmetry of $\Lambda$ baryons in hadronic $Z^0$ decays}},
  \href{https://doi.org/10.1007/s100520050123}{\emph{Eur. Phys. J. C}
  {\bfseries 2} (1998) 49}
  [\href{https://arxiv.org/abs/hep-ex/9708027}{{\ttfamily hep-ex/9708027}}].

\bibitem{Kang:2020yqw}
Z.-B.~Kang, D.Y.~Shao and F.~Zhao, \emph{{QCD resummation on single hadron
  transverse momentum distribution with the thrust axis}},
  \href{https://doi.org/10.1007/JHEP12(2020)127}{\emph{JHEP} {\bfseries 12}
  (2020) 127} [\href{https://arxiv.org/abs/2007.14425}{{\ttfamily
  2007.14425}}].

\bibitem{Gamberg:2021iat}
L.~Gamberg, Z.-B.~Kang, D.Y.~Shao, J.~Terry and F.~Zhao, \emph{{Transverse
  $\Lambda$ polarization in $e^+ e^-$ collisions}},
  \href{https://doi.org/10.1016/j.physletb.2021.136371}{\emph{Phys. Lett. B}
  {\bfseries 818} (2021) 136371}
  [\href{https://arxiv.org/abs/2102.05553}{{\ttfamily 2102.05553}}].

\bibitem{DAlesio:2020wjq}
U.~D'Alesio, F.~Murgia and M.~Zaccheddu, \emph{{First extraction of the
  $\Lambda$ polarizing fragmentation function from Belle $e^+e^-$ data}},
  \href{https://doi.org/10.1103/PhysRevD.102.054001}{\emph{Phys. Rev. D}
  {\bfseries 102} (2020) 054001}
  [\href{https://arxiv.org/abs/2003.01128}{{\ttfamily 2003.01128}}].

\bibitem{Callos:2020qtu}
D.~Callos, Z.-B.~Kang and J.~Terry, \emph{{Extracting the transverse momentum
  dependent polarizing fragmentation functions}},
  \href{https://doi.org/10.1103/PhysRevD.102.096007}{\emph{Phys. Rev. D}
  {\bfseries 102} (2020) 096007}
  [\href{https://arxiv.org/abs/2003.04828}{{\ttfamily 2003.04828}}].

\bibitem{Anselmino:2000vs}
M.~Anselmino, D.~Boer, U.~D'Alesio and F.~Murgia, \emph{{$\Lambda$ polarization
  from unpolarized quark fragmentation}},
  \href{https://doi.org/10.1103/PhysRevD.63.054029}{\emph{Phys. Rev. D}
  {\bfseries 63} (2001) 054029}
  [\href{https://arxiv.org/abs/hep-ph/0008186}{{\ttfamily hep-ph/0008186}}].

\bibitem{Makris:2020ltr}
Y.~Makris, F.~Ringer and W.J.~Waalewijn, \emph{{Joint thrust and TMD
  resummation in electron-positron and electron-proton collisions}},
  \href{https://doi.org/10.1007/JHEP02(2021)070}{\emph{JHEP} {\bfseries 02}
  (2021) 070} [\href{https://arxiv.org/abs/2009.11871}{{\ttfamily
  2009.11871}}].

\bibitem{Boglione:2020auc}
M.~Boglione and A.~Simonelli, \emph{{Factorization of $e^+e^- \to H \, X$ cross
  section, differential in $z_h$, $P_T$ and thrust, in the $2$-jet limit}},
  \href{https://doi.org/10.1007/JHEP02(2021)076}{\emph{JHEP} {\bfseries 02}
  (2021) 076} [\href{https://arxiv.org/abs/2011.07366}{{\ttfamily
  2011.07366}}].

\bibitem{Boglione:2020cwn}
M.~Boglione and A.~Simonelli, \emph{{Universality-breaking effects in $e^+e^-$
  hadronic production processes}},
  \href{https://doi.org/10.1140/epjc/s10052-020-08821-y}{\emph{Eur. Phys. J. C}
  {\bfseries 81} (2021) 96} [\href{https://arxiv.org/abs/2007.13674}{{\ttfamily
  2007.13674}}].

\bibitem{Boglione:2021wov}
M.~Boglione and A.~Simonelli, \emph{{Kinematic regions in the $e^+e^- \to h \,
  X$ factorized cross section in a $2$-jet topology with thrust}},
  \href{https://doi.org/10.1007/JHEP02(2022)013}{\emph{JHEP} {\bfseries 02}
  (2022) } [\href{https://arxiv.org/abs/2109.11497}{{\ttfamily 2109.11497}}].

\bibitem{DAlesio:2021dcx}
U.~D'Alesio, F.~Murgia and M.~Zaccheddu, \emph{{General helicity formalism for
  two-hadron production in $e^+e^-$ annihilation within a TMD
  approach}}, \href{https://doi.org/10.1007/JHEP10(2021)078}{\emph{JHEP}
  {\bfseries 10} (2021) 078}
  [\href{https://arxiv.org/abs/2108.05632}{{\ttfamily 2108.05632}}].

\bibitem{Boer:1997mf}
D.~Boer, R.~Jakob and P.J.~Mulders, \emph{{Asymmetries in polarized hadron
  production in $e^+ e^-$ annihilation up to order $1/Q$}},
  \href{https://doi.org/10.1016/S0550-3213(97)00456-2}{\emph{Nucl. Phys. B}
  {\bfseries 504} (1997) 345}
  [\href{https://arxiv.org/abs/hep-ph/9702281}{{\ttfamily hep-ph/9702281}}].

\bibitem{Pitonyak:2013dsu}
D.~Pitonyak, M.~Schlegel and A.~Metz, \emph{{Polarized hadron pair production
  from electron-positron annihilation}},
  \href{https://doi.org/10.1103/PhysRevD.89.054032}{\emph{Phys. Rev. D}
  {\bfseries 89} (2014) 054032}
  [\href{https://arxiv.org/abs/1310.6240}{{\ttfamily 1310.6240}}].

\bibitem{Bacchetta:2004jz}
A.~Bacchetta, U.~D'Alesio, M.~Diehl and C.A.~Miller, \emph{Single-spin
  asymmetries: The {T}rento conventions}, \href{https://doi.org/10.1103/PhysRevD.70.117504}{\emph{Phys. Rev. D} {\bfseries 70}
  (2004) 117504} [\href{https://arxiv.org/abs/hep-ph/0410050}{{\ttfamily
  hep-ph/0410050}}].

\bibitem{Collins:2016hqq}
J.~Collins, L.~Gamberg, A.~Prokudin, T.C.~Rogers, N.~Sato and B.~Wang,
  \emph{{Relating Transverse Momentum Dependent and Collinear Factorization
  Theorems in a Generalized Formalism}},
  \href{https://doi.org/10.1103/PhysRevD.94.034014}{\emph{Phys. Rev. D}
  {\bfseries 94} (2016) 034014}
  [\href{https://arxiv.org/abs/1605.00671}{{\ttfamily 1605.00671}}].

\bibitem{collins_2011}
J.~Collins, \emph{Foundations of Perturbative QCD}, Cambridge Monographs on
  Particle Physics, Nuclear Physics and Cosmology, Cambridge University Press
  (2011),
  \href{https://doi.org/10.1017/CBO9780511975592}{10.1017/CBO9780511975592}.

\bibitem{Collins:2014jpa}
J.~Collins and T.~Rogers, \emph{{Understanding the large-distance behavior of
  transverse-momentum-dependent parton densities and the Collins-Soper
  evolution kernel}},
  \href{https://doi.org/10.1103/PhysRevD.91.074020}{\emph{Phys. Rev. D}
  {\bfseries 91} (2015) 074020}
  [\href{https://arxiv.org/abs/1412.3820}{{\ttfamily 1412.3820}}].

\bibitem{Boer:2010ya}
D.~Boer, Z.-B.~Kang, W.~Vogelsang and F.~Yuan, \emph{{Test of the Universality
  of Naive-time-reversal-odd Fragmentation Functions}},
  \href{https://doi.org/10.1103/PhysRevLett.105.202001}{\emph{Phys. Rev. Lett.}
  {\bfseries 105} (2010) 202001}
  [\href{https://arxiv.org/abs/1008.3543}{{\ttfamily 1008.3543}}].

\bibitem{Qiu:1991pp}
J.-w.~Qiu and G.F.~Sterman, \emph{{Single transverse spin asymmetries}},
  \href{https://doi.org/10.1103/PhysRevLett.67.2264}{\emph{Phys. Rev. Lett.}
  {\bfseries 67} (1991) 2264}.

\bibitem{Qiu:1991wg}
J.-w.~Qiu and G.F.~Sterman, \emph{{Single transverse spin asymmetries in direct
  photon production}},
  \href{https://doi.org/10.1016/0550-3213(92)90003-T}{\emph{Nucl. Phys. B}
  {\bfseries 378} (1992) 52}.

\bibitem{Collins:1984kg}
J.C.~Collins, D.E.~Soper and G.~Sterman, \emph{{Transverse Momentum
  Distribution in Drell-Yan Pair and $W$ and $Z$ Boson Production}},
  \href{https://doi.org/10.1016/0550-3213(85)90479-1}{\emph{Nucl. Phys. B}
  {\bfseries 250} (1985) 199}.

\bibitem{Aidala:2014hva}
C.A.~Aidala, B.~Field, L.P.~Gamberg and T.C.~Rogers, \emph{{Limits on
  transverse momentum dependent evolution from semi-inclusive deep inelastic
  scattering at moderate $Q$}},
  \href{https://doi.org/10.1103/PhysRevD.89.094002}{\emph{Phys. Rev. D}
  {\bfseries 89} (2014) 094002}
  [\href{https://arxiv.org/abs/1401.2654}{{\ttfamily 1401.2654}}].

\bibitem{Sun:2014dqm}
P.~Sun, J.~Isaacson, C.P.~Yuan and F.~Yuan, \emph{{Nonperturbative functions
  for SIDIS and Drell\textendash{}Yan processes}},
  \href{https://doi.org/10.1142/S0217751X18410063}{\emph{Int. J. Mod. Phys. A}
  {\bfseries 33} (2018) 1841006}
  [\href{https://arxiv.org/abs/1406.3073}{{\ttfamily 1406.3073}}].

\bibitem{Landry:2002ix}
F.~Landry, R.~Brock, P.M.~Nadolsky and C.P.~Yuan, \emph{{Tevatron Run-1 $Z$
  boson data and Collins-Soper-Sterman resummation formalism}},
  \href{https://doi.org/10.1103/PhysRevD.67.073016}{\emph{Phys. Rev. D}
  {\bfseries 67} (2003) 073016}
  [\href{https://arxiv.org/abs/hep-ph/0212159}{{\ttfamily hep-ph/0212159}}].

\bibitem{Konychev:2005iy}
A.V.~Konychev and P.M.~Nadolsky, \emph{{Universality of the
  Collins-Soper-Sterman nonperturbative function in gauge boson production}},
  \href{https://doi.org/10.1016/j.physletb.2005.12.063}{\emph{Phys. Lett. B}
  {\bfseries 633} (2006) 710}
  [\href{https://arxiv.org/abs/hep-ph/0506225}{{\ttfamily hep-ph/0506225}}].

\bibitem{Bacchetta:2017gcc}
A.~Bacchetta, F.~Delcarro, C.~Pisano, M.~Radici and A.~Signori,
  \emph{{Extraction of partonic transverse momentum distributions from
  semi-inclusive deep-inelastic scattering, Drell-Yan and Z-boson production}},
  \href{https://doi.org/10.1007/JHEP06(2017)081}{\emph{JHEP} {\bfseries 06}
  (2017) 081} [\href{https://arxiv.org/abs/1703.10157}{{\ttfamily
  1703.10157}}].

\bibitem{Bacchetta:2022awv}
A.~Bacchetta, V.~Bertone, C.~Bissolotti, G.~Bozzi, M.~Cerutti, F.~Piacenza
  et~al., \emph{{Unpolarized Transverse Momentum Distributions from a global
  fit of Drell-Yan and Semi-Inclusive Deep-Inelastic Scattering data}},
  [\href{https://arxiv.org/abs/2206.07598}{{\ttfamily 2206.07598}}].

\bibitem{Boglione:2017jlh}
M.~Boglione, J.O.~Gonzalez-Hernandez and R.~Taghavi, \emph{{Transverse parton
  momenta in single inclusive hadron production in ${e^ + }{e^ - }$
  annihilation processes}},
  \href{https://doi.org/10.1016/j.physletb.2017.06.034}{\emph{Phys. Lett.}
  {\bfseries B772} (2017) 78}
  [\href{https://arxiv.org/abs/1704.08882}{{\ttfamily 1704.08882}}].

\bibitem{Boglione:2022nzq}
M.~Boglione, J.O.~Gonzalez-Hernandez and A.~Simonelli, \emph{{Transverse
  Momentum Dependent Fragmentation Functions from recent BELLE data}},
  [\href{https://arxiv.org/abs/2206.08876}{{\ttfamily 2206.08876}}].

\bibitem{Becher:2015hka}
T.~Becher, M.~Neubert, L.~Rothen and D.Y.~Shao, \emph{{Effective Field Theory
  for Jet Processes}},
  \href{https://doi.org/10.1103/PhysRevLett.116.192001}{\emph{Phys. Rev. Lett.}
  {\bfseries 116} (2016) 192001}
  [\href{https://arxiv.org/abs/1508.06645}{{\ttfamily 1508.06645}}].

\bibitem{Becher:2016mmh}
T.~Becher, M.~Neubert, L.~Rothen and D.Y.~Shao, \emph{{Factorization and
  Resummation for Jet Processes}},
  \href{https://doi.org/10.1007/JHEP11(2016)019}{\emph{JHEP} {\bfseries 11}
  (2016) 019} [\href{https://arxiv.org/abs/1605.02737}{{\ttfamily
  1605.02737}}].

\bibitem{Becher:2016omr}
T.~Becher, B.D.~Pecjak and D.Y.~Shao, \emph{{Factorization for the light-jet
  mass and hemisphere soft function}},
  \href{https://doi.org/10.1007/JHEP12(2016)018}{\emph{JHEP} {\bfseries 12}
  (2016) 018} [\href{https://arxiv.org/abs/1610.01608}{{\ttfamily
  1610.01608}}].

\bibitem{Becher:2017nof}
T.~Becher, R.~Rahn and D.Y.~Shao, \emph{{Non-global and rapidity logarithms in
  narrow jet broadening}},
  \href{https://doi.org/10.1007/JHEP10(2017)030}{\emph{JHEP} {\bfseries 10}
  (2017) 030} [\href{https://arxiv.org/abs/1708.04516}{{\ttfamily
  1708.04516}}].

\bibitem{Caron-Huot:2015bja}
S.~Caron-Huot, \emph{{Resummation of non-global logarithms and the BFKL
  equation}}, \href{https://doi.org/10.1007/JHEP03(2018)036}{\emph{JHEP}
  {\bfseries 03} (2018) 036}
  [\href{https://arxiv.org/abs/1501.03754}{{\ttfamily 1501.03754}}].

\bibitem{Nagy:2016pwq}
Z.~Nagy and D.E.~Soper, \emph{{Summing threshold logs in a parton shower}},
  \href{https://doi.org/10.1007/JHEP10(2016)019}{\emph{JHEP} {\bfseries 10}
  (2016) 019} [\href{https://arxiv.org/abs/1605.05845}{{\ttfamily
  1605.05845}}].

\bibitem{Nagy:2017ggp}
Z.~Nagy and D.E.~Soper, \emph{{What is a parton shower?}},
  \href{https://doi.org/10.1103/PhysRevD.98.014034}{\emph{Phys. Rev. D}
  {\bfseries 98} (2018) 014034}
  [\href{https://arxiv.org/abs/1705.08093}{{\ttfamily 1705.08093}}].

\bibitem{Dasgupta:2001sh}
M.~Dasgupta and G.P.~Salam, \emph{{Resummation of nonglobal QCD observables}},
  \href{https://doi.org/10.1016/S0370-2693(01)00725-0}{\emph{Phys. Lett. B}
  {\bfseries 512} (2001) 323}
  [\href{https://arxiv.org/abs/hep-ph/0104277}{{\ttfamily hep-ph/0104277}}].

\bibitem{Albino:2008fy}
S.~Albino, B.A.~Kniehl and G.~Kramer, \emph{{AKK Update: Improvements from New
  Theoretical Input and Experimental Data}},
  \href{https://doi.org/10.1016/j.nuclphysb.2008.05.017}{\emph{Nucl. Phys. B}
  {\bfseries 803} (2008) 42} [\href{https://arxiv.org/abs/0803.2768}{{\ttfamily
  0803.2768}}].

\bibitem{deFlorian:2007aj}
D.~de~Florian, R.~Sassot and M.~Stratmann, \emph{Global analysis of
  fragmentation functions for pions and kaons and their uncertainties},
  \href{https://doi.org/10.1103/PhysRevD.75.114010}{\emph{Phys. Rev. D} {\bfseries 75} (2007) 114010}
  [\href{https://arxiv.org/abs/hep-ph/0703242}{{\ttfamily hep-ph/0703242}}].

\bibitem{Anselmino:2005nn}
M.~Anselmino, M.~Boglione, U.~D'Alesio, A.~Kotzinian, F.~Murgia and
  A.~Prokudin, \emph{{The Role of Cahn and Sivers effects in deep inelastic
  scattering}}, \href{https://doi.org/10.1103/PhysRevD.71.074006}{\emph{Phys.
  Rev. D} {\bfseries 71} (2005) 074006}
  [\href{https://arxiv.org/abs/hep-ph/0501196}{{\ttfamily hep-ph/0501196}}].

\bibitem{iminuit}
H.~Dembinski, P.~Ongmongkolkul, et~al., \emph{scikit-hep/iminuit}, .
\href{http://dx.doi.org/10.5281/zenodo.3949207}{\emph{10.5281/zenodo.3949207} (2020)}.


\bibitem{DBLP:journals/cphysics/KangPST21}
Z.~Kang, A.~Prokudin, N.~Sato and J.~Terry, \emph{Efficient {F}ourier
  transforms for transverse momentum dependent distributions},
  \href{https://doi.org/10.1016/j.cpc.2020.107611}{\emph{Comput. Phys. Commun.}
  {\bfseries 258} (2021) 107611}.

\bibitem{Anselmino:2008sga}
M.~Anselmino, M.~Boglione, U.~D'Alesio, A.~Kotzinian, S.~Melis, F.~Murgia
  et~al., \emph{{Sivers Effect for Pion and Kaon Production in Semi-Inclusive
  Deep Inelastic Scattering}},
  \href{https://doi.org/10.1140/epja/i2008-10697-y}{\emph{Eur. Phys. J. A}
  {\bfseries 39} (2009) 89} [\href{https://arxiv.org/abs/0805.2677}{{\ttfamily
  0805.2677}}].

\bibitem{Chen:2021hdn}
K.-B.~Chen, Z.-T.~Liang, Y.-L.~Pan, Y.-K.~Song and S.-Y.~Wei, \emph{{Isospin
  Symmetry of Fragmentation Functions}},
  \href{https://doi.org/10.1016/j.physletb.2021.136217}{\emph{Phys. Lett. B}
  {\bfseries 816} (2021) 136217}
  [\href{https://arxiv.org/abs/2102.00658}{{\ttfamily 2102.00658}}].

\end{thebibliography}

\providecommand{\href}[2]{#2}\begingroup\raggedright\endgroup

\end{document}